\PassOptionsToPackage{pdfpagelabels=false}{hyperref}
\newcommand{\aref}[1]{\hyperref[#1]{Appendix \ref*{#1}}}
\documentclass{JFM-FLM_Au}

\usepackage{graphicx}
\usepackage{epstopdf,epsfig}
\usepackage{newtxtext}
\usepackage{newtxmath}
\usepackage{natbib}
\usepackage{xpatch}

\usepackage{orcidlink}
\usepackage{comment}
\usepackage{soul}
\usepackage[nameinlink,noabbrev]{cleveref}
\usepackage{xkeyval}
\usepackage{xcolor}
\usepackage{bigints}

\definecolor{ggreen}{rgb}{0.13,0.55,0.13}
\definecolor{ropurp}{rgb}{0.6,0.4,0.8}
\definecolor{cbpink}{rgb}{01,0.72,0.77}
\newcommand{\copyedit}[2][green]{{\colorlet{foo}{#1}\sethlcolor{foo}\hl{#2}}}
\newcommand{\rev}[1]{{{\color{red}#1}}}
\renewcommand\copyedit[1]{#1}
\renewcommand\rev[1]{#1}

\newcommand{\ol}[1]{\overline{#1}}
\newcommand{\A}[3]{{#1}^{#2}_{#3}}
\newcommand{\w}{\omega}
\newcommand{\iPi}{\mathit{\Pi}}
\newcommand{\iW}{\mathit{\Omega}}

\makeatletter
\xpatchcmd\NAT@citex
 {%
  \@citea\NAT@hyper@{%
    \NAT@nmfmt{\NAT@nm}%
    \hyper@natlinkbreak{\NAT@aysep\NAT@spacechar}{\@citeb\@extra@b@citeb}%
    \NAT@date
  }%
 }
 {%
  \@citea
  \NAT@nmfmt{\NAT@nm}%
  \NAT@aysep\NAT@spacechar
  \NAT@hyper@{\NAT@date}%
 }
 {}{}
\xpatchcmd\NAT@citex
 {%
  \@citea\NAT@hyper@{%
    \NAT@nmfmt{\NAT@nm}%
    \hyper@natlinkbreak{\NAT@spacechar\NAT@@open\if*#1*\else#1\NAT@spacechar\fi}%
    {\@citeb\@extra@b@citeb}%
    \NAT@date
  }%
 }
 {
  \@citea
    \NAT@nmfmt{\NAT@nm}%
    \NAT@spacechar\NAT@@open\if*#1*\else#1\NAT@spacechar\fi
    \NAT@hyper@{\NAT@date}%
 }
 {}{}
\makeatother

\makeatletter
\newlength{\sfp@hseplen}\newlength{\sfp@vseplen}
\define@cmdkey{subfigpos}[sfp@]{pos}[ul]{}
\define@cmdkey{subfigpos}[sfp@]{font}[\small]{}
\define@cmdkey{subfigpos}[sfp@]{vsep}[2\baselineskip]{\setlength{\sfp@vseplen}{\sfp@vsep}}
\define@cmdkey{subfigpos}[sfp@]{hsep}[10pt]{\setlength{\sfp@hseplen}{\sfp@hsep}}
\newcommand{\subfigimg}[4][,]{%
  \setkeys{Gin,subfigpos}{pos,font,vsep,hsep,#1}
  \setbox1=\hbox{\includegraphics[#4]{#3}}
  \ifnum\pdfstrcmp{\sfp@pos}{ul}=0
    \leavevmode\rlap{\usebox1}
    \rlap{\hspace*{\sfp@hsep}\raisebox{\dimexpr\ht1-\sfp@vsep}{\sfp@font{#2}}}
    \phantom{\usebox1}
  \else\ifnum\pdfstrcmp{\sfp@pos}{ur}=0
    \leavevmode\usebox1
    \llap{\raisebox{\dimexpr\ht1-\sfp@vsep}{\sfp@font{#2}}\hspace*{\sfp@hsep}}
  \else\ifnum\pdfstrcmp{\sfp@pos}{lr}=0
    \leavevmode\usebox1
    \llap{\raisebox{\sfp@vsep}{\sfp@font{#2}}\hspace*{\sfp@hsep}}
  \else
    \leavevmode\rlap{\usebox1}
    \rlap{\hspace*{\sfp@hseplen}\raisebox{\sfp@vsep}{\sfp@font{#2}}}
    \phantom{\usebox1}
  \fi\fi\fi
}
\makeatother

\lefttitle{R. Arun, M. Kamal, T. Colonius and P.L. Johnson}
\righttitle{Journal of Fluid Mechanics}

\title{Normality-based analysis of multiscale velocity gradients and energy transfer in direct and large-eddy simulations of isotropic turbulence}

\author{Rahul Arun\hyperlink{aff1}{\textsuperscript{\normalfont\small\aff{1}}}\orcidlink{0000-0002-5942-169X}, Mostafa Kamal\hyperlink{aff2}{\textsuperscript{\normalfont\small\aff{2}}}\orcidlink{0009-0000-3892-4987}, Tim Colonius\hyperlink{aff3}{\textsuperscript{\normalfont\small\aff{3}}}\orcidlink{0000-0003-0326-3909} \and Perry L. Johnson\hyperlink{aff2}{\textsuperscript{\normalfont\small\aff{2}}}\orcidlink{0000-0002-7929-9396}}

\affiliation{
\hypertarget{aff1}{\aff{1}}Graduate Aerospace Laboratories, California Institute of Technology, Pasadena, CA 91125, USA
\hypertarget{aff2}{\aff{2}}Department of Mechanical and Aerospace Engineering, University of California, Irvine, CA 92697, USA
\hypertarget{aff3}{\aff{3}}Department of Mechanical and Civil Engineering, California Institute of Technology, Pasadena, \\ CA 91125, USA
}

\corresau{Rahul Arun, \href{mailto:rarun@caltech.edu}{rarun@caltech.edu}}

\begin{document}
\maketitle

\begin{abstract}
Symmetry-based analyses of multiscale velocity gradients highlight that strain self-amplification \copyedit{(SS)} and vortex stretching \copyedit{(VS)} drive forward energy transfer in turbulent flows. By contrast, a strain--vorticity covariance mechanism produces backscatter that contributes to the bottleneck effect in the subinertial range of the energy cascade. We extend these analyses by using a normality-based decomposition of filtered velocity gradients in forced isotropic turbulence to distinguish contributions from normal straining, pure shearing and rigid rotation at a given scale. Our analysis of direct numerical simulation (DNS) data illuminates the importance of shear layers in the inertial range and (especially) the subinertial range of the cascade. Shear layers contribute significantly to \copyedit{SS} and \copyedit{VS} and play a dominant role in the backscatter mechanism responsible for the bottleneck effect. Our concurrent analysis of large-eddy simulation (LES) data characterizes how different closure models affect the flow structure and energy transfer throughout the resolved scales. We thoroughly demonstrate that the multiscale flow features produced by a mixed model closely resemble those in a filtered DNS, whereas the features produced by an eddy viscosity model resemble those in an unfiltered DNS at a lower Reynolds number. This analysis helps explain how small-scale shear layers, whose imprint is mitigated upon filtering, amplify the artificial bottleneck effect produced by the eddy viscosity model in the inertial range of the cascade. Altogether, the present results provide a refined interpretation of the flow structures and mechanisms underlying the energy cascade and insight for designing and evaluating LES closure models.
\end{abstract}



\newpage


\section{Introduction}\label{sec:intro}

Small-scale flow features form a cornerstone of efforts to understand and model turbulent flows \citep{Kol1941,Kol1962,Obo1962,Sre1997,Men2000}. The velocity gradient tensor (VGT), $\A{A}{}{ij} = \partial u_i^{} \big/ \partial x_j^{}$, is central to our understanding of small-scale turbulence since it encodes a linear approximation of the local flow structure about critical points \citep{Cho1990} and underpins descriptions of fundamental dynamical mechanisms \citep{Men2011,Joh2024}.

\subsection{Normality-based analysis of velocity gradients in turbulent flows}\label{sec:intro:normality}

Conventionally, the VGT is decomposed as
\begin{equation}\label{eq:VGT_sym}
    \A{A}{}{ij} = \A{S}{}{ij} + \A{\iW}{}{ij},
\end{equation}
where $\A{S}{}{ij} = \tfrac{1}{2} (\A{A}{}{ij} + \A{A}{}{ji})$ is the (symmetric) strain-rate tensor and $\A{\iW}{}{ij} = \tfrac{1}{2} (\A{A}{}{ij} - \A{A}{}{ji})$ is the (antisymmetric) vorticity tensor. This symmetry-based decomposition forms the basis for mechanisms like strain self-amplification \copyedit{(SS)} and vortex stretching \copyedit{(VS)}, vortex identification criteria like $Q$ \citep{Hun1988} and $\lambda_2$ \citep{Jeo1995}, and alignment analyses in the strain-rate eigenframe \citep{Els2010,Els2017}. However, despite its ubiquity, the symmetry-based decomposition provides relatively limited insight into local flow structure.

A more detailed description can be obtained by \copyedit{analysing} the normality properties of the VGT. These properties can be identified by considering a `principal' reference frame, denoted by $(\cdot)^*$, in which the VGT obtains a quasi-triangular form. This form can be expressed as $\A{A}{*}{ij} = \A{U}{}{ik} \A{A}{}{km} \A{U}{}{jm}$, where $\A{U}{}{ij}$ is an orthogonal matrix that can be interpreted as a rotation of the local coordinate axes in physical space. In this principal frame, the normality-based decomposition of the VGT can be expressed as 
\begin{equation}\label{eq:VGT_nor}
    \A{A}{*}{ij} = \underbrace{\begin{bmatrix} \dot{\epsilon}_1^* & 0 & 0 \\ 0 & \dot{\epsilon}_2^* & 0 \\ 0 & 0 & \dot{\epsilon}_3^* \end{bmatrix}}_{\textstyle \A{S}{\epsilon*}{ij}} + \underbrace{\begin{bmatrix} 0 & \dot{\gamma}_3^* & \dot{\gamma}_2^* \\ 0 & 0 & \dot{\gamma}_1^* \\ 0 & 0 & 0 \end{bmatrix}}_{\textstyle \A{A}{\gamma*}{ij}} + \underbrace{\begin{bmatrix} 0 & 0 & 0 \\ 0 & 0 & \dot{\varphi}_1^* \\ 0 & -\dot{\varphi}_1^* & 0 \end{bmatrix}}_{\textstyle \A{\iW}{\varphi*}{ij}},
\end{equation}
where $\A{S}{\epsilon*}{ij}$, $\A{A}{\gamma*}{ij}$ and $\A{\iW}{\varphi*}{ij}$ denote the normal straining, pure shearing and rigid rotation tensors, respectively. The normal straining tensor is symmetric and normal whereas the rigid rotation tensor is antisymmetric and normal. The pure shearing tensor is non-normal and it can be further decomposed into the (symmetric) shear straining tensor, $\A{S}{\gamma}{ij} = \tfrac{1}{2} (\A{A}{\gamma}{ij} + \A{A}{\gamma}{ji})$, and the (antisymmetric) shear vorticity tensor, $\A{\iW}{\gamma}{ij} = \tfrac{1}{2} (\A{A}{\gamma}{ij} - \A{A}{\gamma}{ji})$. Therefore, (\ref{eq:VGT_nor}) can be viewed as a refinement of (\ref{eq:VGT_sym}) since $\A{S}{}{ij} = \A{S}{\epsilon}{ij} + \A{S}{\gamma}{ij}$ and $\A{\iW}{}{ij} = \A{\iW}{\varphi}{ij} + \A{\iW}{\gamma}{ij}$. The tensors in (\ref{eq:VGT_nor}) can be determined and transformed to the original coordinate system using the ordered real Schur form of the VGT \citep{Kro2023}. \citet{Aru2024b} provided an instructional code for implementing the normality-based decomposition, which is available at \href{https://doi.org/10.22002/17h15-gr910}{https://doi.org/10.22002/17h15-gr910}.

Identifying a frame in which normal straining, pure shearing and rigid rotation can be distinguished is a key advantage of the normality-based decomposition. \citet{Kol2004,Kol2007} originally distinguished these motions by identifying a `basic' reference frame in which pure shearing is extracted as a purely asymmetric tensor. While that approach typically produces results similar to the present normality-based decomposition, it requires solving a more challenging pointwise optimization problem that may yield non-unique solutions. Some related studies \citep{Key2018,Key2019,Key2025,Bea2019,Bil2025} \copyedit{analyse} the normality properties of the VGT using its complex Schur form, which is triangular (as opposed to quasi-triangular). While that approach resembles our approach for locally non-rotational points, where the VGT has three real eigenvalues, it differs from our approach for locally rotational points, where the VGT has a pair of complex conjugate eigenvalues. In the latter case, the complex Schur form requires a complex transformation matrix that cannot be interpreted in physical space. Regardless of the differences in these approaches, the results produced by the original triple decomposition and the complex Schur form motivate and complement those produced by the present normality-based framework.

The flow statistics associated with the normality-based decomposition of the VGT concisely express key features of small-scale turbulence. For example, the strength of velocity gradients can be partitioned as
\begin{equation}\label{eq:VGT_part}
    A^2 = \A{A}{}{ij}\A{A}{}{ij} = \underbrace{\A{S}{\epsilon}{ij}\A{S}{\epsilon}{ij}}_{\textstyle S_\epsilon^2} + \overbrace{\underbrace{\A{S}{\gamma}{ij}\A{S}{\gamma}{ij}}_{\textstyle S_\gamma^2} + \underbrace{\A{\iW}{\gamma}{ij}\A{\iW}{\gamma}{ij}}_{\textstyle \iW_\gamma^2}}^{\textstyle A_\gamma^2} + \underbrace{\A{\iW}{\varphi}{ij}\A{\iW}{\varphi}{ij}}_{\textstyle \iW_\varphi^2} + \underbrace{2\A{\iW}{\varphi}{ij}\A{\iW}{\gamma}{ij}}_{\textstyle \iW_{\varphi\gamma}^2},
\end{equation}
where $S_\epsilon^2$, $A_\gamma^2$ and $\iW_\varphi^2$ represent the strengths of the constituents in (\ref{eq:VGT_nor}), $\iW_{\varphi\gamma}^2$ represents shear--rotation correlations, and $S_\gamma^2 = \iW_\gamma^2 = \tfrac{1}{2} A_\gamma^2$. The relative contributions of $S_\epsilon^2$, $A_\gamma^2$, and $\iW_\varphi^2 + \iW_{\varphi\gamma}^2$ to $A^2$ can be expressed as exact algebraic functions of the normalized invariants of the VGT, whereas distinguishing the contributions of $\iW_\varphi^2$ and $\iW_{\varphi\gamma}^2$ requires an additional parameter \citep{Aru2024}. An averaged form of this partitioning can be expressed as
\begin{equation}\label{eq:VGT_part_avg}
   \left\langle A^2 \right\rangle = \left\langle S_\epsilon^2 \right\rangle + \underbrace{\left\langle S_\gamma^2 \right\rangle + \left\langle \iW_\gamma^2 \right\rangle}_{\textstyle \left\langle A_\gamma^2 \right\rangle} + \left\langle \iW_\varphi^2 \right\rangle + \left\langle \iW_{\varphi\gamma}^2 \right\rangle,
\end{equation}
where $\langle (\cdot) \rangle$ denotes ensemble averaging. This averaged partitioning distinguishes the contributions of normal straining and shear straining to $\left\langle S^2 \right\rangle$ and the contributions of shear vorticity, rigid rotation and their correlations to $\left\langle \iW^2 \right\rangle$ \citep{Das2020}. Recent work has shown that, for a broad class of flows, it is expressive of various flow features and regimes that are obscured by the symmetry-based partitioning into $\left\langle S^2 \right\rangle$ and $\left\langle \iW^2 \right\rangle$. For wall-bounded flows, \citet{Aru2024b} showed that it can distinguish between near-wall turbulence, which is dominated by shearing, and turbulence far from walls, which is more reminiscent of isotropic turbulence. Furthermore, \citet{Aru2024} showed that it can distinguish between the initial, transitional and turbulent regimes of a vortex ring collision. The normality-based analysis of that flow also identified enhanced shear--rotation correlations as an imprint of the elliptic instability that reflects relevant structural features of local streamlines.

Beyond statistical flow features, the normality-based decomposition can also distinguish tube-like vortical structures, which are associated with rigid rotation, from sheet-like vortical structures, which are associated with shear vorticity. This distinction underpins recently-developed vortex identification criteria that preferentially identify tubular vortical structures \citep{Gao2018,Liu2018,Liu2019}. However, recent evidence highlights that shear layers are also critical to the structure and dynamics of small-scale turbulence \citep{Nag2020,Wat2020,Wat2023}. The Burgers vortex layer forms a reasonable model for these (strained) small-scale shear layers, which typically have widths of $9\eta-11\eta$ \citep{Els2017,Nag2020,Fis2021} and half-widths of $4.5\eta$ \citep{Wat2020,Wat2023}, where $\eta$ is the Kolmogorov scale. Interestingly, the typical diameters of intense vortex tubes at small scales, which are often \copyedit{modelled} as Burgers vortex tubes, lie in a similar range \citep{Jim1993,Ghi2022}. In \aref{sec:app:Burgers}, we use the normality-based partitioning of Burgers vortex tubes and layers to illustrate how rigid rotation and shear vorticity are associated with vortex cores and shear layers, respectively.

Although vortex tubes and shear layers capture many essential features of Kolmogorov-scale flow structures, a complete `recipe' for turbulence must effectively capture its \textit{multiscale} structure. Therefore, motivated by the discussion in this section, the present study aims to characterize the structure of isotropic turbulence in the subinertial and inertial ranges using the enhanced statistical and structural expressivity of the normality-based velocity gradient analysis. 

\subsection{Multiscale velocity gradients and interscale energy transfer}\label{sec:intro:multiscale}

Spatial filtering frameworks enable tailored consideration of multiscale flow features by specifying a filter width, $\ell$, that controls the range of resolved scales. They also facilitate insights into closure models for large-eddy simulation (LES), which are often formulated in terms of the filtered (i.e. resolved) flow field \citep{Joh2024}. We denote filtered quantities as $\ol{(\cdot)}^\ell$ = $G_\ell \star (\cdot)$, where $G_\ell$ represents the filter kernel and $\star$ represents the spatial convolution operator. 

Applying a uniform filtering operation to the incompressible Navier--Stokes equations yields
\begin{equation}\label{eq:NSE_filt}
    \frac{\partial \ol{u}^\ell_i}{\partial t} + \ol{u}^\ell_j \frac{\partial \ol{u}^\ell_i}{\partial x_j^{}} = -\frac{1}{\rho} \frac{\partial \ol{p}^\ell}{\partial x_i^{}} + \nu \frac{\partial^2 \ol{u}^\ell_i}{\partial x_j^2} + \ol{f}^\ell_i - \frac{\partial \sigma^\ell_{ij}}{\partial x_j^{}}, \quad \frac{\partial \ol{u}^\ell_i}{\partial x_i^{}} = 0,
\end{equation}
where $\rho$, $\nu$ and $p$ denote the density, kinematic viscosity and pressure, respectively. The forcing, $f_i$, is typically designed to sustain the flow by injecting energy at large scales (e.g. in forced isotropic turbulence). The residual stress tensor, $\sigma^\ell_{ij}$, represents the effective stress imposed on the resolved motions by the unresolved motions. It is the subject of closure models in LES \citep{Men2000} and plays a key role in the interpretation of interscale energy transfer. The kinetic energy equation associated with the filtered velocity field is given by
\begin{equation}\label{eq:energy_filt}
    \frac{\partial E^\ell}{\partial t} + \frac{\partial T^\ell_i}{\partial x_i^{}} = \ol{u}^\ell_i \ol{f}^\ell_i - \iPi^\ell - \mathit{\Phi}^\ell,
\end{equation}
where $\partial T^\ell_i / \partial x_i^{}$, $\ol{u}^\ell_i \ol{f}^\ell_i$ and $\mathit{\Phi}^\ell = 2\nu\A{\ol{S}}{\ell}{ij}\A{\ol{S}}{\ell}{ij}$ represent spatial redistribution, energy injection by the forcing and the resolved dissipation rate, respectively. The term $\iPi^\ell = -\A{\ol{S}}{\ell}{ij} \sigma^\ell_{ij}$ represents the interscale energy transfer (or cascade rate) across scale $\ell$. By convention, $\iPi^\ell > 0$ and $\iPi^\ell < 0$ correspond to downscale and upscale energy transfer, respectively. For incompressible flows, the isotropic part of $\sigma^\ell_{ij}$ does not contribute to interscale energy transfer since $\A{\ol{S}}{\ell}{ij}$ is traceless.

The filtering framework for interscale energy transfer has provided sustained insight into salient features of the energy cascade \citep{Leo1975,Bor1998,Eyi2009,Bal2018}. Recent studies have employed conditional averaging to identify a statistical imprint of small-scale flow structures that contribute to forward energy transfer. Remarkably, in both isotropic \citep{Par2025} and shear \citep{Don2020} turbulence, this imprint manifests as a localized region of energy transfer located between hairpin-like vortical structures with opposite orientations. The filtering perspective of energy transfer has been extended to the settings of compressible turbulence \citep{Wan2018} and magnetohydrodynamic turbulence \citep{Alu2017,Ale2022,Cap2025}, among others, and it has aided the development of global maps of energy transfer in the ocean \citep{Alu2018,Sto2023}. Beyond filtering, energy transfer can also be \copyedit{analysed} using a structure function approach, which is formulated in terms of the scale-integrated local Kolmogorov--Hill equation \citep{Yao2024}. This approach provides a less ambiguous interpretation of upscale energy transfer (i.e. backscatter) than the filtering formulation. However, since both formulations typically produce similar results \citep{Car2020}, we focus on the filtering framework as it provides pragmatic implications for LES \copyedit{modelling}.

Recent work has begun to refine our understanding of interscale energy transfer by decomposing the VGT into contributions from normal straining, pure shearing and rigid rotation. \citet{Eno2023} found that the interaction of shearing with the residual stresses dominates interscale energy transfer over a broad range of scales in isotropic turbulence. Using a different formulation, \citet{Fat2024} concluded that the contributions of shearing and normal straining dominate spectral energy transfer in isotropic turbulence. Both of these results were obtained by decomposing the velocity gradient term in the expressions used to represent interscale energy transfer. \citet{Eno2023} also decomposed the residual stress tensor based on contributions from the shearing and non-shearing velocity fields, which were identified using the Biot--Savart law. Beyond these studies, there remains a significant gap in our understanding of how interscale energy transfer at scale $\ell$ is influenced by flow features at smaller scales. We address this gap, which has significant implications for LES \copyedit{modelling}, by \copyedit{analysing} multiscale contributions to the residual stress tensor using the normality-based decomposition of the VGT.

The present study utilizes Gaussian filtering, for which the filter kernel is given by
\begin{equation}\label{eq:Gaussian}
    G_\ell(\boldsymbol{r}) = \frac{1}{\left( 2\pi\ell^2 \right)^{3/2}}{\rm exp}\left( -\frac{\lvert \boldsymbol{r} \rvert^2}{2\ell^2} \right), \quad \mathcal{F}\left\{ G_\ell \right\}(\boldsymbol{k}) = {\rm exp}\left( -\frac{\lvert \boldsymbol{k}\rvert^2 \ell^2}{2} \right),
\end{equation}
where $\boldsymbol{r}$ and $\boldsymbol{k}$ represent the spatial offset and wavenumber vectors, respectively, and $\mathcal{F}\{\cdot\}$ represents the spatial Fourier transform. The Gaussian filtering framework allows the residual stress tensor to be expressed in terms of multiscale velocity gradients as
\begin{equation}\label{eq:res_VGT}
    \sigma^\ell_{ij} = \bigintsss_{\;0}^{\ell^2} {\rm d}\theta^2 \left( \ol{\A{\ol{A}}{\theta}{ik}\A{\ol{A}}{\theta}{jk}}^{\phi} \right),
\end{equation}
where $\phi = \sqrt{\ell^2 - \theta^2}$. As discussed in detail by \citet{Joh2020,Joh2021}, this expression is derived by treating the Gaussian-filtered velocity field as the solution to a diffusion equation where $\ell^2$ represents a time-like variable. This formulation leads to a forced diffusion equation for the residual stress tensor, for which (\ref{eq:res_VGT}) is the formal solution. While we limit our attention to Gaussian filtering, the results we cite have been shown to be insensitive to the filter shape for non-negative filter kernels \citep{Joh2020}.

Using (\ref{eq:res_VGT}), a multiscale velocity gradient expansion for interscale energy transfer can be expressed as
\begin{equation}\label{eq:iPi_VGT}
    \iPi^\ell = -\A{\ol{S}}{\ell}{ij} \bigintsss_{\;0}^{\ell^2} {\rm d}\theta^2 \left( \ol{\A{\ol{A}}{\theta}{ik}\A{\ol{A}}{\theta}{jk}}^{\phi} \right).
\end{equation}
Following \citet{Joh2020,Joh2021}, this expression can be decomposed in terms of familiar energy transfer mechanisms by inserting the symmetry-based decomposition of the VGT to obtain
\begin{equation}\label{eq:iPi_sym}
    \iPi^\ell = \iPi^{\ell,s} + \iPi^{\ell,\w} + \iPi^{\ell,c},
\end{equation}
where
\begin{alignat}{100}
    &\iPi^{\ell,s} &&= -\A{\ol{S}}{\ell}{ij} \bigintsss_{\;0}^{\ell^2} {\rm d}\theta^2 \left( \ol{\A{\ol{S}}{\theta}{ik}\A{\ol{S}}{\theta}{jk}}^{\phi} \right), \label{eq:iPi_s} \\
    &\iPi^{\ell,\w} &&= -\A{\ol{S}}{\ell}{ij} \bigintsss_{\;0}^{\ell^2} {\rm d}\theta^2 \left( \ol{\A{\ol{\iW}}{\theta}{ik}\A{\ol{\iW}}{\theta}{jk}}^{\phi} \right), \label{eq:iPi_w} \\
    &\iPi^{\ell,c} &&= -\A{\ol{S}}{\ell}{ij} \bigintsss_{\;0}^{\ell^2} {\rm d}\theta^2 \left( \ol{\A{\ol{S}}{\theta}{ik}\A{\ol{\iW}}{\theta}{jk}}^{\phi} + \ol{\A{\ol{\iW}}{\theta}{ik} \A{\ol{S}}{\theta}{jk}}^{\phi} \right). \label{eq:iPi_c}
\end{alignat}
Here, $\iPi^{\ell,s}$, $\iPi^{\ell,\w}$ and $\iPi^{\ell,c}$ represent the contributions of \copyedit{SS}, \copyedit{VS} and strain--vorticity covariance at scales $\theta \leq \ell$ to the interscale energy transfer at scale $\ell$. These terms can be further decomposed as
\begin{alignat}{100}
    &\iPi^{\ell,s} &&= \iPi^{\ell,s1} &&+ \iPi^{\ell,s2}, \label{eq:iPi_s_12_intro} \\
    &\iPi^{\ell,\w} &&= \iPi^{\ell,\w1} &&+ \iPi^{\ell,\w2}, \label{eq:iPi_w_12_intro} \\
    &\iPi^{\ell,c} &&= \iPi^{\ell,c1} &&+ \iPi^{\ell,c2}, \label{eq:iPi_c_12_intro}
\end{alignat}
where terms with superscripts of $(\cdot)^1$ and $(\cdot)^2$ represent scale-local and scale-\copyedit{non-local} contributions, respectively. \rev{These contributions are given by
\begin{alignat}{100}
    &\iPi^{\ell,s1} &&= -\ell^2 \A{\ol{S}}{\ell}{ij}\A{\ol{S}}{\ell}{ik}\A{\ol{S}}{\ell}{jk}, \quad &&\iPi^{\ell,s2} &&= \iPi^{\ell,s} &&- \iPi^{\ell,s1}, \label{eq:iPi_s_12} \\
    &\iPi^{\ell,\w1} &&= -\ell^2 \A{\ol{S}}{\ell}{ij}\A{\ol{\iW}}{\ell}{ik}\A{\ol{\iW}}{\ell}{jk}, \quad &&\iPi^{\ell,\w2} &&= \iPi^{\ell,\w} &&- \iPi^{\ell,\w1}, \label{eq:iPi_w_12} \\
    &\iPi^{\ell,c1} &&= -\ell^2 \A{\ol{S}}{\ell}{ij} \left( \A{\ol{S}}{\ell}{ik}\A{\ol{\iW}}{\ell}{jk} +  \A{\ol{\iW}}{\ell}{ik}\A{\ol{S}}{\ell}{jk} \right) = 0, \quad &&\iPi^{\ell,c2} &&= \iPi^{\ell,c} &&- \iPi^{\ell,c1} = \iPi^{\ell,c}, \label{eq:iPi_c_12}
\end{alignat}
and the scale-\copyedit{non-local} contributions can equivalently be expressed using scale-space integrals of generalized second moments of velocity gradient fields \citep{Joh2020,Joh2021}.} The scale-local strain--vorticity covariance term vanishes since it can be expressed as the contraction of symmetric tensors with antisymmetric tensors. Therefore, only \copyedit{SS} and \copyedit{VS} contribute to scale-local energy transfer, and their combined contribution resembles that produced by the nonlinear gradient model for the residual stress tensor \citep{Cla1979,Bor1998,Men2000}.

\begin{figure}
    \centering
    \subfigimg[width=0.49\linewidth,pos=ul,vsep=7pt,hsep=1.5pt]{(\textit{a})}{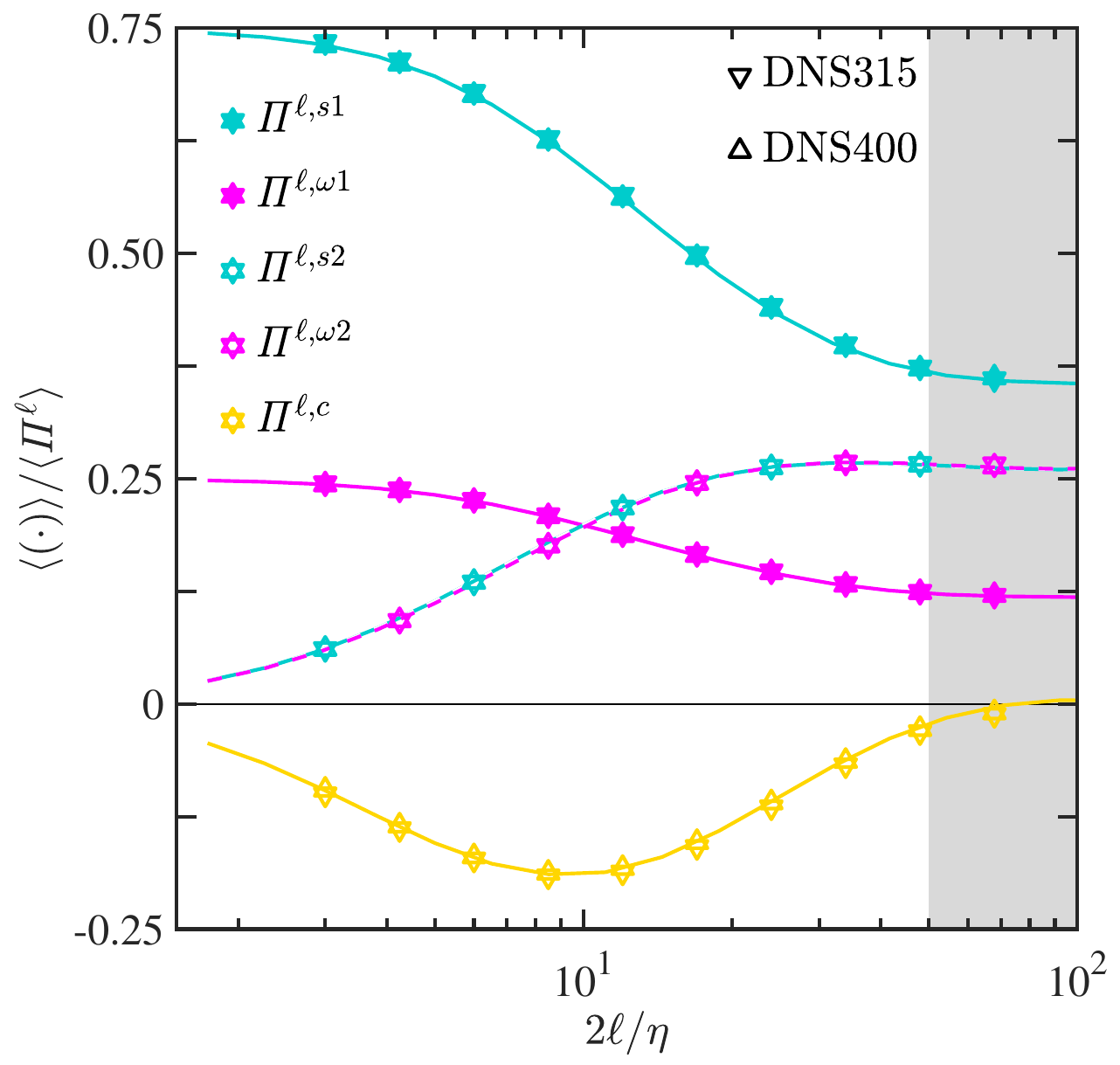}{}
    \subfigimg[width=0.49\linewidth,pos=ul,vsep=7pt,hsep=1.5pt]{(\textit{b})}{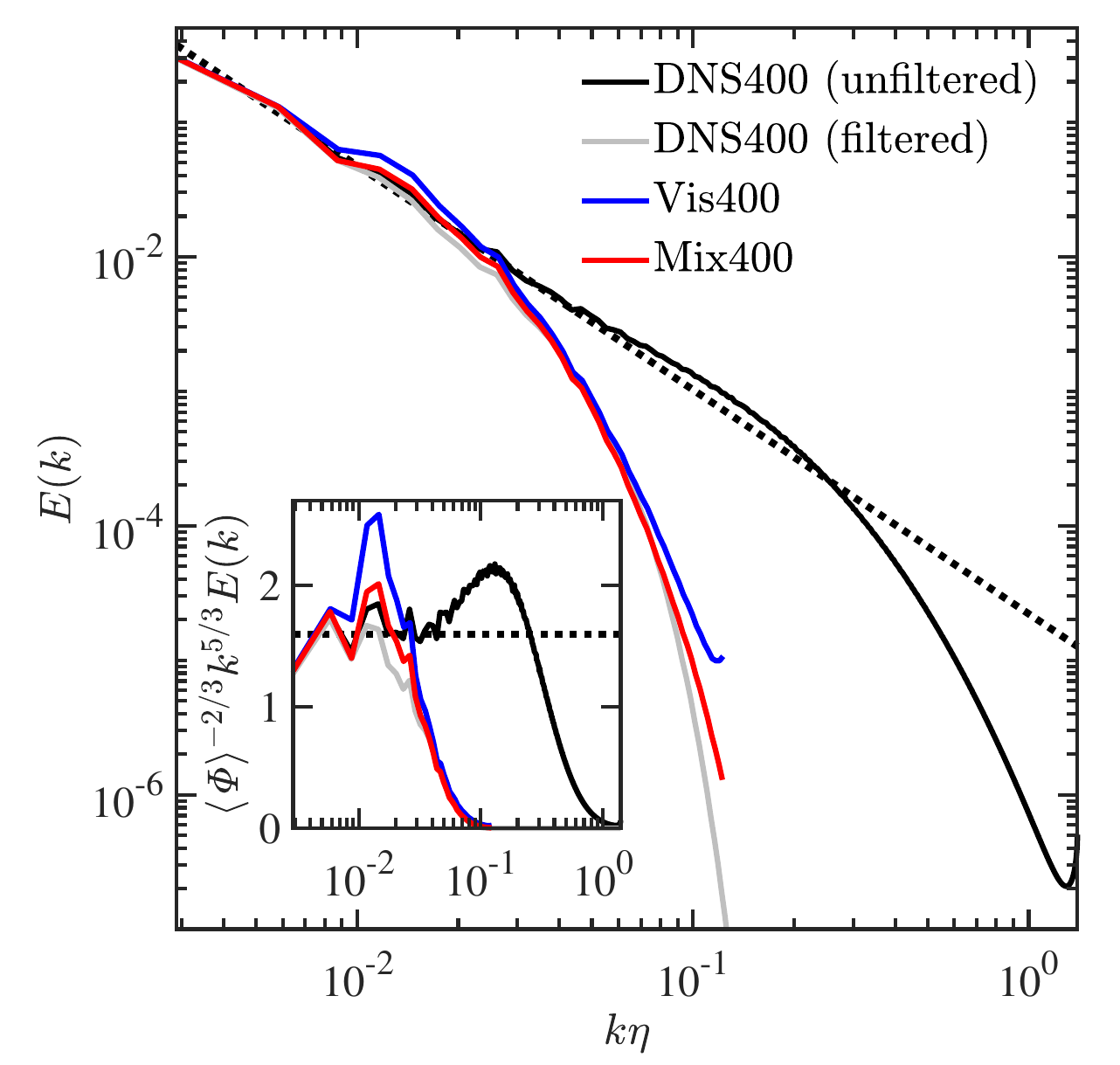}{}
    \caption{(\textit{a}) Symmetry-based scale-local and scale-\copyedit{non-local} contributions to interscale energy transfer in forced isotropic turbulence. The symbols represent \copyedit{direct numerical simulation (DNS)} datasets at Taylor-scale Reynolds numbers of $Re_\lambda \approx 315$ (DNS315) and $Re_\lambda \approx 400$ (DNS400) and the curves represent the $Re_\lambda \approx 400$ results of \citet{Joh2020,Joh2021}. The shaded region captures the bottom of the inertial range for DNS400. (\textit{b}) Energy spectra for the unfiltered and filtered velocity fields in DNS400 as well as LES cases that employ eddy viscosity (Vis400) and mixed (Mix400) models at $Re_\lambda \approx 400$. The filtered DNS and LES cases employ a filter width of $2\ell/\eta = 48$. The dotted line represents the inertial range scaling, $E(k) = 1.6 \left\langle \mathit{\Phi} \right\rangle^{2/3} k^{-5/3}$, and the inset depicts a linear--log plot of the compensated energy spectra. Technical details of the simulations are described in \textsection \ref{sec:sims}.}
    \label{fig:spectra}
\end{figure}

The resulting decomposition of interscale energy transfer,
\begin{equation}\label{eq:iPi_sym_12}
    \iPi^\ell = \iPi^{\ell,s1} + \iPi^{\ell,\w1} + \iPi^{\ell,s2} + \iPi^{\ell,\w2} + \iPi^{\ell,c},
\end{equation}
provides insight into mechanisms underlying the energy cascade. As shown in \cref{fig:spectra}(\textit{a}), the relative contributions of the constituents are approximately $\left\langle \iPi^{\ell,s1} \right\rangle : \left\langle \iPi^{\ell,\w1} \right\rangle : \left\langle \iPi^{\ell,s2} \right\rangle : \left\langle \iPi^{\ell,\w2} \right\rangle : \left\langle \iPi^{\ell,c} \right\rangle \approx 3:1:2:2:0$ in the inertial range for isotropic turbulence \citep{Joh2020,Joh2021}. The implication that $\left\langle \iPi^{\ell,s} \right\rangle : \left\langle \iPi^{\ell,\w} \right\rangle = 5:3$ supports the claim that \copyedit{SS} contributes more to the cascade than \copyedit{VS}, extending prior results that focused on single-scale contributions \citep{Car2020}. While the covariance term, $\iPi^{\ell,c}$, has a negligible net contribution in the inertial range of the cascade, it produces significant upscale energy transfer (i.e. backscatter) in the subinertial range. This feature bears resemblance to two-dimensional turbulence, where the covariance term is responsible for the inverse cascade \citep{Joh2021}.

This backscatter occurs at scales that coincide with the bottleneck effect in the subinertial range of the cascade. As shown in \cref{fig:spectra}(\textit{b}), the bottleneck effect in DNS manifests as a bump in the energy spectrum that exceeds the $k^{-5/3}$ scaling from the inertial range, where $k = \lvert \boldsymbol{k} \rvert$. This spectral bump is \copyedit{centred} around $k\eta \approx 0.13$ and, while its location and width are relatively insensitive to $Re_\lambda$, its height decreases with increasing $Re_\lambda$ \citep{Don2010}. There have been various interpretations of the bottleneck effect, including as a result of quenched local interactions due to viscous effects \citep{Fal1994}, helicity dynamics \citep{Kur2004}, incomplete thermalization \citep{Fri2008} and insufficient width of the inertial range \citep{Ver2007}. More recently, inspired by the backscatter produced by $\iPi^{\ell,c}$ in two-dimensional turbulence, \citet{Joh2021} proposed vortex thinning \citep{Kra1976,Che2006,Xia2009} as a plausible mechanism responsible for the subinertial bottleneck effect in three-dimensional turbulence. Despite significant attention, a complete understanding of the origins of the bottleneck effect remains elusive. The present study aims to clarify its structural and statistical imprints using our normality-based analysis of filtered velocity gradients.

While the bottleneck effect is an expected feature of a fully-resolved DNS, it is mitigated when the flow field is filtered at scales typically employed for LES  (e.g. $2\ell/\eta = 48$). However, as shown in \cref{fig:spectra}(\textit{b}), the dynamic eddy viscosity model introduced by \citet{Kam2024} produces a significant `artificial' bottleneck effect at scales that are larger than the LES filter scale and, thus, larger than the scales associated with the true bottleneck effect. By contrast, the dynamic mixed model introduced by \citet{Kam2024}, which synthesizes the nonlinear gradient model with an eddy viscosity model, mitigates this artificial bottleneck to more accurately represent the spectrum of the corresponding filtered DNS. These results reflect that, while the scale-\copyedit{non-local} terms in (\ref{eq:iPi_sym_12}) are reasonably \copyedit{modelled} by eddy viscosity physics, explicitly capturing the scale-local terms, which are associated with the nonlinear gradient model, produces more realistic energy transfer. Statistical analyses more broadly suggest that eddy viscosity and mixed models tend to replicate flow features associated with unfiltered and filtered DNS data, respectively \citep{Joh2022,Kam2024}. Further, preliminary two-dimensional visualizations suggest that the unfiltered DNS and eddy viscosity model LES cases produce sheet-like vorticity structures that are not observed in the filtered DNS and mixed model LES cases \citep{Kam2024}. In the present study, our normality-based velocity gradient analysis definitively links these observations and captures their relationship to the artificial bottleneck effect.

\subsection{Contributions}\label{sec:intro:contributions}

Motivated by the preceding discussion, the present study aims to refine our understanding of multiscale flow structures and energy transfer mechanisms in turbulent flows. In \textsection \ref{sec:theory}, we formulate the normality-based analysis of filtered velocity gradients and use it to develop a novel decomposition of interscale energy transfer. We apply this normality-based analysis to filtered velocity gradients obtained from DNS and LES data that represent forced isotropic turbulence. The technical details of these simulations are described in \textsection \ref{sec:sims}. The results of our analysis are presented in \textsection \ref{sec:results}, including identifying the effect of filtering and LES \copyedit{modelling} on the velocity gradient partitioning (\textsection \ref{sec:results:VGT}), vortical flow structures (\textsection \ref{sec:results:structures}) and interscale energy transfer (\textsection \ref{sec:results:energy}). Finally, we summarize the implications of these results for energy cascade physics, LES \copyedit{modelling} and future prospects in \textsection \ref{sec:conc}.

\section{Theoretical framework}\label{sec:theory}

The analysis described in \textsection \ref{sec:intro:normality} can be reformulated in a multiscale setting by applying it to filtered velocity gradients. The corresponding normality-based decomposition of the filtered VGT can be expressed as
\begin{equation}\label{eq:VGT_nor_filt}
    \A{\ol{A}}{\ell}{ij} = \A{\ol{S}}{\ell,\epsilon}{ij} + \A{\ol{A}}{\ell,\gamma}{ij} + \A{\ol{\iW}}{\ell,\varphi}{ij},
\end{equation}
where the order of the superscripts indicates that filtering is performed \textit{prior} to the decomposition. The order of these operations is important since, unlike the symmetry-based decomposition, the normality-based decomposition does not commute with the filtering operation. We choose to decompose the filtered velocity gradients since this approach retains the normality properties of each tensor in the decomposition. It also enables consistent analysis of LES data, for which only the resolved (i.e. filtered) flow field is available. Using this filtering-first approach, the normality-based partitioning of the filtered velocity gradients can be expressed as
\begin{equation}\label{eq:VGT_part_filt}
    A_\ell^2 = \A{\ol{A}}{\ell}{ij}\A{\ol{A}}{\ell}{ij} = \underbrace{\A{\ol{S}}{\ell,\epsilon}{ij}\A{\ol{S}}{\ell,\epsilon}{ij}}_{\textstyle S_{\ell,\epsilon}^2} + \overbrace{\underbrace{\A{\ol{S}}{\ell,\gamma}{ij}\A{\ol{S}}{\ell,\gamma}{ij}}_{\textstyle S_{\ell,\gamma}^2} + \underbrace{\A{\ol{\iW}}{\ell,\gamma}{ij}\A{\ol{\iW}}{\ell,\gamma}{ij}}_{\textstyle \iW_{\ell,\gamma}^2}}^{\textstyle A_{\ell,\gamma}^2} + \underbrace{\A{\ol{\iW}}{\ell,\varphi}{ij}\A{\ol{\iW}}{\ell,\varphi}{ij}}_{\textstyle \iW_{\ell,\varphi}^2} + \underbrace{2\A{\ol{\iW}}{\ell,\varphi}{ij}\A{\ol{\iW}}{\ell,\gamma}{ij}}_{\textstyle \iW_{\ell,\varphi\gamma}^2},
\end{equation}
where $S_{\ell,\gamma}^2 = \iW_{\ell,\gamma}^2 = \tfrac{1}{2}A_{\ell,\gamma}^2$.

The normality-based decomposition of the filtered VGT can also be applied to the multiscale velocity gradient expansions of the residual stress tensor and interscale energy transfer discussed in \textsection \ref{sec:intro:multiscale}. Previous studies employing related approaches have primarily decomposed terms analogous to $\A{\ol{S}}{\ell}{ij}$ in the expression, $\iPi^\ell = -\A{\ol{S}}{\ell}{ij} \sigma^\ell_{ij}$, for interscale energy transfer \citep{Eno2023,Fat2024}. The analysis of \citet{Eno2023} briefly considered the residual stress tensor using the Biot--Savart law, but it did not address its multiscale composition and did not employ a filtering-first framework. Therefore, the following normality-based formulation of interscale energy transfer provides novel insight by explicitly capturing multiscale contributions to the residual stress tensor in a manner amenable to the assessment and development of LES models.

When applied to the expression in (\ref{eq:iPi_VGT}), the normality-based decomposition of interscale energy transfer yields
\begin{equation}\label{eq:iPi_nor}
    \iPi^\ell = \iPi^\ell_{\epsilon\epsilon} + \iPi^\ell_{\varphi\varphi} + \iPi^\ell_{\gamma\gamma} + \iPi^\ell_{\epsilon\gamma} + \iPi^\ell_{\varphi\gamma},
\end{equation}
where
\begin{alignat}{100}
    &\iPi^\ell_{\epsilon\epsilon} &&= -\A{\ol{S}}{\ell}{ij} \bigintsss_{\;0}^{\ell^2} {\rm d}\theta^2 \left( \ol{\A{\ol{S}}{\theta,\epsilon}{ik}\A{\ol{S}}{\theta,\epsilon}{jk}}^{\phi} \right), \label{eq:iPi_nor_eps_eps} \\
    &\iPi^\ell_{\varphi\varphi} &&= -\A{\ol{S}}{\ell}{ij} \bigintsss_{\;0}^{\ell^2} {\rm d}\theta^2 \left( \ol{\A{\ol{\iW}}{\theta,\varphi}{ik}\A{\ol{\iW}}{\theta,\varphi}{jk}}^{\phi} \right), \label{eq:iPi_nor_phi_phi} \\
    &\iPi^\ell_{\gamma\gamma} &&= -\A{\ol{S}}{\ell}{ij} \bigintsss_{\;0}^{\ell^2} {\rm d}\theta^2 \left( \ol{\A{\ol{A}}{\theta,\gamma}{ik}\A{\ol{A}}{\theta,\gamma}{jk}}^{\phi} \right), \label{eq:iPi_nor_gam_gam} \\
    &\iPi^\ell_{\epsilon\gamma} &&= -\A{\ol{S}}{\ell}{ij} \bigintsss_{\;0}^{\ell^2} {\rm d}\theta^2 \left( \ol{\A{\ol{S}}{\theta,\epsilon}{ik}\A{\ol{A}}{\theta,\gamma}{jk}}^{\phi} + \ol{\A{\ol{A}}{\theta,\gamma}{ik}\A{\ol{S}}{\theta,\epsilon}{jk}}^{\phi} \right), \label{eq:iPi_nor_eps_gam} \\
    &\iPi^\ell_{\varphi\gamma} &&= -\A{\ol{S}}{\ell}{ij} \bigintsss_{\;0}^{\ell^2} {\rm d}\theta^2 \left( \ol{\A{\ol{\iW}}{\theta,\varphi}{ik}\A{\ol{A}}{\theta,\gamma}{jk}}^{\phi} + \ol{\A{\ol{A}}{\theta,\gamma}{ik}\A{\ol{\iW}}{\theta,\varphi}{jk}}^{\phi} \right), \label{eq:iPi_nor_phi_gam}
\end{alignat}
and $\phi = \sqrt{\ell^2 - \theta^2}$. Here, $\iPi^\ell_{\epsilon\epsilon}$, $\iPi^\ell_{\varphi\varphi}$ and $\iPi^\ell_{\gamma\gamma}$ represent contributions from normal straining, rigid rotation and pure shearing, respectively, at scales $\theta \leq \ell$. Similarly, $\iPi^\ell_{\epsilon\gamma}$ and $\iPi^\ell_{\varphi\gamma}$ represent contributions from the covariance of normal straining with pure shearing and the covariance of rigid rotation with pure shearing, respectively, at these scales.

The normality-based approach can also be used to refine the symmetry-based analysis of interscale energy transfer discussed in \textsection \ref{sec:intro:multiscale}. The cascade rates associated with \copyedit{SS}, \copyedit{VS} and strain--vorticity covariance, as defined in (\ref{eq:iPi_s}), (\ref{eq:iPi_w}) and (\ref{eq:iPi_c}), respectively, can be decomposed as
\begin{alignat}{100}
    &\iPi^{\ell,s} &&= \iPi^{\ell,s}_{\epsilon\epsilon} &&+ \iPi^{\ell,s}_{\gamma\gamma} &&+ \iPi^{\ell,s}_{\epsilon\gamma}, \label{eq:iPi_s_nor} \\
    &\iPi^{\ell,\w} &&= \iPi^{\ell,\w}_{\varphi\varphi} &&+ \iPi^{\ell,\w}_{\gamma\gamma} &&+ \iPi^{\ell,\w}_{\varphi\gamma}, \label{eq:iPi_w_nor} \\
    &\iPi^{\ell,c} &&= \iPi^{\ell,c}_{\epsilon\gamma} &&+ \iPi^{\ell,c}_{\gamma\gamma} &&+ \iPi^{\ell,c}_{\varphi\gamma}, \label{eq:iPi_c_nor}
\end{alignat}
respectively. The \copyedit{SS} terms in (\ref{eq:iPi_s_nor}), defined as
\begin{alignat}{100}
    &\iPi^{\ell,s}_{\epsilon\epsilon} &&= -\A{\ol{S}}{\ell}{ij} \bigintsss_{\;0}^{\ell^2} {\rm d}\theta^2 \left( \ol{\A{\ol{S}}{\theta,\epsilon}{ik}\A{\ol{S}}{\theta,\epsilon}{jk}}^{\phi} \right), \label{eq:iPi_s_eps_eps} \\
    &\iPi^{\ell,s}_{\gamma\gamma} &&= -\A{\ol{S}}{\ell}{ij} \bigintsss_{\;0}^{\ell^2} {\rm d}\theta^2 \left( \ol{\A{\ol{S}}{\theta,\gamma}{ik}\A{\ol{S}}{\theta,\gamma}{jk}}^{\phi} \right), \label{eq:iPi_s_gam_gam} \\
    &\iPi^{\ell,s}_{\epsilon\gamma} &&= -\A{\ol{S}}{\ell}{ij} \bigintsss_{\;0}^{\ell^2} {\rm d}\theta^2 \left( \ol{\A{\ol{S}}{\theta,\epsilon}{ik}\A{\ol{S}}{\theta,\gamma}{jk}}^{\phi} + \ol{\A{\ol{S}}{\theta,\gamma}{ik}\A{\ol{S}}{\theta,\epsilon}{jk}}^{\phi} \right), \label{eq:iPi_s_eps_gam}
\end{alignat}
represent the contributions associated with normal straining, shear straining and their covariance, respectively, at scales $\theta \leq \ell$. The \copyedit{VS} terms in (\ref{eq:iPi_w_nor}), defined as
\begin{alignat}{100}
    &\iPi^{\ell,\w}_{\varphi\varphi} &&= -\A{\ol{S}}{\ell}{ij} \bigintsss_{\;0}^{\ell^2} {\rm d}\theta^2 \left( \ol{\A{\ol{\iW}}{\theta,\varphi}{ik}\A{\ol{\iW}}{\theta,\varphi}{jk}}^{\phi} \right), \label{eq:iPi_w_phi_phi} \\
    &\iPi^{\ell,\w}_{\gamma\gamma} &&= -\A{\ol{S}}{\ell}{ij} \bigintsss_{\;0}^{\ell^2} {\rm d}\theta^2 \left( \ol{\A{\ol{\iW}}{\theta,\gamma}{ik}\A{\ol{\iW}}{\theta,\gamma}{jk}}^{\phi} \right), \label{eq:iPi_w_gam_gam} \\
    &\iPi^{\ell,\w}_{\varphi\gamma} &&= -\A{\ol{S}}{\ell}{ij} \bigintsss_{\;0}^{\ell^2} {\rm d}\theta^2 \left( \ol{\A{\ol{\iW}}{\theta,\varphi}{ik}\A{\ol{\iW}}{\theta,\gamma}{jk}}^{\phi} + \ol{\A{\ol{\iW}}{\theta,\gamma}{ik}\A{\ol{\iW}}{\theta,\varphi}{jk}}^{\phi} \right), \label{eq:iPi_w_phi_gam}
\end{alignat}
represent the contributions associated with rigid rotation, shear vorticity and their covariance, respectively, at scales $\theta \leq \ell$. The strain--vorticity covariance terms in (\ref{eq:iPi_c_nor}), defined as
\begin{alignat}{100}
    &\iPi^{\ell,c}_{\epsilon\gamma} &&= -\A{\ol{S}}{\ell}{ij} \bigintsss_{\;0}^{\ell^2} {\rm d}\theta^2 \left( \ol{\A{\ol{S}}{\theta,\epsilon}{ik}\A{\ol{\iW}}{\theta,\gamma}{jk}}^{\phi} + \ol{\A{\ol{\iW}}{\theta,\gamma}{ik}\A{\ol{S}}{\theta,\epsilon}{jk}}^{\phi} \right), \label{eq:iPi_c_eps_gam} \\
    &\iPi^{\ell,c}_{\gamma\gamma} &&= -\A{\ol{S}}{\ell}{ij} \bigintsss_{\;0}^{\ell^2} {\rm d}\theta^2 \left( \ol{\A{\ol{S}}{\theta,\gamma}{ik}\A{\ol{\iW}}{\theta,\gamma}{jk}}^{\phi} + \ol{\A{\ol{\iW}}{\theta,\gamma}{ik}\A{\ol{S}}{\theta,\gamma}{jk}}^{\phi} \right), \label{eq:iPi_c_gam_gam} \\
    &\iPi^{\ell,c}_{\varphi\gamma} &&= -\A{\ol{S}}{\ell}{ij} \bigintsss_{\;0}^{\ell^2} {\rm d}\theta^2 \left( \ol{\A{\ol{S}}{\theta,\gamma}{ik}\A{\ol{\iW}}{\theta,\varphi}{jk}}^{\phi} + \ol{\A{\ol{\iW}}{\theta,\varphi}{ik}\A{\ol{S}}{\theta,\gamma}{jk}}^{\phi} \right), \label{eq:iPi_c_phi_gam}
\end{alignat}
represent the contributions associated with the covariance of normal straining with shear vorticity, the covariance of shear straining with shear vorticity and the covariance of shear straining with rigid rotation, respectively, at scales $\theta \leq \ell$. Using the form of the VGT in (\ref{eq:VGT_nor}), it can be shown that the analogous covariance term associated with normal straining and rigid rotation is identically zero. The contributions from the terms in (\ref{eq:iPi_s_nor}), (\ref{eq:iPi_w_nor}) and (\ref{eq:iPi_c_nor}) can be used to reconstruct the terms in (\ref{eq:iPi_nor}) as
\begin{alignat}{100}
    &\iPi^\ell_{\epsilon\epsilon} &&= \iPi^{\ell,s}_{\epsilon\epsilon} &&, \label{eq:iPi_nor_eps_eps_sym} \\
    &\iPi^\ell_{\varphi\varphi} &&= \iPi^{\ell,\w}_{\varphi\varphi} &&, \label{eq:iPi_nor_phi_phi_sym} \\
    &\iPi^\ell_{\gamma\gamma} &&= \iPi^{\ell,s}_{\gamma\gamma} &&+ \iPi^{\ell,\w}_{\gamma\gamma} + \iPi^{\ell,c}_{\gamma\gamma}, \label{eq:iPi_nor_gam_gam_sym} \\
    &\iPi^\ell_{\epsilon\gamma} &&= \iPi^{\ell,s}_{\epsilon\gamma} &&+ \iPi^{\ell,c}_{\epsilon\gamma}, \label{eq:iPi_nor_eps_gam_sym} \\
    &\iPi^\ell_{\varphi\gamma} &&= \iPi^{\ell,\w}_{\varphi\gamma} &&+ \iPi^{\ell,c}_{\varphi\gamma}. \label{eq:iPi_nor_phi_gam_sym}
\end{alignat}

A further refinement can be obtained by identifying scale-local and scale-\copyedit{non-local} contributions to the \copyedit{SS} terms in (\ref{eq:iPi_s_nor}) and the \copyedit{VS} terms in (\ref{eq:iPi_w_nor}) as
\begin{alignat}{100}
    &\iPi^{\ell,s}_{\epsilon\epsilon} &&= \iPi^{\ell,s1}_{\epsilon\epsilon} &&+ \iPi^{\ell,s2}_{\epsilon\epsilon}, \quad &&\iPi^{\ell,s}_{\gamma\gamma} &&= \iPi^{\ell,s1}_{\gamma\gamma} &&+ \iPi^{\ell,s2}_{\gamma\gamma}, \quad &&\iPi^{\ell,s}_{\epsilon\gamma} &&= \iPi^{\ell,s1}_{\epsilon\gamma} &&+ \iPi^{\ell,s2}_{\epsilon\gamma}, \label{eq:iPi_s_nor_2term} \\
    &\iPi^{\ell,\w}_{\varphi\varphi} &&= \iPi^{\ell,\w1}_{\varphi\varphi} &&+ \iPi^{\ell,\w2}_{\varphi\varphi}, \quad &&\iPi^{\ell,\w}_{\gamma\gamma} &&= \iPi^{\ell,\w1}_{\gamma\gamma} &&+ \iPi^{\ell,\w2}_{\gamma\gamma}, \quad &&\iPi^{\ell,\w}_{\varphi\gamma} &&= \iPi^{\ell,\w1}_{\varphi\gamma} &&+ \iPi^{\ell,\w2}_{\varphi\gamma}, \label{eq:iPi_w_nor_2term}
\end{alignat}
respectively. We do not identify scale-local and scale-\copyedit{non-local} contributions to the strain--vorticity covariance decomposition in (\ref{eq:iPi_c_nor}) since, as expressed in (\ref{eq:iPi_c_12}), the scale-local contributions identically sum to zero. The scale-local terms in (\ref{eq:iPi_s_nor_2term}) and (\ref{eq:iPi_w_nor_2term}) contribute to scale-local \copyedit{SS} and \copyedit{VS} as 
\begin{alignat}{100}
    &\iPi^{\ell,s1} &&= \underbrace{-\ell^2 \A{\ol{S}}{\ell}{ij}\A{\ol{S}}{\ell,\epsilon}{ik}\A{\ol{S}}{\ell,\epsilon}{jk}}_{\textstyle \iPi^{\ell,s1}_{\epsilon\epsilon}} \underbrace{-\;\ell^2 \A{\ol{S}}{\ell}{ij}\A{\ol{S}}{\ell,\gamma}{ik}\A{\ol{S}}{\ell,\gamma}{jk}}_{\textstyle \iPi^{\ell,s1}_{\gamma\gamma}} \underbrace{-\;2\ell^2 \A{\ol{S}}{\ell}{ij}\A{\ol{S}}{\ell,\epsilon}{ik}\A{\ol{S}}{\ell,\gamma}{jk}}_{\textstyle \iPi^{\ell,s1}_{\epsilon\gamma}}, \label{eq:iPi_s1_2term} \\
    &\iPi^{\ell,\w1} &&= \underbrace{-\ell^2 \A{\ol{S}}{\ell}{ij}\A{\ol{\iW}}{\ell,\varphi}{ik}\A{\ol{\iW}}{\ell,\varphi}{jk}}_{\textstyle \iPi^{\ell,\w1}_{\varphi\varphi}} \underbrace{-\;\ell^2 \A{\ol{S}}{\ell}{ij}\A{\ol{\iW}}{\ell,\gamma}{ik}\A{\ol{\iW}}{\ell,\gamma}{jk}}_{\textstyle \iPi^{\ell,\w1}_{\gamma\gamma}} \underbrace{-\;2\ell^2 \A{\ol{S}}{\ell}{ij}\A{\ol{\iW}}{\ell,\varphi}{ik}\A{\ol{\iW}}{\ell,\gamma}{jk}}_{\textstyle \iPi^{\ell,\w1}_{\varphi\gamma}}, \label{eq:iPi_w1_2term}
\end{alignat}
respectively. Here, the leading strain-rate tensor, $\A{\ol{S}}{\ell}{ij}$, is not decomposed since the normality-based analysis is applied to the multiscale velocity gradient expansion of the residual stress tensor. However, the mechanisms underlying scale-local \copyedit{SS} and \copyedit{VS} can be more clearly expressed by further decomposing $\A{\ol{S}}{\ell}{ij}$, which leads to
\begin{alignat}{100}
    &\iPi^{\ell,s1} &&= \underbrace{-\ell^2 \A{\ol{S}}{\ell,\epsilon}{ij}\A{\ol{S}}{\ell,\epsilon}{ik}\A{\ol{S}}{\ell,\epsilon}{jk}}_{\textstyle \iPi^{\ell,s1}_{\epsilon\epsilon\epsilon}} \underbrace{-\;3\ell^2 \A{\ol{S}}{\ell,\epsilon}{ij}\A{\ol{S}}{\ell,\gamma}{ik}\A{\ol{S}}{\ell,\gamma}{jk}}_{\textstyle \iPi^{\ell,s1}_{\epsilon\gamma\gamma}} \underbrace{-\;\ell^2 \A{\ol{S}}{\ell,\gamma}{ij}\A{\ol{S}}{\ell,\gamma}{ik}\A{\ol{S}}{\ell,\gamma}{jk}}_{\textstyle \iPi^{\ell,s1}_{\gamma\gamma\gamma}}, \label{eq:iPi_s1_3term} \\
    &\iPi^{\ell,\w1} &&= \underbrace{-\ell^2 \A{\ol{S}}{\ell,\epsilon}{ij}\A{\ol{\iW}}{\ell,\varphi}{ik}\A{\ol{\iW}}{\ell,\varphi}{jk}}_{\textstyle \iPi^{\ell,\w1}_{\epsilon\varphi\varphi}} \underbrace{-\;\ell^2 \A{\ol{S}}{\ell,\epsilon}{ij}\A{\ol{\iW}}{\ell,\gamma}{ik}\A{\ol{\iW}}{\ell,\gamma}{jk}}_{\textstyle \iPi^{\ell,\w1}_{\epsilon\gamma\gamma}} \underbrace{-\;\ell^2 \A{\ol{S}}{\ell,\gamma}{ij}\A{\ol{\iW}}{\ell,\gamma}{ik}\A{\ol{\iW}}{\ell,\gamma}{jk}}_{\textstyle \iPi^{\ell,\w1}_{\gamma\gamma\gamma}} \underbrace{-\;2\ell^2 \A{\ol{S}}{\ell,\epsilon}{ij}\A{\ol{\iW}}{\ell,\varphi}{ik}\A{\ol{\iW}}{\ell,\gamma}{jk}}_{\textstyle \iPi^{\ell,\w1}_{\epsilon\varphi\gamma}}. \label{eq:iPi_w1_3term}
\end{alignat}
These terms are related to those in (\ref{eq:iPi_s1_2term}) and (\ref{eq:iPi_w1_2term}) in that
\begin{alignat}{100}
    &\iPi^{\ell,s1}_{\epsilon\epsilon} &&= \iPi^{\ell,s1}_{\epsilon\epsilon\epsilon}, \quad &&\iPi^{\ell,s1}_{\gamma\gamma} &&= \iPi^{\ell,s1}_{\gamma\gamma\gamma} + \frac{1}{3}&&\iPi^{\ell,s1}_{\epsilon\gamma\gamma}, \quad &&\iPi^{\ell,s1}_{\epsilon\gamma} &&= \frac{2}{3}&&\iPi^{\ell,s1}_{\epsilon\gamma\gamma}, \label{eq:2to3_s1} \\
    &\iPi^{\ell,\w1}_{\varphi\varphi} &&= \iPi^{\ell,\w1}_{\epsilon\varphi\varphi}, \quad &&\iPi^{\ell,\w1}_{\gamma\gamma} &&= \iPi^{\ell,\w1}_{\gamma\gamma\gamma} + &&\iPi^{\ell,\w1}_{\epsilon\gamma\gamma}, \quad &&\iPi^{\ell,\w1}_{\varphi\gamma} &&= &&\iPi^{\ell,\w1}_{\epsilon\varphi\gamma}. \label{eq:2to3_w1}
\end{alignat}

In (\ref{eq:iPi_s1_3term}), $\iPi^{\ell,s1}_{\epsilon\epsilon\epsilon}$, $\iPi^{\ell,s1}_{\gamma\gamma\gamma}$ and $\iPi^{\ell,s1}_{\epsilon\gamma\gamma}$ represent scale-local \copyedit{SS} due to normal straining, shear straining and their interaction, respectively. In (\ref{eq:iPi_w1_3term}), $\iPi^{\ell,\w1}_{\epsilon\varphi\varphi}$, $\iPi^{\ell,\w1}_{\epsilon\gamma\gamma}$ and $\iPi^{\ell,\w1}_{\epsilon\varphi\gamma}$ represent scale-local \copyedit{VS} due to the normal straining of rigid rotation, shear vorticity and shear--rotation interactions, respectively, and $\iPi^{\ell,\omega1}_{\gamma\gamma\gamma}$ is due to the shear straining of shear vorticity. All terms omitted from these equations, including the $\iPi^{\ell,s1}_{\epsilon\epsilon\gamma}$ term in (\ref{eq:iPi_s1_3term}) and the $\iPi^{\ell,\w1}_{\gamma\varphi\varphi}$ and $\iPi^{\ell,\w1}_{\gamma\varphi\gamma}$ terms in (\ref{eq:iPi_w1_3term}), can be shown to be identically zero using the filtered VGT in its principal frame, which is of the form in (\ref{eq:VGT_nor}). It can also be shown that $\iPi^{\ell,s1}_{\epsilon\gamma\gamma} = 3\iPi^{\ell,\w1}_{\epsilon\gamma\gamma}$ and $\iPi^{\ell,s1}_{\gamma\gamma\gamma} = 3\iPi^{\ell,\w1}_{\gamma\gamma\gamma}$ hold as exact pointwise relationships. When combined with the fact that $\left\langle \iPi^{\ell,s1} \right\rangle$ = 3$\left\langle \iPi^{\ell,\w1} \right\rangle$ for homogeneous turbulence \citep{Bet1956}, these relationships imply that the relative contributions of the shear straining terms, $\iPi^{\ell,s1}_{\epsilon\gamma\gamma}$ and $\iPi^{\ell,s1}_{\gamma\gamma\gamma}$, to $\left\langle \iPi^{\ell,s1} \right\rangle$ are equivalent to the relative contributions of the shear vorticity terms, $\iPi^{\ell,\w1}_{\epsilon\gamma\gamma}$ and $\iPi^{\ell,\w1}_{\gamma\gamma\gamma}$, to $\left\langle \iPi^{\ell,\w1} \right\rangle$ in the absence of inhomogeneities. This, in turn, implies that the relative contribution of the normal straining term, $\iPi^{\ell,s1}_{\epsilon\epsilon\epsilon}$, to $\left\langle \iPi^{\ell,s1} \right\rangle$ is equivalent to the relative contribution of the rigid rotation terms, $\iPi^{\ell,\w1}_{\epsilon\varphi\varphi} + \iPi^{\ell,\w1}_{\epsilon\varphi\gamma}$, to $\left\langle \iPi^{\ell,\w1} \right\rangle$. These relationships typify insights that can be garnered from the present normality-based formulation of interscale energy transfer.

\section{Simulation details}\label{sec:sims}

\begin{table}
\begin{center}
\def~{\hphantom{0}}
\begin{tabular}{lccccll}
Case   & Symbol                        & Type & Model          & Reynolds number          & \multicolumn{1}{c}{Resolution}  & Reference       \\[0pt]
       &                               &                       &                          &                                 &                 \\[-4pt]
DNS315 & $\boldsymbol{\triangledown}$  & DNS  & $-$            & $Re_\lambda \approx$ 315 & $k_{max}\eta \approx 2.0$       & \citet{Car2017} \\
DNS400 & $\boldsymbol{\triangle}$      & DNS  & $-$            & $Re_\lambda \approx$ 400 & $k_{max}\eta \approx 1.4$       & \citet{Kam2024} \\
Vis400 & $\boldsymbol{\triangleleft}$  & LES  & Eddy viscosity & $Re_\lambda \approx$ 400 & $k_{max}\ell_{LES} \approx 3.0$ & \citet{Kam2024} \\
Mix400 & $\boldsymbol{\triangleright}$ & LES  & Mixed          & $Re_\lambda \approx$ 400 & $k_{max}\ell_{LES} \approx 3.0$ & \citet{Kam2024} \\
\end{tabular}
\caption{Primary simulations considered in the present study. The DNS cases employ $N_x^3 = 1024^3$ collocation points and are dealiased using the $2\sqrt{2}/3$ truncation rule with phase-shifting \citep{Pat1971} such that $k_{max} = N_x\sqrt{2}/3$. The LES cases employ $N_x^3 = 128^3$ collocation points and filter widths of $2\ell_{LES}/\eta = 48$. They are dealiased using the $2/3$ truncation rule such that $k_{max} = N_x/3$. Additional cases, including random velocity gradients, DNS cases at lower Reynolds numbers and LES cases with different filter widths are described and \copyedit{analysed} in \aref{sec:app:collapse}.}
\label{tab:sims}
\end{center}
\end{table}

We apply the multiscale normality-based analysis formulated in \textsection \ref{sec:theory} to filtered velocity gradients produced by simulations of forced isotropic turbulence. All simulations solve the incompressible Navier--Stokes equations a triply-periodic box using pseudo-spectral methods. \autoref{tab:sims} summarizes the primary cases we consider, which include DNS cases at two different Reynolds numbers and LES cases that employ two different models at the higher Reynolds number. The references for these simulations further expound their numerical implementations, forcing schemes and other relevant parameters and results. Each DNS dataset includes 21 temporal snapshots that are spaced one large-eddy turnover time apart and each LES dataset includes 68 temporal snapshots that are spaced half of a large-eddy turnover time apart. All snapshots are obtained from a statistically stationary regime (i.e. after initial transients).

For the LES cases, the eddy viscosity model (Vis400) and the mixed model (Mix400) were formulated and discussed by \citet{Kam2024} using the Stokes flow regularization (SFR) framework \citep{Joh2022}. They are dynamic models that assume local equilibrium for the residual kinetic energy and employ local clipping of the eddy viscosity to satisfy this assumption. We utilize these SFR-based models since they do not require the specification of a free model parameter and do not require test filter calculations. However, since they perform similarly to other formulations \citep{Kam2024}, we do not expect our insights to be limited to the SFR-based approach.

For the eddy viscosity model, the residual stress tensor at the LES filter scale, $\ell = \ell_{LES}$, is \copyedit{modelled} as 
\begin{equation}\label{eq:model_vis}
    \sigma^\ell_{ij} = -2\nu_T \A{\ol{S}}{\ell}{ij},
\end{equation}
where the eddy viscosity, $\nu_T$, is the solution to
\begin{equation}\label{eq:nuT_vis}
    \nu_T = \frac{3\ell^2}{4}\left( \nabla^2 \nu_T - \frac{\A{\ol{S}}{\ell}{ij}\A{\ol{A}}{\ell}{ik}\A{\ol{A}}{\ell}{jk}}{S_\ell^2} \right).
\end{equation}
Therefore, the interscale energy transfer at the LES filter scale is \copyedit{modelled} as
\begin{equation}\label{eq:transfer_vis}
    \iPi^{\ell} = 2\nu_T S_\ell^2.
\end{equation}
For the mixed model, the residual stress tensor at the LES filter scale is \copyedit{modelled} as 
\begin{equation}\label{eq:model_mix}
    \sigma^\ell_{ij} = \ell^2 \A{\ol{A}}{\ell}{ik}\A{\ol{A}}{\ell}{jk} - 2\nu_T \A{\ol{S}}{\ell}{ij}.
\end{equation}
The first term in this model resembles the nonlinear gradient model \citep{Cla1979,Bor1998,Men2000}. Only its deviatoric part is relevant since its isotropic part, which can be lumped with pressure in the momentum equations, does not contribute to the dynamics of the resolved velocity field. The second term in (\ref{eq:model_mix}) resembles an eddy viscosity model, for which $\nu_T$ is the solution to
\begin{equation}\label{eq:nuT_mix}
    \nu_T = \frac{3\ell^2}{4}\left( \nabla^2 \nu_T - \ell^2 \frac{\A{\ol{S}}{\ell}{ij}\A{\ol{B}}{\ell}{imn}\A{\ol{B}}{\ell}{jmn}}{S_\ell^2} \right),
\end{equation}
where $\A{B}{}{ijk} = \partial^2 u_i^{} \big/ \partial x_j^{} \partial x_k^{}$ is the velocity Hessian tensor. The corresponding interscale energy transfer at the LES filter scale is \copyedit{modelled} as
\begin{equation}\label{eq:transfer_mix}
    \iPi^{\ell} = -\ell^2 \A{\ol{S}}{\ell}{ij}\A{\ol{A}}{\ell}{ik}\A{\ol{A}}{\ell}{jk} + 2\nu_T S_{\ell}^2.
\end{equation}
Here, the first term explicitly captures the contributions from scale-local \copyedit{VS} and \copyedit{SS} (i.e. $\iPi^{\ell,s1} + \iPi^{\ell,\w1}$) and the second term models the residual contribution from the scale-\copyedit{non-local} terms (i.e. $\iPi^{\ell,s2} + \iPi^{\ell,\w2} + \iPi^{\ell,c2})$. This contrasts with the eddy viscosity model, for which the scale-local and scale-\copyedit{non-local} contributions are \copyedit{modelled} together. Explicitly capturing the scale-local energy transfer allows the mixed model to more accurately reproduce the statistics of a DNS filtered at scale $\ell_{LES}$, whereas the eddy viscosity model instead mimics an unfiltered DNS \citep{Joh2022,Kam2024}. This theme is further explored in \textsection \ref{sec:results} using the normality-based analysis.

\section{Results}\label{sec:results}

\subsection{Multiscale velocity gradient partitioning}\label{sec:results:VGT}

The normality-based partitioning of filtered velocity gradients defined in (\ref{eq:VGT_part_filt}) identifies the contributions of normal straining, pure shearing, rigid rotation and shear--rotation correlations to the strength of filtered velocity gradients. We consider the averaged form of this partitioning to characterize how the velocity gradients in forced isotropic turbulence vary as a function of scale. For unfiltered velocity gradients ($\ell = 0$), the partitioning is given by $\big\langle S_\epsilon^2 \big\rangle \big/ \big\langle A^2 \big\rangle = 0.240$, $\big\langle A_\gamma^2 \big\rangle \big/ \big\langle A^2 \big\rangle = 0.520$, $\big\langle \iW_\varphi^2 \big\rangle \big/ \big\langle A^2 \big\rangle = 0.106$ and $\big\langle \iW_{\varphi\gamma}^2 \big\rangle \big/ \big\langle A^2 \big\rangle = 0.134$, and it is relatively insensitive to $Re_\lambda$ for $Re_\lambda \gtrsim 200$ \citep{Das2020,Aru2024b}. This unfiltered partitioning serves as a baseline for our analysis of the effects of filtering and LES \copyedit{modelling} in this section.

\subsubsection{\copyedit{The} DNS cases: the effect of filtering}\label{sec:results:VGT:DNS}

\begin{figure}
    \centering
    \includegraphics[width=0.49\textwidth]{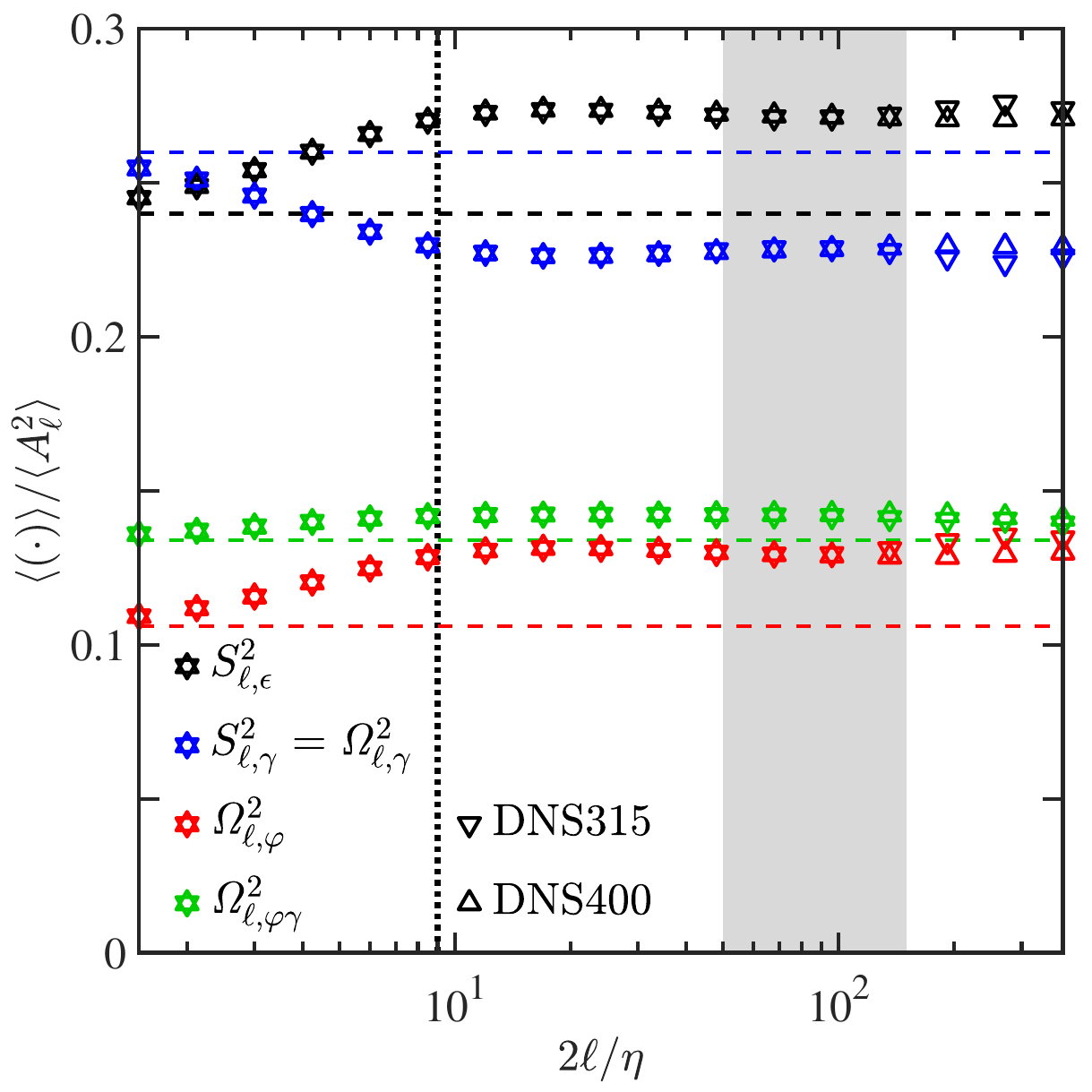}
    \caption{Partitioning of filtered velocity gradients for the DNS cases, where shearing is represented using $S_{\ell,\gamma}^2 = \iW_{\ell,\gamma}^2 = \tfrac{1}{2}A_{\ell,\gamma}^2$. The horizontal dashed  lines represent the unfiltered partitioning in the high-$Re_\lambda$ limit, the vertical dotted line represents the typical thickness of small-scale shear layers, $\delta_\gamma = 9\eta$, and the shaded region approximates the inertial range for DNS400 as $50 \leq 2\ell/\eta \leq 150$.}
    \label{fig:partFiltDNS}
\end{figure}

\autoref{fig:partFiltDNS} shows how the partitioning varies as a function of the filter width, $2\ell/\eta$, for the DNS cases. Increasing the filter width mitigates the relative contribution of pure shearing and increases the relative contributions of normal straining, rigid rotation, and, to a lesser extent, shear--rotation correlations. This \copyedit{behaviour} occurs primarily for $2\ell \lesssim \delta_\gamma$, where $\delta_\gamma = 9 \eta$ represents the typical thickness of small-scale shear layers \citep{Els2017,Nag2020,Wat2020,Fis2021,Wat2023}. It highlights that filtering mitigates the imprint of small-scale shear layers until the filter width surpasses their thickness, whereafter filtering at larger scales does not significantly affect the partitioning statistics, including in the inertial range. For $2\ell \gtrsim \delta_\gamma$, this scale-invariant partitioning is given by $\big\langle S_{\ell,\epsilon}^2 \big\rangle \big/ \big\langle A^2 \big\rangle \approx 0.27$, $\big\langle A_{\ell,\gamma}^2 \big\rangle \big/ \big\langle A^2 \big\rangle \approx 0.46$, $\big\langle \iW_{\ell,\varphi}^2 \big\rangle \big/ \big\langle A^2 \big\rangle \approx 0.13$ and $\big\langle \iW_{\ell,\varphi\gamma}^2 \big\rangle \big/ \big\langle A^2 \big\rangle \approx 0.14$. The distinction between this and the unfiltered partitioning would be obscured by an analogous symmetry-based analysis since $\big\langle S_\ell^2 \big\rangle = \big\langle \iW_\ell^2 \big\rangle$ at all scales for homogeneous turbulence \citep{Bet1956}. This highlights the ability of the normality-based decomposition to capture how flow structures in the viscous range differ from those in the inertial range.

The comparison between the filter width, $2\ell$, and the shear layer thickness, $\delta_\gamma$, can be interpreted in terms of velocity increments. \citet{Joh2021} highlighted that filtered velocity gradients can be expressed in terms of spatially integrated velocity increments that are weighted by the gradient of the filter kernel \citep{Eyi1995}. For a Gaussian kernel, the weighting is an odd function of the spatial offset, $\boldsymbol{r}$, and its magnitude is maximized when $\lvert \boldsymbol{r} \rvert = \ell$. These features imply that velocity increments given by $\boldsymbol{u}\left( \boldsymbol{x} + \ell\boldsymbol{\hat{e}} \right) - \boldsymbol{u}\left( \boldsymbol{x} - \ell\boldsymbol{\hat{e}} \right)$, where $\boldsymbol{\hat{e}}$ represents an arbitrary unit vector, have the strongest weighting in the construction of the filtered velocity gradients, $\A{\ol{A}}{\ell}{ij}(\boldsymbol{x})$. Hence, the filtered velocity gradients at a given point are most strongly related to velocity increments of size $2\ell$ that are \copyedit{centred} at the same point. This result provides a direct relationship to $\delta_\gamma$, which is often formulated in a similar manner using the velocity jump across a shear layer \citep{Els2017,Fis2021}. Even alternative definitions of $\delta_\gamma$, e.g. in terms of vorticity \citep{Nag2020,Wat2020,Wat2023}, are empirically associated with this velocity jump. Therefore, the comparison of $2\ell$ and $\delta_\gamma$ is appropriate since it has strong physical foundations based on velocity increments.

For the Reynolds numbers shown in \cref{fig:partFiltDNS}, the partitioning collapses as a function of the filter width except at very large scales, where it becomes more sensitive to the details of the forcing and the number of samples. \aref{sec:app:collapse} further investigates the collapse of the partitioning statistics in terms of $Re_\lambda$ and $\ell$. It highlights that the partitioning at small filter widths collapses well for $Re_\lambda \gtrsim 200$, consistent with the collapse of the unfiltered partitioning \citep{Das2020}. As $Re_\lambda$ increases, the collapse extends to larger filter widths, including within the inertial range. The collapse observed in the present study also complements the collapsed power law scalings for the intensities of shearing and rigid rotation obtained by \citet{Wat2024}, which were normalized using the Kolmogorov time scale instead of $\left\langle A_\ell^2 \right\rangle$. Based on the excellent collapse we observe for DNS315 and DNS400, we use $Re_\lambda \approx 400$ to represent the partitioning statistics at moderately high Reynolds numbers and as the primary subject of our LES analyses.

\subsubsection{\copyedit{The} LES cases: the effect of closure \copyedit{modelling}}\label{sec:results:VGT:LES}

\begin{figure}
    \centering
    \subfigimg[width=0.49\linewidth,pos=ll,vsep=30pt,hsep=29pt]{(\textit{a})}{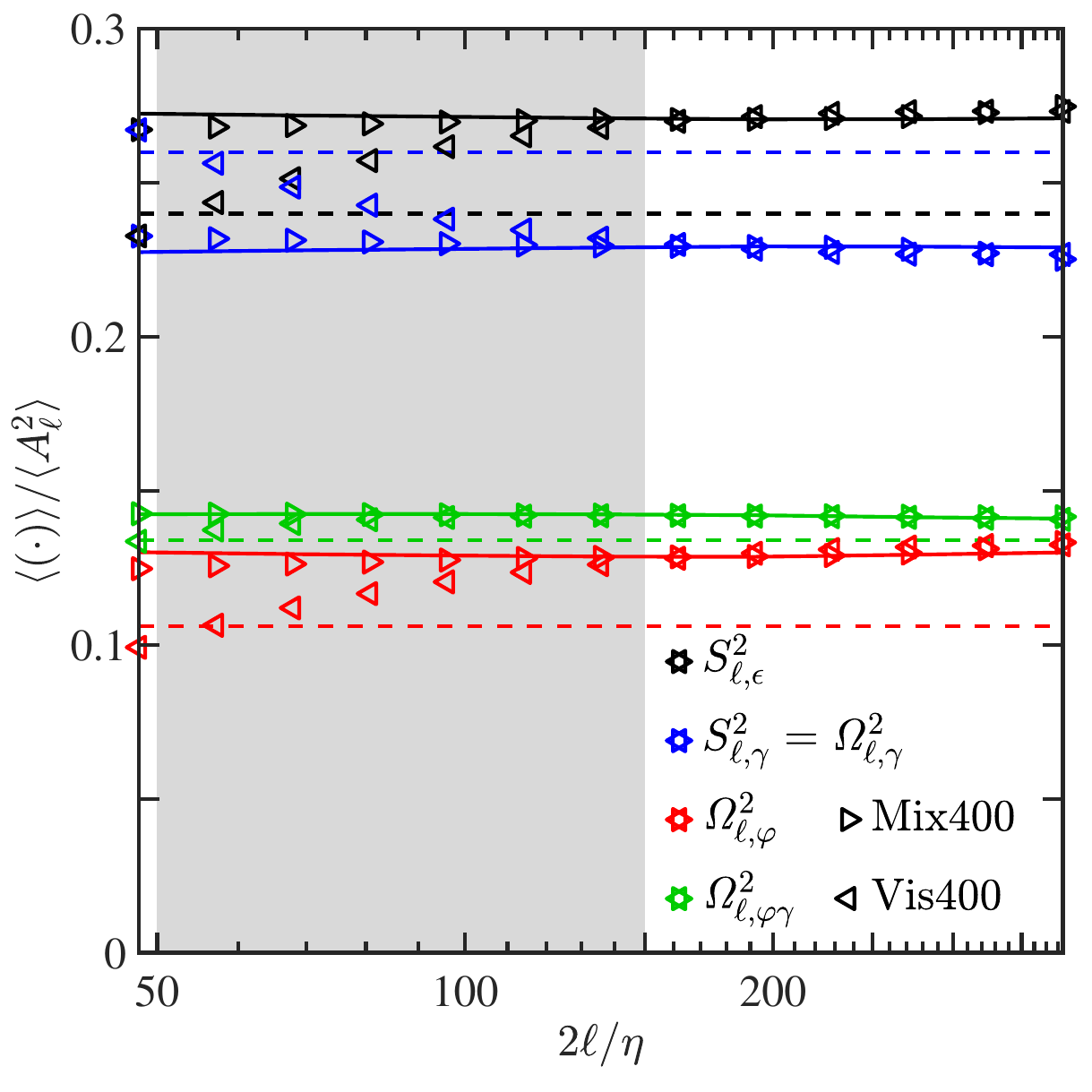}{}
    \subfigimg[width=0.49\linewidth,pos=ll,vsep=30pt,hsep=29pt]{(\textit{b})}{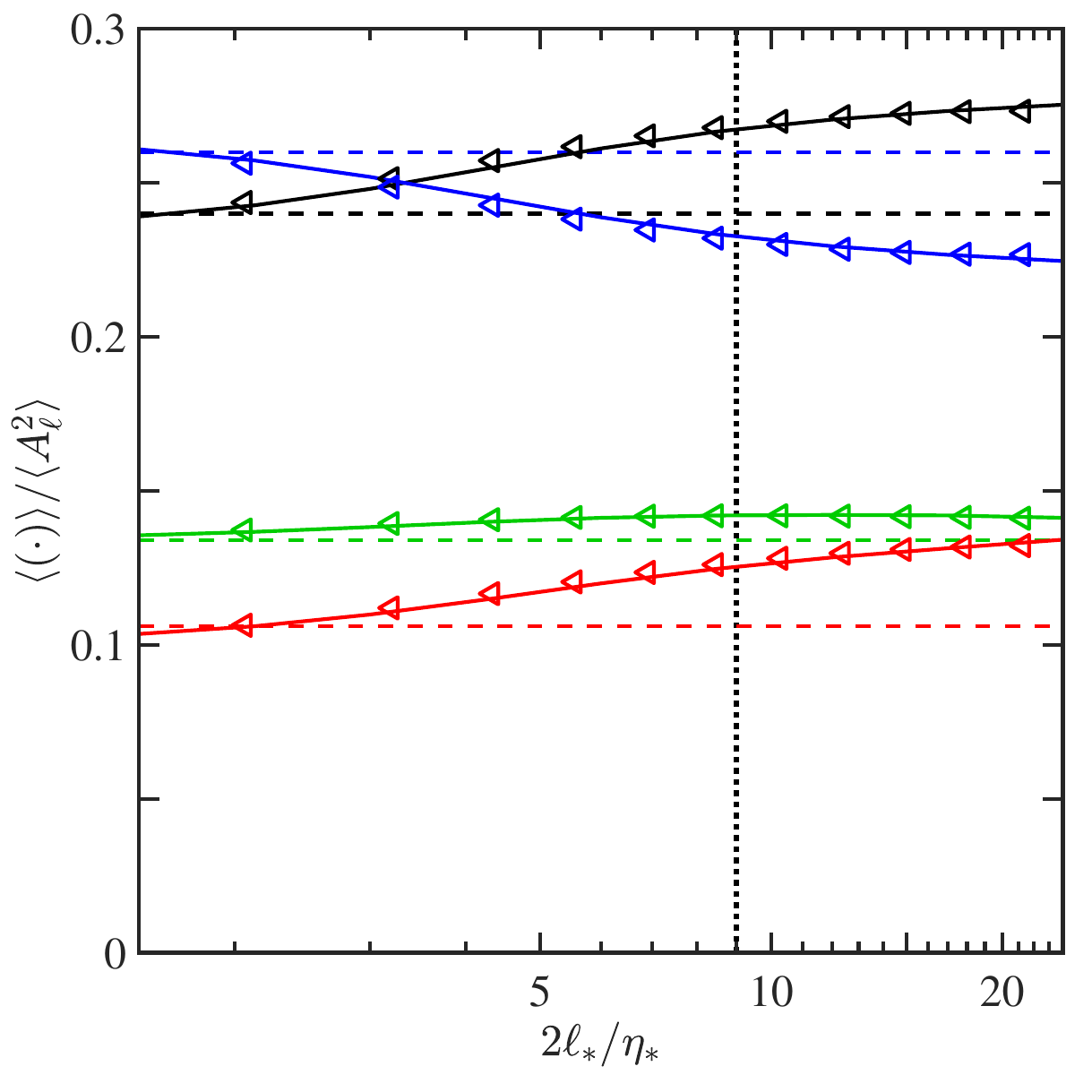}{}
    \caption{(\textit{a}) Partitioning of filtered velocity gradients for the LES cases. The solid curves represent the filtered DNS400 partitioning and the horizontal dashed lines represent the unfiltered partitioning in the high-$Re_\lambda$ limit. The lower limit of the filter width axis represents the LES filter width, $2\ell_{LES}/\eta = 48$, and the shaded region approximates the inertial range. (\textit{b}) Partitioning for Vis400 replotted as a function of $2\ell_*/\eta_*$, where $\ell_* = \sqrt{\ell^2 - \ell_{LES}^2}$ is the complementary filter width and $\eta_* \approx 15 \eta$ is an effective Kolmogorov scale. The solid curves represent the partitioning produced by a DNS at $Re_\lambda \approx 61$, which has a Kolmogorov scale of approximately $\eta_*$. The vertical dotted line represents the typical thickness of small-scale shear layers, $\delta_\gamma = 9\eta$.}
    \label{fig:partFiltLES}
\end{figure}

Previous investigations have demonstrated that the eddy viscosity and mixed models we consider produce flow statistics at the LES filter scale, $\ell_{LES}$, that resemble the statistics of unfiltered and filtered DNS data, respectively \citep{Joh2022,Kam2024}. We consider the multiscale partitioning statistics of the resolved velocity gradients for each model, thereby providing a more holistic assessment than the previous single-scale analyses. \autoref{fig:partFiltLES}(\textit{a}) shows how the partitioning produced by each LES model varies as a function of the filter width for $\ell \geq \ell_{LES}$. The filtered velocity gradients at scale $\ell$ are obtained by applying a complementary filter, of width $\ell_* = \sqrt{\ell^2 - \ell_{LES}^2}$, to the resolved velocity gradients, which are filtered at scale $\ell_{LES}$. Consistent with the previous results, the mixed model accurately reproduces the multiscale partitioning statistics of the filtered DNS across all resolved scales. By contrast, the partitioning produced by the eddy viscosity model deviates from the filtered DNS near the LES filter width, where it instead approaches the unfiltered DNS partitioning. This result qualitatively agrees with the notion that the flow field produced by the eddy viscosity model tends to behave like an unfiltered DNS at a lower Reynolds number \citep{Joh2022,Kam2024}.

In \cref{fig:partFiltLES}(\textit{b}), we quantitatively evaluate this claim by replotting the Vis400 partitioning statistics as a function of $2\ell_*/\eta_*$ instead of $2\ell/\eta$, where $\eta_* \approx 15\eta$ represents an effective Kolmogorov scale that is determined empirically. In effect, replacing $\ell$ with $\ell_*$ treats the resolved velocity gradients as if they are unfiltered and replacing $\eta$ with $\eta_*$ treats the flow statistics as if they were produced by a simulation at a lower Reynolds number. Using these modifications, we compare the multiscale partitioning statistics for Vis400 with those produced by a DNS at $Re_\lambda \approx 61$, which has a Kolmogorov scale of approximately $\eta_*$ \rev{(see \aref{sec:app:collapse} for computational details)}. This comparison shows that, when plotted as a function of $2\ell_*/\eta_*$, the Vis400 partitioning statistics collapse well onto those of the lower-$Re_\lambda$ DNS. It thus quantitatively validates the claim that the eddy viscosity model produces multiscale flow statistics that resemble an unfiltered DNS at a lower Reynolds number. In \aref{sec:app:collapse}, we show that the partitioning statistics produced by LES cases that employ different filter widths, $\ell_{LES}$, yield a similar conclusion.

Whereas $\eta_*$ is determined empirically in the present study, developing techniques to predict it \copyedit{\textit{a priori}} could provide valuable insight into the \copyedit{behaviour} of eddy viscosity models. A logical approach would be to hypothesize that $\eta_* = \left(  \nu_*^3 \big/ \big\langle \mathit{\Phi} \big\rangle \right)^{1/4}$, where $\nu_* = \nu \big\langle \mathit{\Phi} \big\rangle \big/ \big\langle\mathit{\Phi^{\ell_{LES}}} \big\rangle$ represents an effective viscosity that captures the statistical imprint of the dynamic eddy viscosity field in terms of the total dissipation rate, $\mathit{\Phi}$, and the resolved dissipation rate, $\mathit{\Phi^{\ell_{LES}}}$. However, this approach leads to a prediction of $\eta_*/\eta \approx 10.6$ for Vis400, which does not satisfactorily agree with the empirical value of $\eta_*/\eta \approx 15$. Furthermore, a similar analysis for Mix400 would yield $\eta_*/\eta \approx 8.6$, which does not reflect that the mixed model statistics resemble those of the filtered DNS. We leave formulating and validating improved predictions of $\eta_*$, which would likely require a more nuanced approach, to future work.

\subsection{Tube-like and sheet-like vortical structures}\label{sec:results:structures}

The partitioning statistics \copyedit{analysed} in \textsection \ref{sec:results:VGT} encode information about the flow structures that produce them. We \copyedit{analyse} how these structures manifest in the DNS and LES cases to characterize how filtering and LES \copyedit{modelling} affect their spatial organization. Our analysis focuses on tube-like and sheet-like vortical structures, which can be distinguished using the contributions of rigid rotation and shear vorticity and, as shown in \aref{sec:app:Burgers}, are associated with vortex cores and shear layers, respectively.

\begin{figure}
    \centering
    \subfigimg[width=0.40\linewidth,pos=ul,vsep=10pt,hsep=0pt]{(\textit{a})}{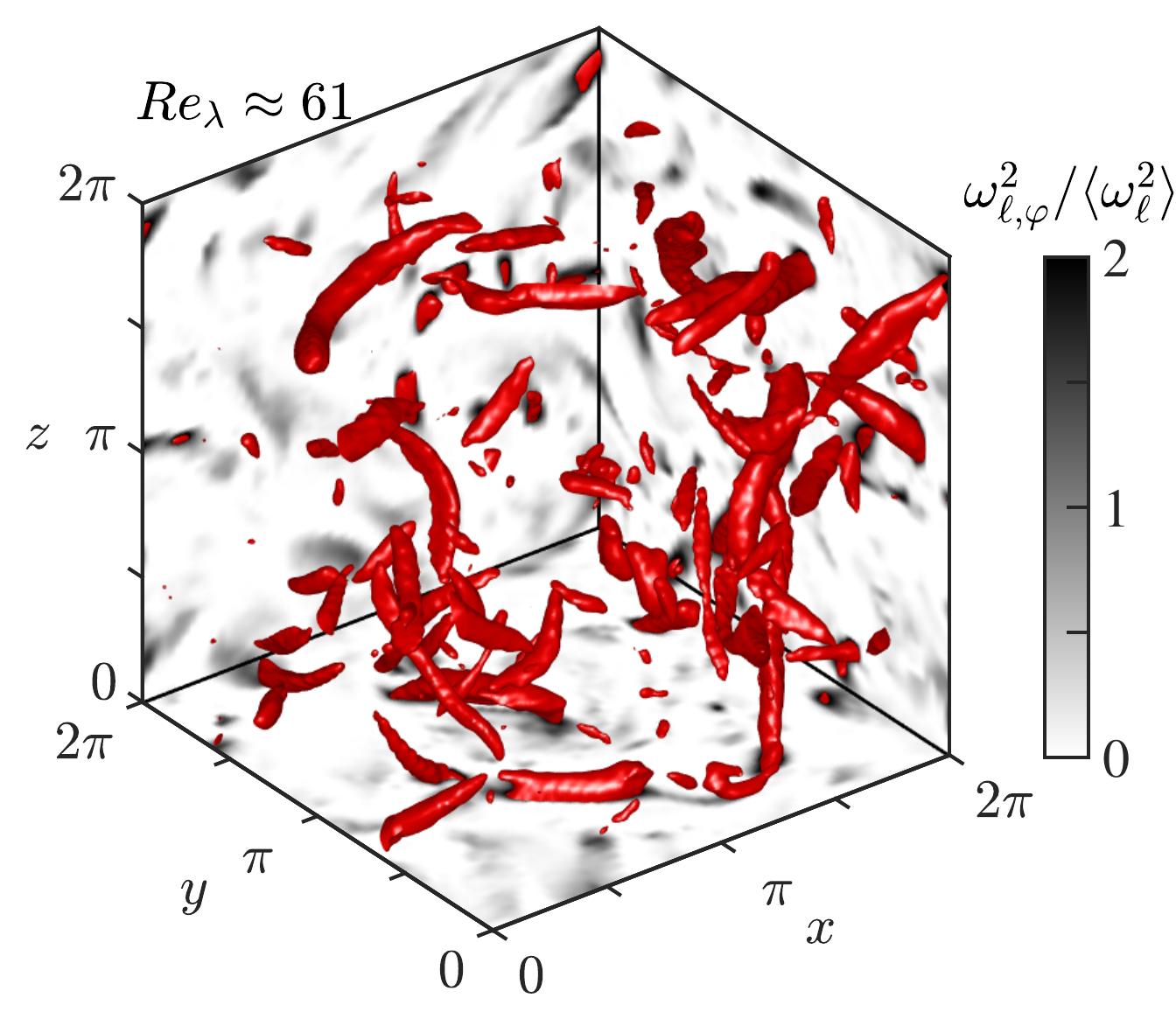}{}
    \subfigimg[width=0.40\linewidth,pos=ul,vsep=10pt,hsep=0pt]{(\textit{b})}{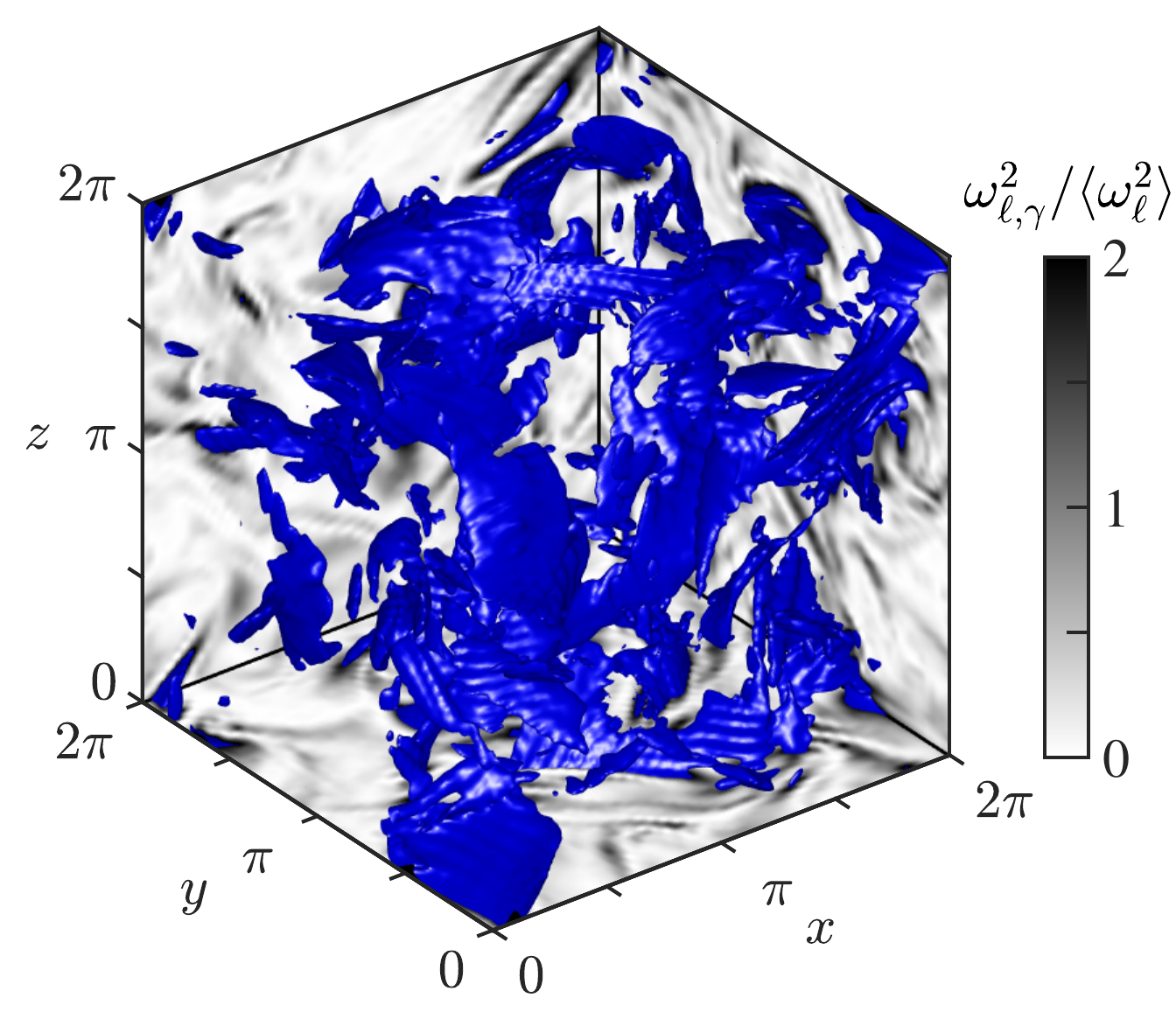}{}
    \subfigimg[width=0.40\linewidth,pos=ul,vsep=10pt,hsep=0pt]{(\textit{c})}{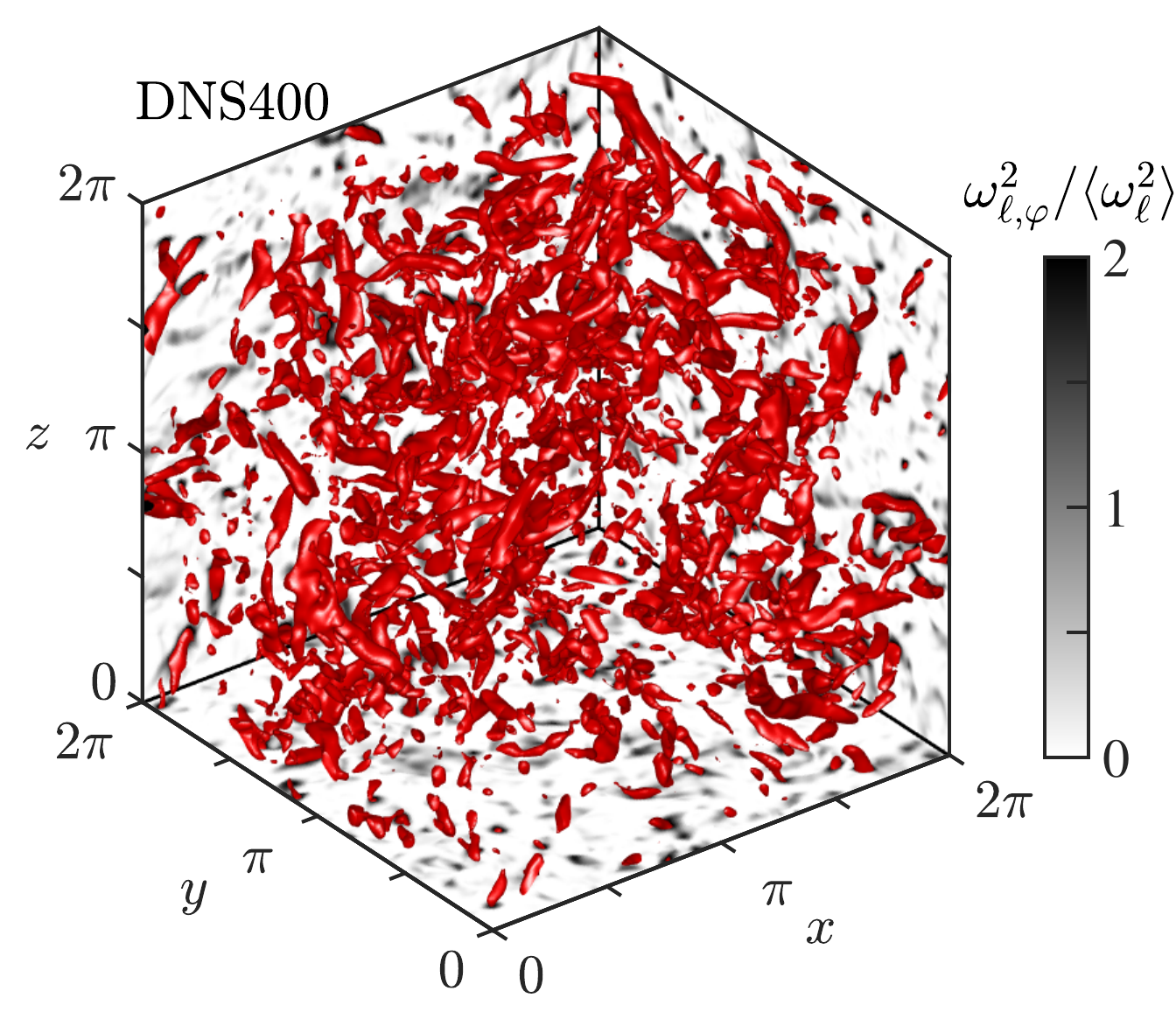}{}
    \subfigimg[width=0.40\linewidth,pos=ul,vsep=10pt,hsep=0pt]{(\textit{d})}{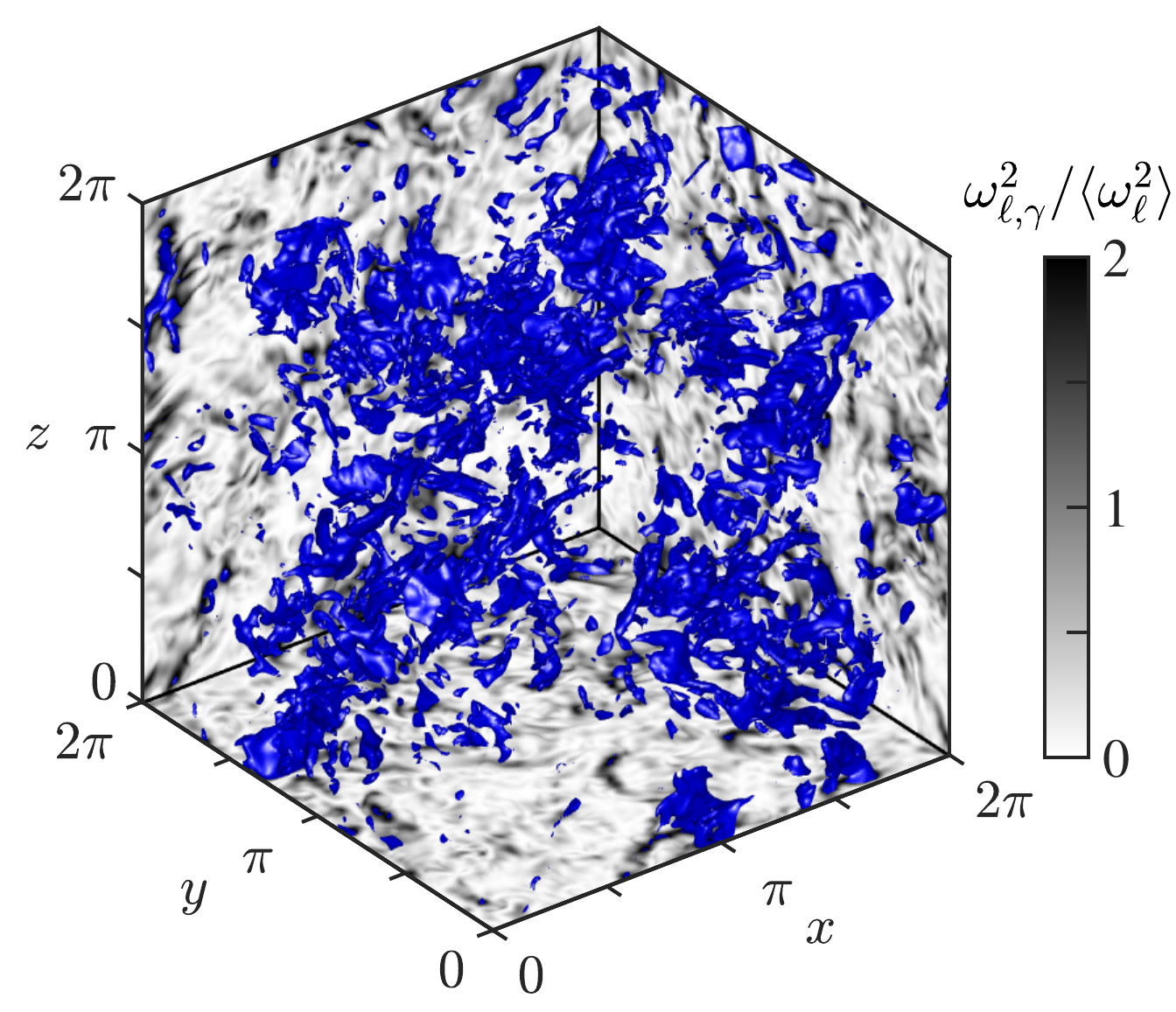}{}
    \subfigimg[width=0.40\linewidth,pos=ul,vsep=10pt,hsep=0pt]{(\textit{e})}{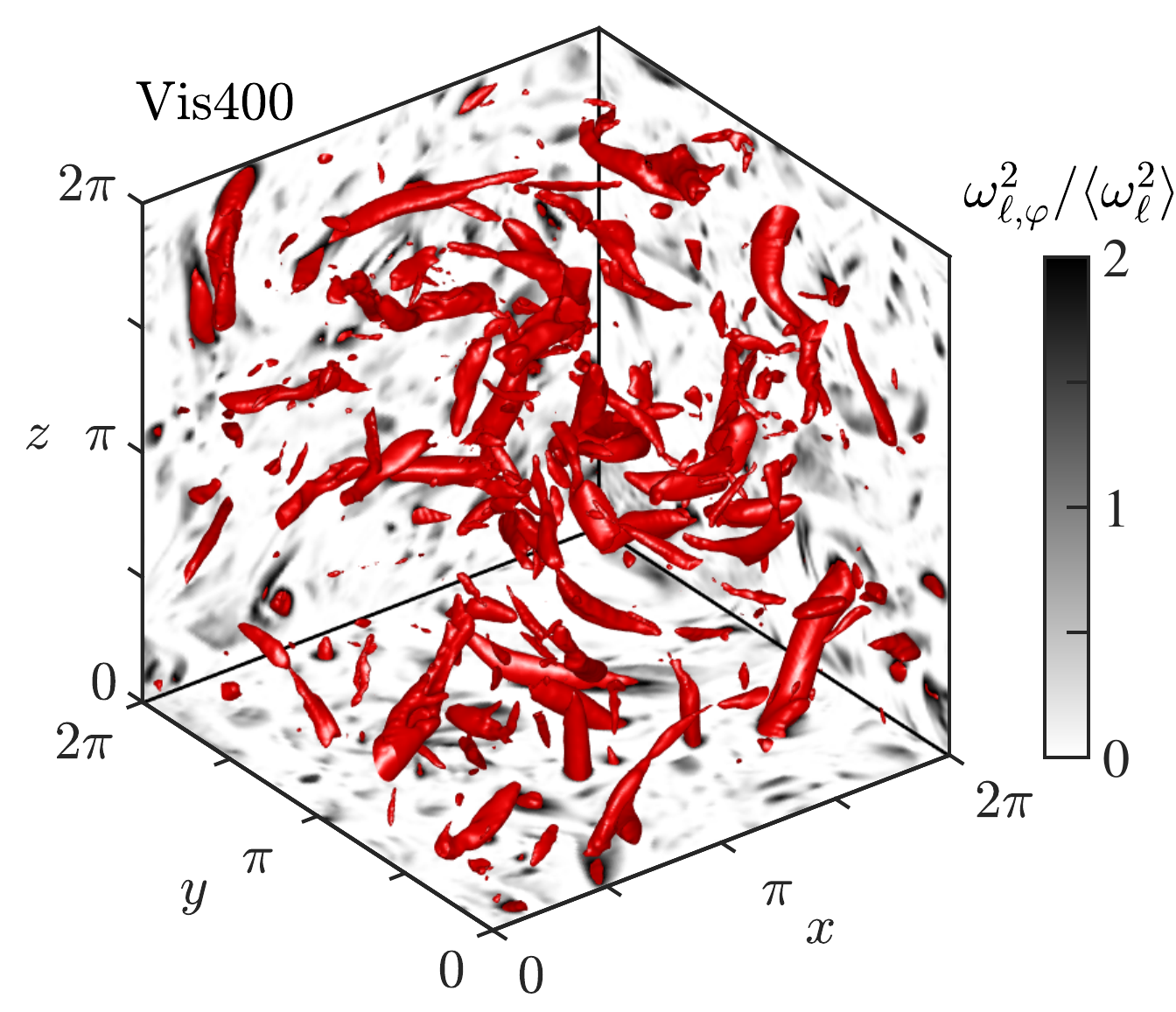}{}
    \subfigimg[width=0.40\linewidth,pos=ul,vsep=10pt,hsep=0pt]{(\textit{f})}{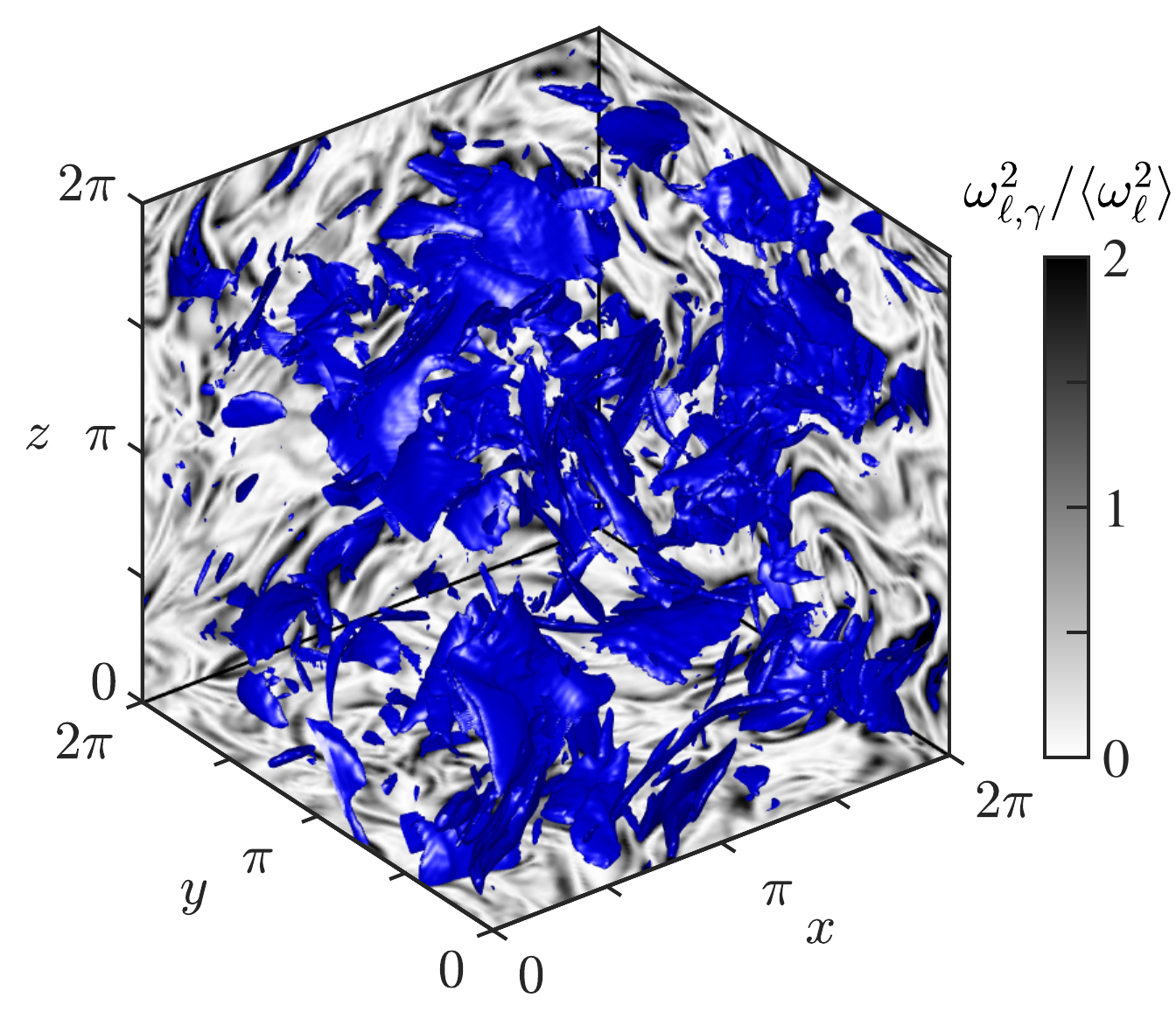}{}
    \subfigimg[width=0.40\linewidth,pos=ul,vsep=10pt,hsep=0pt]{(\textit{g})}{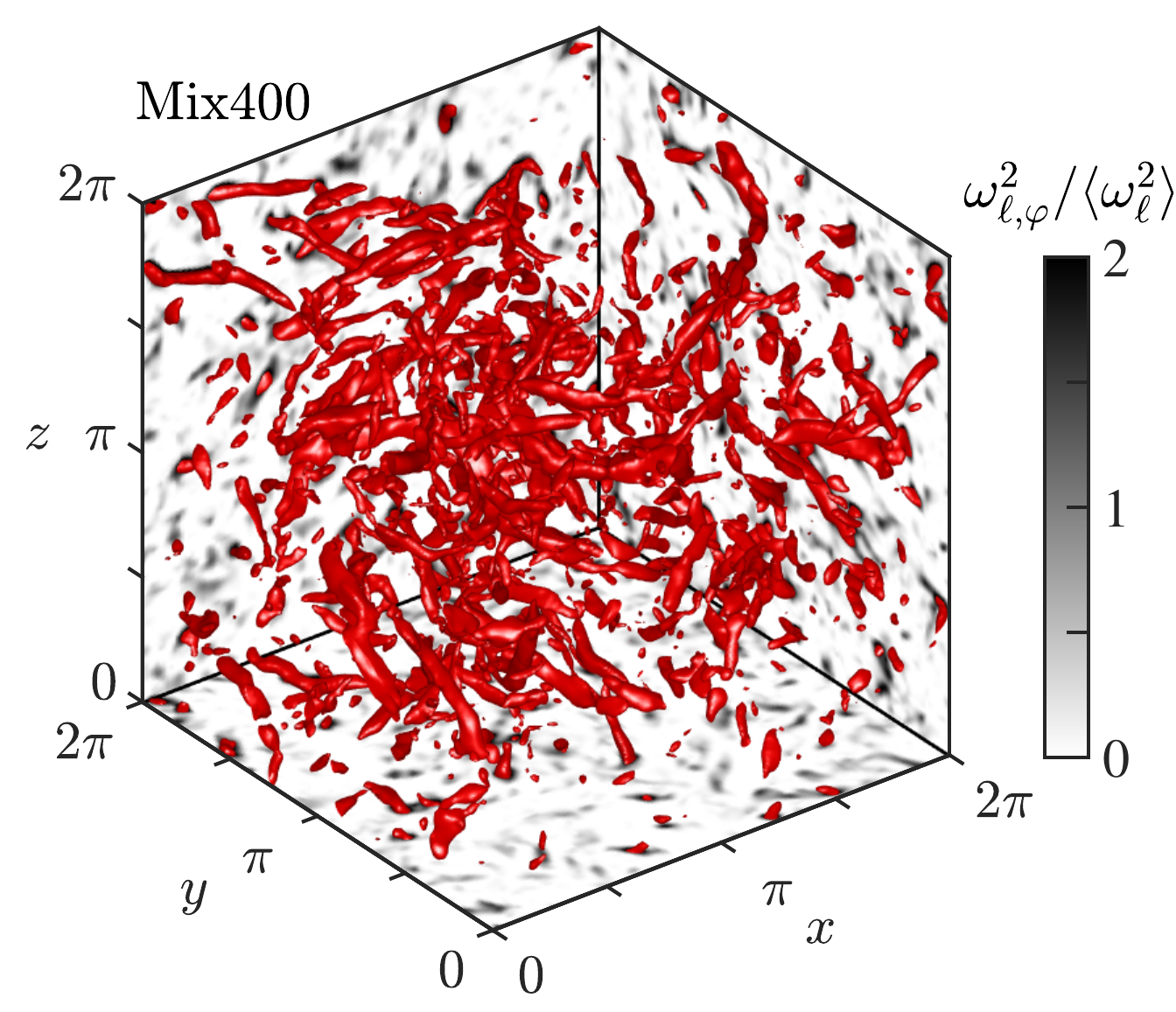}{}
    \subfigimg[width=0.40\linewidth,pos=ul,vsep=10pt,hsep=0pt]{(\textit{h})}{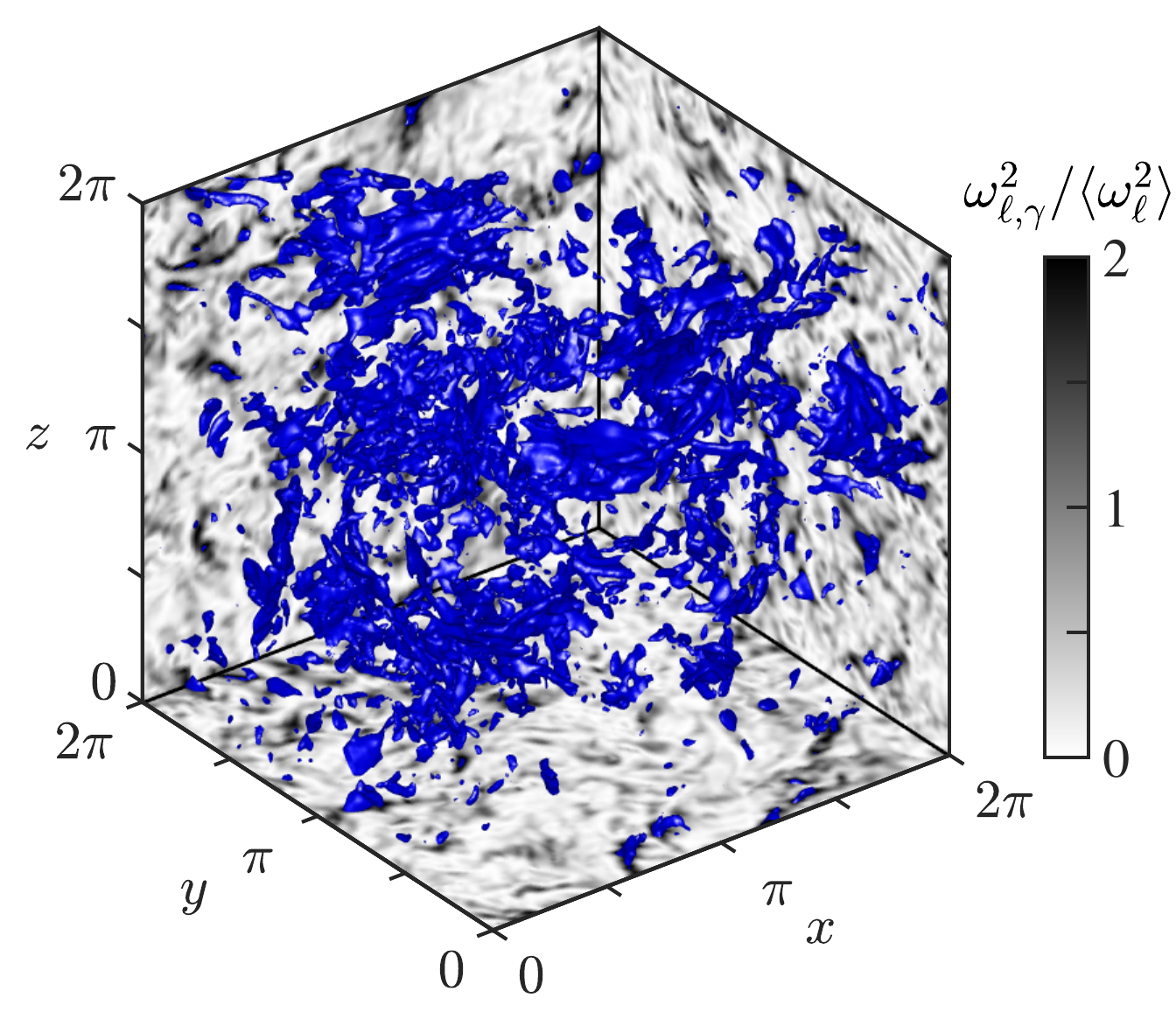}{}
    \caption{Vortical flow structures associated with rigid rotation and shear vorticity for an unfiltered ($\ell = 0$) DNS at $Re_\lambda \approx 61$ (\textit{a},\textit{b}), DNS400 filtered at $2\ell/\eta = 48$ (\textit{c},\textit{d}) and Vis400 (\textit{e},\textit{f}) and Mix400 (\textit{g},\textit{h}) at the LES filter width, $2\ell_{LES}/\eta = 48$. The \copyedit{greyscale} visualizations depict the strengths of rigid rotation, $\omega_{\ell,\varphi}^2$, and shear vorticity, $\omega_{\ell,\gamma}^2$, normalized by the spatially averaged vorticity strength, $\big\langle \omega_\ell^2 \big\rangle$, and the isosurfaces represent $\omega_{\ell,\varphi}^2 \big/ \big\langle \omega_\ell^2 \big\rangle = 2$ (red) and $\omega_{\ell,\gamma}^2 \big/ \big\langle \omega_\ell^2 \big\rangle = 2$ (blue).}
    \label{fig:vort_RR_SH}
\end{figure}

\autoref{fig:vort_RR_SH} depicts the vortical structures associated with rigid rotation and shear vorticity. The unfiltered DNS at $Re_\lambda \approx 61$ confirms that rigid rotation and shear vorticity are associated with tube-like and sheet-like vortical structures, respectively. The structures for this case qualitatively resemble typical structures from DNS400 (not shown) in a subdomain of size $(2\pi/15)^3$. This resemblance reflects that the Kolmogorov scale for $Re_\lambda \approx 61$ is roughly \copyedit{15} times larger than the Kolmogorov scale for $Re_\lambda \approx 400$. When filtered at $2\ell/\eta = 48$, DNS400 still produces rigid rotation structures that resemble vortex tubes. However, the shear vorticity structures appear less sharp (i.e. more blob-like) since the filter scale is larger than the typical thickness of small-scale shear layers (i.e. $2\ell > \delta_\gamma$). This highlights that the decreasing relative contribution from shearing with increasing filter width observed in \cref{fig:partFiltDNS} reflects the progressive smoothing of the vorticity profiles across shear layers in the subinertial range. Regardless of the filter scale, shear vorticity tends to be more space-filling than rigid rotation, which is consistent with previous findings \citep{Wat2024}.

The LES visualizations in \cref{fig:vort_RR_SH} provide insight into the structural implications of the partitioning statistics in \cref{fig:partFiltLES}. Just as the multiscale partitioning for Mix400 resembles that of DNS400 filtered at $2\ell/\eta = 48$, the corresponding vortical flow structures are also qualitatively similar. Likewise, the vortical flow structures for Vis400 resemble those of the unfiltered DNS at $Re_\lambda \approx 61$, consistent with the collapse of their multiscale partitioning statistics in \cref{fig:partFiltLES}(\textit{b}), where the Vis400 partitioning is replotted as a function of $2\ell_*/\eta_*$. The distinctive sheet-like vortical flow structures observed for these cases are significantly less prominent in the Mix400 and filtered DNS400 visualizations. These observations augment previous observations made using two-dimensional snapshots of out-of-plane vorticity \citep{Kam2024} by unambiguously identifying and distinguishing the three-dimensional imprint of vortex tubes and shear layers in a principled manner.

The visualizations \copyedit{analysed} in this section provide structural validation for the observations made in \textsection \ref{sec:results:VGT} using the multiscale partitioning statistics. Specifically, they highlight that the enhanced shearing at small scales reflects contributions from sharp velocity gradients across small-scale shear layers, which are softened upon filtering. They also highlight that the flow structures produced by the eddy viscosity model resemble those of an unfiltered DNS at a lower Reynolds number, whereas the structures produced by the mixed model resemble those of a filtered DNS at the same Reynolds number. These results emphasize that the ability of an LES to produce realistic multiscale flow structures can be highly sensitive to the choice of closure model. Together, the DNS and LES results highlight that a key advantage of the normality-based analysis is its ability to distinguish vortex cores from shear layers in a manner that would be obscured by an analogous symmetry-based analysis (e.g. of vorticity alone). In \textsection \ref{sec:results:energy}, we use this expressivity to provide insight into flow structures that contribute to various mechanisms of interscale energy transfer.

\subsection{Interscale energy transfer analysis}\label{sec:results:energy}

As formulated in \textsection \ref{sec:theory}, we characterize mechanisms responsible for interscale energy transfer by identifying how the normality-based constituents of the filtered VGT contribute to the multiscale velocity gradient expansion of $\iPi^\ell$. Numerically evaluating this expansion at a given scale, $\ell$, involves computing scale-space integrals over scales $0 \leq \theta \leq \ell$. Following previous studies \citep{Joh2020,Joh2021}, the scale-space integrals for each $\ell$ are discretized using the trapezoidal rule with filter widths that are logarithmically spaced such that $\Delta{\rm log}_{10}\theta = 0.376$ for the DNS cases and $\Delta{\rm log}_{10}\theta = 0.188$ for the LES cases. We have confirmed that our results are insensitive to the details of these discretization schemes, including the smallest non-zero filter width ($2\theta/\eta = 0.75$) for the DNS cases. They are also insensitive to the number of snapshots used to compute the energy transfer statistics. To reduce computational costs, we only use two snapshots to compute these statistics for the DNS cases. This is sufficient since, as demonstrated in \aref{sec:app:collapse}, the velocity gradient statistics for these cases are approximately converged even with a single snapshot. As depicted in \cref{fig:spectra}(\textit{a}), the collapse of our symmetry-based cascade rate computations onto those of \citet{Joh2020,Joh2021} further validates the statistical convergence of our results.

 \subsubsection{\copyedit{The} DNS cases: multiscale mechanisms in the energy cascade}\label{sec:results:energy:DNS}

\begin{figure}
    \centering
    \subfigimg[width=0.49\linewidth,pos=ul,vsep=7pt,hsep=1.5pt]{(\textit{a})}{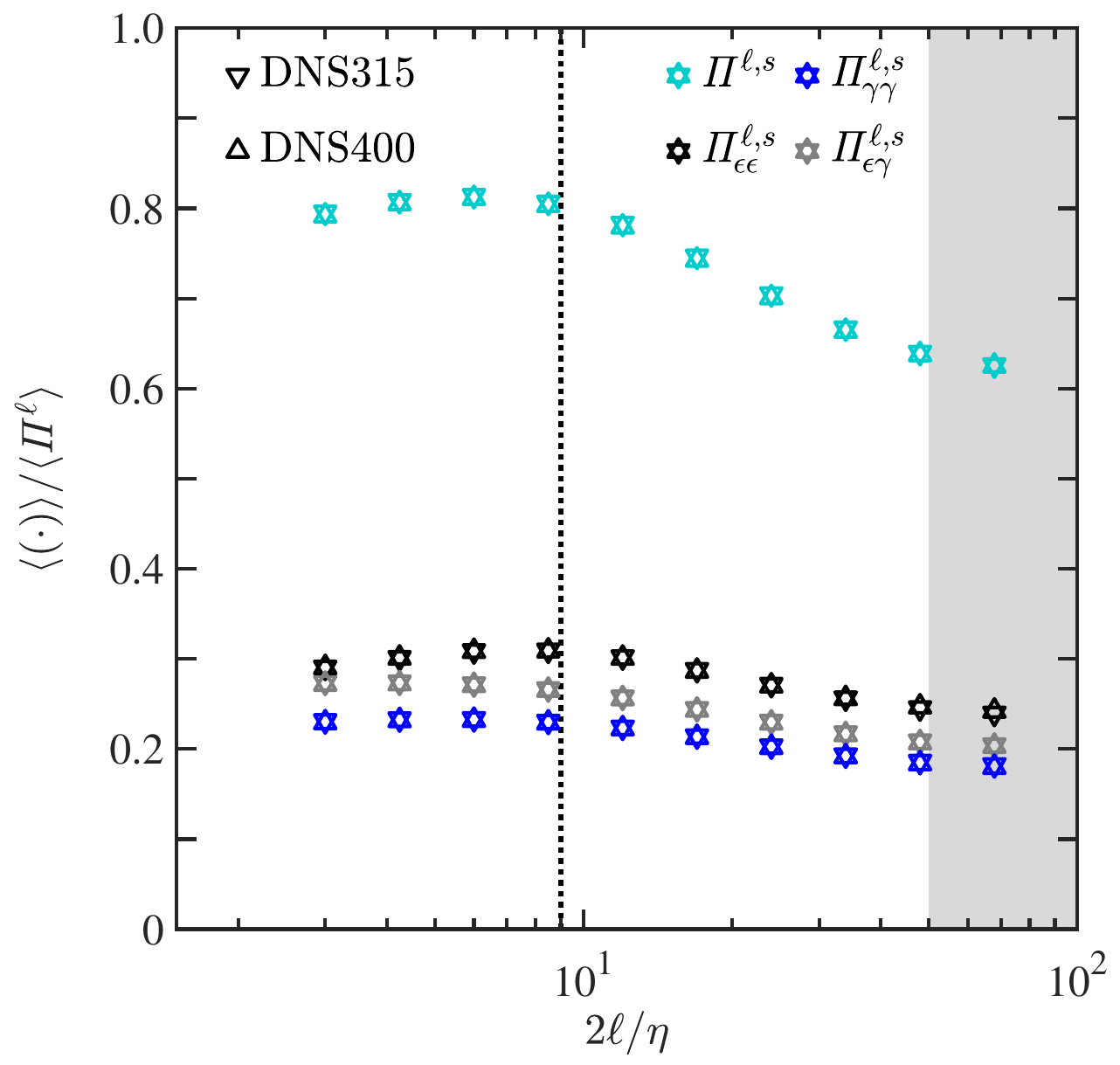}{}
    \subfigimg[width=0.49\linewidth,pos=ul,vsep=7pt,hsep=1.5pt]{(\textit{b})}{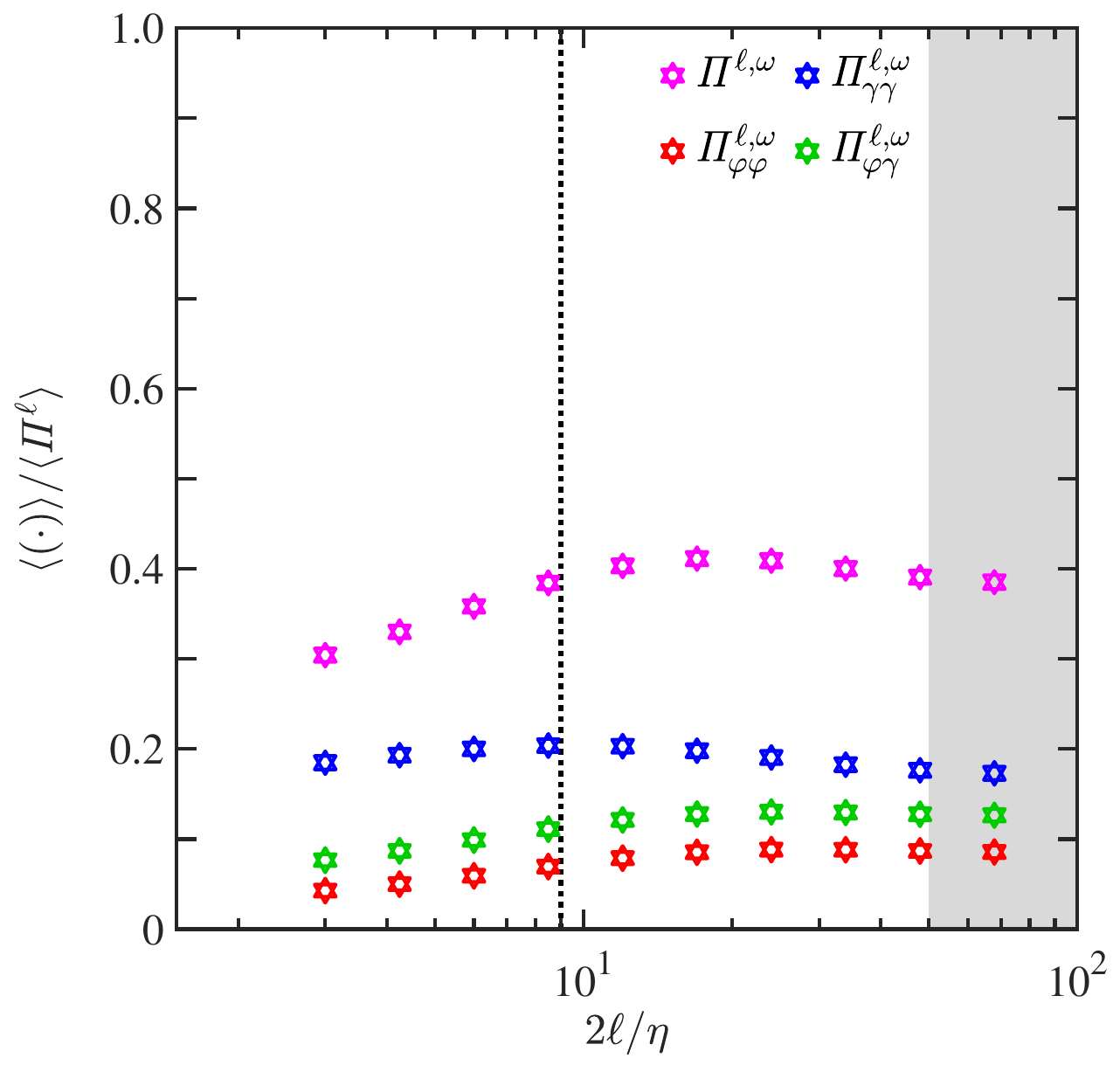}{}
    \subfigimg[width=0.49\linewidth,pos=ul,vsep=7pt,hsep=1.5pt]{(\textit{c})}{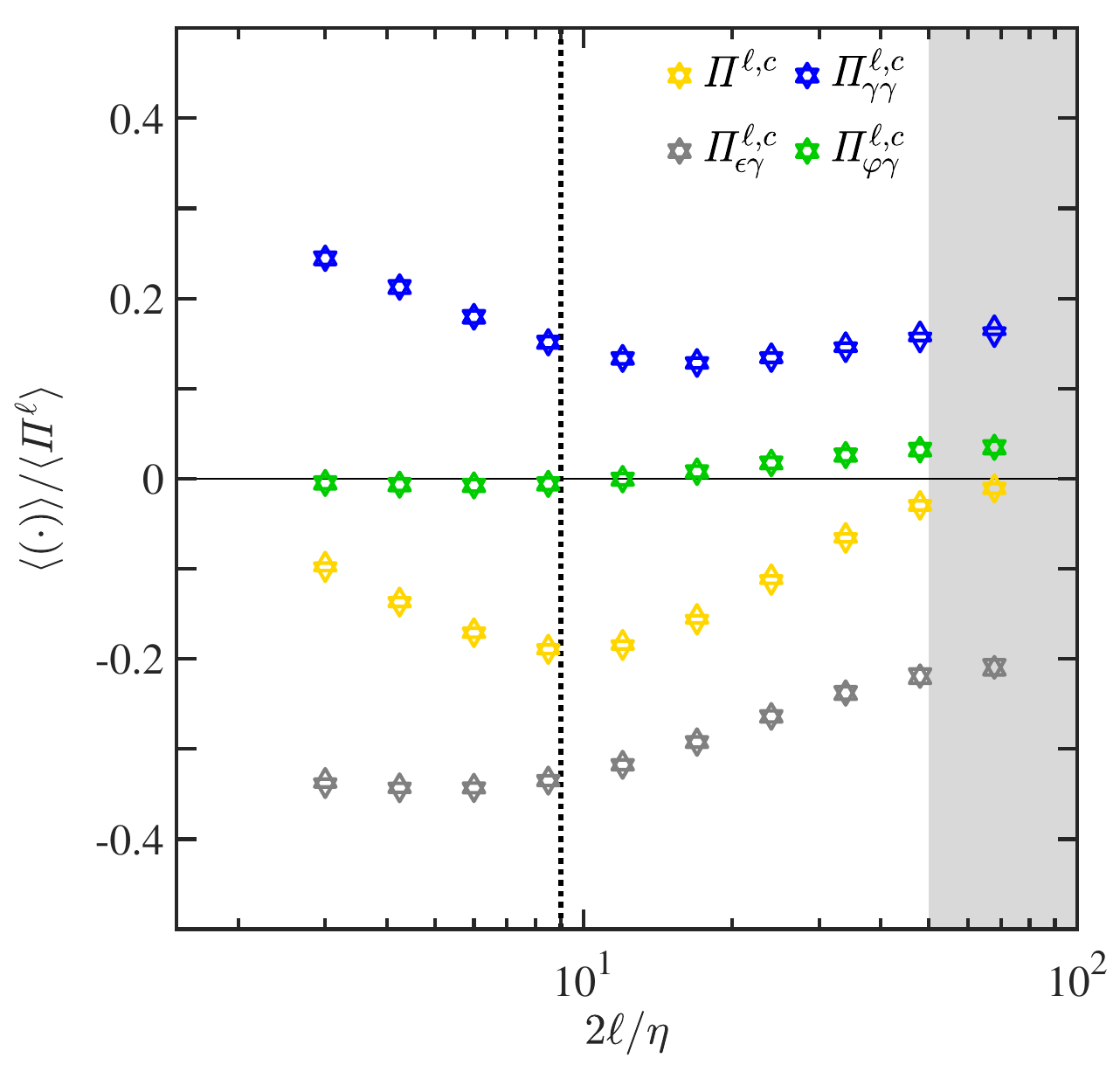}{}
    \subfigimg[width=0.49\linewidth,pos=ul,vsep=7pt,hsep=1.5pt]{(\textit{d})}{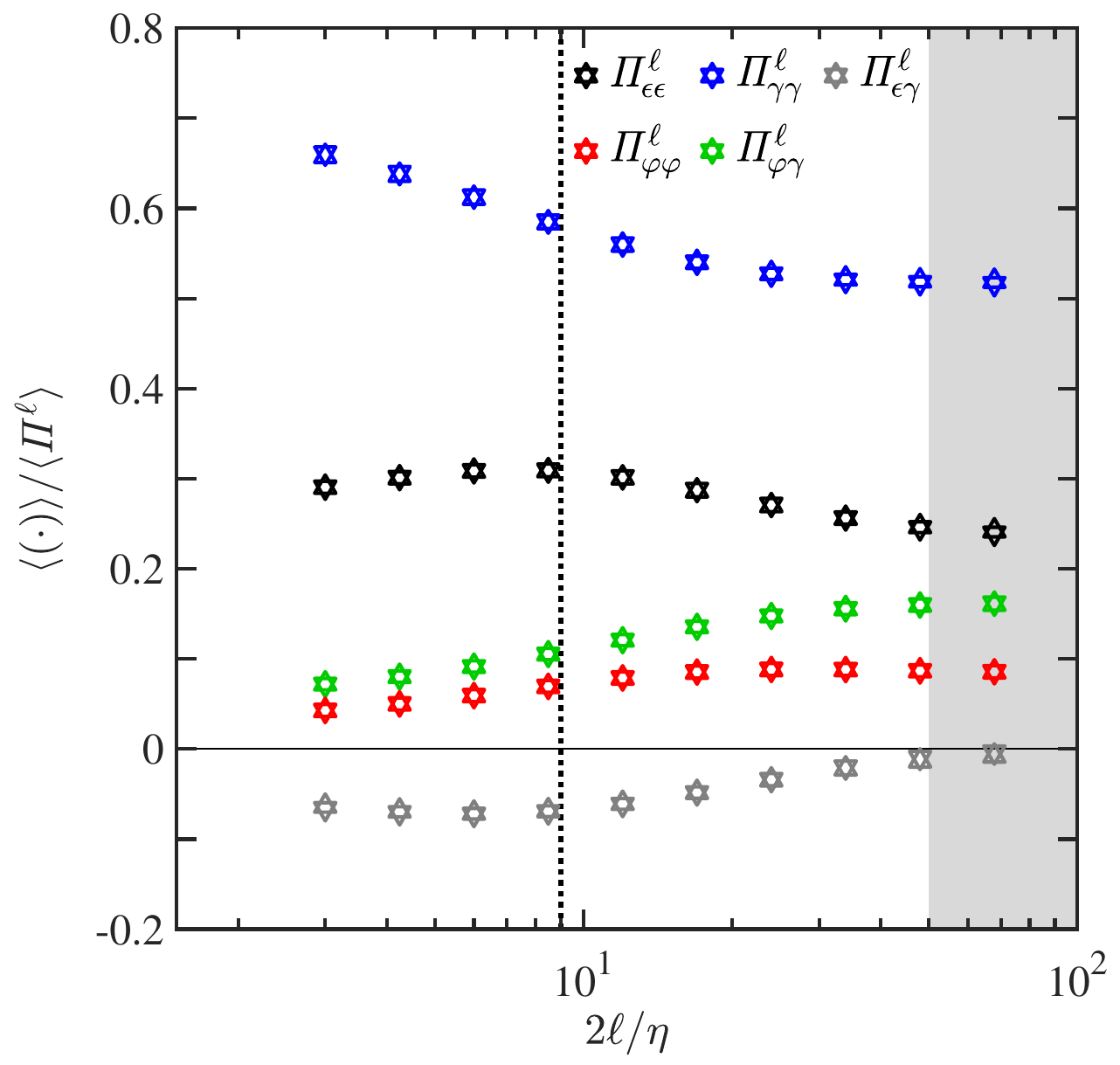}{}
    \caption{Normality-based contributions to interscale energy transfer for the DNS cases. The contributions represent multiscale \copyedit{SS} (\textit{a}), \copyedit{VS} (\textit{b}), strain--vorticity covariance (\textit{c}) and aggregates across these three mechanisms (\textit{d}). The vertical dotted lines represent the typical thickness of small-scale shear layers, $\delta_\gamma = 9\eta$, and the shaded regions capture the bottom of the inertial range for DNS400.}
    \label{fig:partCascDNS}
\end{figure}

\autoref{fig:partCascDNS} summarizes the normality-based contributions to interscale energy transfer for the DNS cases. Since these contributions are formulated by decomposing the multiscale velocity gradient expansion of the residual stress tensor, they represent the straining of flow features at scales $\theta \leq \ell$ by the flow at scale $\ell$. We specifically identify contributions associated with multiscale \copyedit{SS}, \copyedit{VS} and strain--vorticity covariance, as expressed in (\ref{eq:iPi_s_nor}), (\ref{eq:iPi_w_nor}) and (\ref{eq:iPi_c_nor}), respectively.

The multiscale \copyedit{SS} term, $\iPi^{\ell,s}$, is known to provide the strongest contribution to the energy cascade from the perspective of the symmetry-based analysis \citep{Joh2020,Joh2021}. As shown in \cref{fig:partCascDNS}(\textit{a}), the contributions of normal straining, shear straining and their interaction at scales $\theta \leq \ell$ to this term are all significant, with normal straining providing the strongest contribution. The multiscale \copyedit{VS} term, $\iPi^{\ell,\w}$, also contributes significantly to forward energy transfer across scale $\ell$. As shown in \cref{fig:partCascDNS}(\textit{b}), the stretching of shear vorticity at scales $\theta \leq \ell$ by the flow at scale $\ell$ provides the strongest contribution to this term, exceeding the contributions associated with rigid rotation and shear--rotation interactions. Since shear vorticity and rigid rotation at a given scale are associated with shear layers and vortex cores, respectively, our results highlight the stretching of small-scale shear layers as a significant mechanism underlying multiscale \copyedit{VS}. However, the significant contribution from shear--rotation interactions, which are typically strongest between the cores and boundaries of vortex tubes (see \aref{sec:app:Burgers} and \citet{Aru2024}), suggests that \copyedit{VS} near the boundaries of vortex tubes may also contribute to some fraction of $\iPi^{\ell,\w}_{\gamma\gamma}$. Nevertheless, the strong association of shear vorticity with sheet-like vortical structures in the visualizations in \textsection \ref{sec:results:structures} highlights that, in addition to vortex tubes, shear layers are essential to the description of multiscale \copyedit{VS}.

In contrast to the forward energy transfer produced by $\iPi^{\ell,s}$ and $\iPi^{\ell,\w}$, the net contribution of the strain--vorticity covariance term, $\iPi^{\ell,c}$, produces upscale energy transfer that drives the bottleneck effect in the subinertial range (see \cref{fig:spectra}). Therefore, the normality-based decomposition of $\iPi^{\ell,c}$, as shown in \cref{fig:partCascDNS}(\textit{c}), provides insight into the flow features that contribute to this backscatter. The $\iPi^{\ell,c}_{\epsilon\gamma}$ and $\iPi^{\ell,c}_{\gamma\gamma}$ terms represent the covariances of normal straining and shear straining (respectively) with shear vorticity at scales $\theta \leq \ell$ and can thus be associated with small-scale shear layers. The $\iPi^{\ell,c}_{\varphi\gamma}$ term represents the covariance of shear straining with rigid rotation at scales $\theta \leq \ell$ and can thus be associated with small-scale vortex tubes. \rev{Since $\big\langle \iPi^{\ell,c}_{\varphi\gamma} \big\rangle$ is negligible throughout the subinertial range, the shear vorticity terms dominate $\big\langle \iPi^{\ell,c} \big\rangle \approx \big\langle \iPi^{\ell,c}_{\epsilon\gamma} \big\rangle + \big\langle \iPi^{\ell,c}_{\gamma\gamma} \big\rangle < 0$ at these scales.} This suggests that small-scale shear layers \rev{(rather than vortex tubes)} are primarily responsible for the backscatter that produces the bottleneck effect in the subinertial range of the energy cascade. It is also consistent with the fact that the filter width associated with the peak relative contribution of the backscatter coincides with the empirical shear layer thickness (at $2\ell \approx \delta_\gamma = 9\eta$). In particular, the covariance of normal straining with shear vorticity is responsible for the backscatter, and its contribution is tempered by a net positive contribution from the covariance of shear straining with shear vorticity. Further work is required to determine whether these findings are consistent with the hypothesis that a mechanism reminiscent of vortex thinning in two-dimensional turbulence drives the backscatter associated with $\iPi^{\ell,c}$ in the subinertial range of three-dimensional turbulence \citep{Joh2021}. One such mechanism might involve the flattening of a vortex into a shear layer that results in a misalignment between the strain rates at small and large scales.

Combining the contributions from \cref{fig:partCascDNS}(\textit{a}--\textit{c}), \cref{fig:partCascDNS}(\textit{d}) summarizes the normality-based contributions to the total interscale energy transfer, as expressed in (\ref{eq:iPi_nor}). The contribution of $\iPi^\ell_{\gamma\gamma}$ accounts for more than half of the energy transfer in the subinertial and inertial ranges, with significant contributions to multiscale \copyedit{SS}, \copyedit{VS} and strain--vorticity covariance. This highlights that the energy transfer across scale $\ell$ is largely attributed to the straining of shear layers at scales $\theta \leq \ell$. The next strongest contribution is due to $\iPi^\ell_{\epsilon\epsilon}$, which represents \copyedit{SS} associated with normal straining at scales $\theta \leq \ell$. The combined contribution of these two terms becomes increasingly dominant at small scales and accounts for more than 90\% of the energy transfer when $2\ell/\eta \sim O(1)$. The \copyedit{VS} associated with rigid rotation, $\iPi^\ell_{\varphi\varphi}$, and the cascade rate associated with shear--rotation interactions, $\iPi^\ell_{\varphi\gamma}$, which is primarily due to \copyedit{VS}, also account for significant (albeit smaller) forward energy transfer. From the normality-based perspective, the $\iPi^\ell_{\epsilon\gamma}$ term, which represents the interaction between normal straining and pure shearing at scales $\theta \leq \ell$, is the only term with a negative net contribution to energy transfer in the subinertial range. This negative contribution is produced by the strain--vorticity covariance term, $\iPi^{\ell,c}_{\epsilon\gamma}$, and tempered by the \copyedit{SS} term, $\iPi^{\ell,s}_{\epsilon\gamma}$. It becomes negligible in the inertial range, which is consistent with our claim that it is associated with the subinertial bottleneck effect. From strongest to weakest, the relative contributions of the normality-based mechanisms in the inertial range can be summarized as $\big\langle \iPi^\ell_{\gamma\gamma} \big\rangle \big/ \big\langle \iPi^\ell \big\rangle \approx 0.52$, $\big\langle \iPi^\ell_{\epsilon\epsilon} \big\rangle \big/ \big\langle \iPi^\ell \big\rangle \approx 0.24$, $\big\langle \iPi^\ell_{\varphi\gamma} \big\rangle \big/ \big\langle \iPi^\ell \big\rangle \approx 0.16$, $\big\langle \iPi^\ell_{\varphi\varphi} \big\rangle \big/ \big\langle \iPi^\ell \big\rangle \approx 0.08$ and $\big\langle \iPi^\ell_{\epsilon\gamma} \big\rangle \big/ \big\langle \iPi^\ell \big\rangle \approx 0$. These relative contributions are qualitatively similar to the corresponding partitioning contributions in the inertial and subinertial ranges, which highlights that the energy transfer across a given scale is related to the strength of the filtered velocity gradients that produce it.

\begin{figure}
    \centering
    \subfigimg[width=0.49\linewidth,pos=ul,vsep=7pt,hsep=1.5pt]{(\textit{a})}{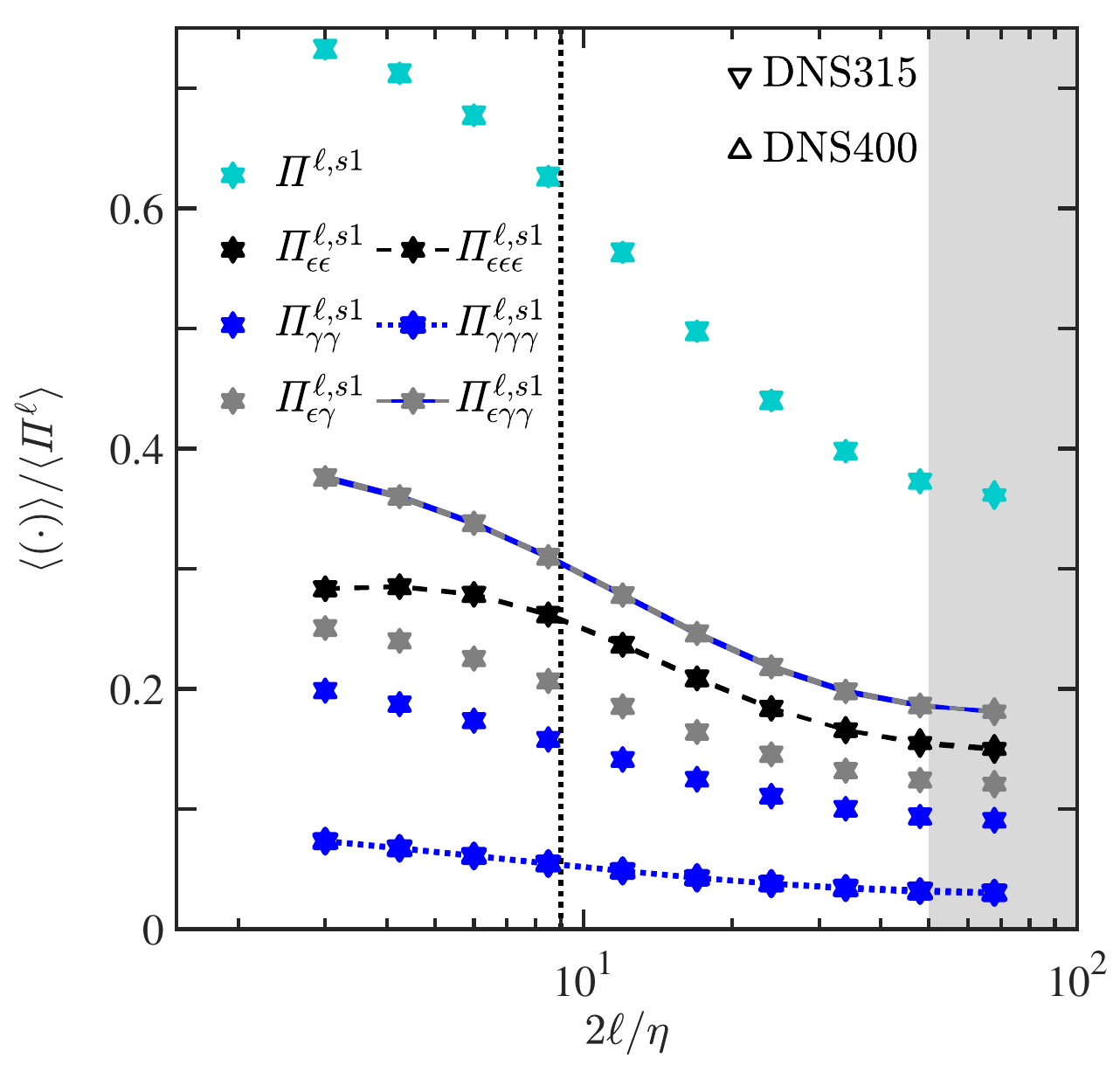}{}
    \subfigimg[width=0.49\linewidth,pos=ul,vsep=7pt,hsep=1.5pt]{(\textit{b})}{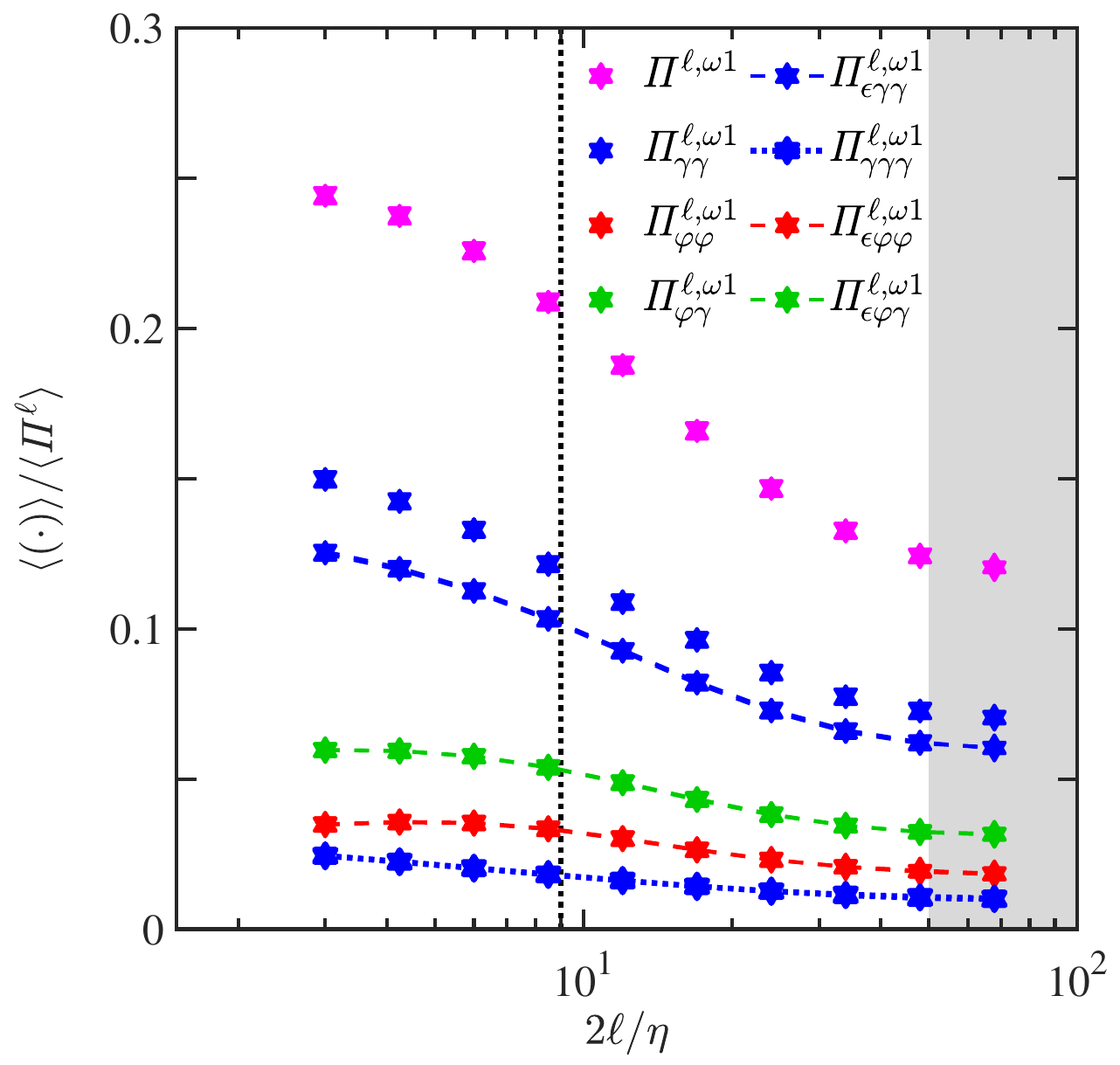}{}
    \subfigimg[width=0.49\linewidth,pos=ul,vsep=7pt,hsep=1.5pt]{(\textit{c})}{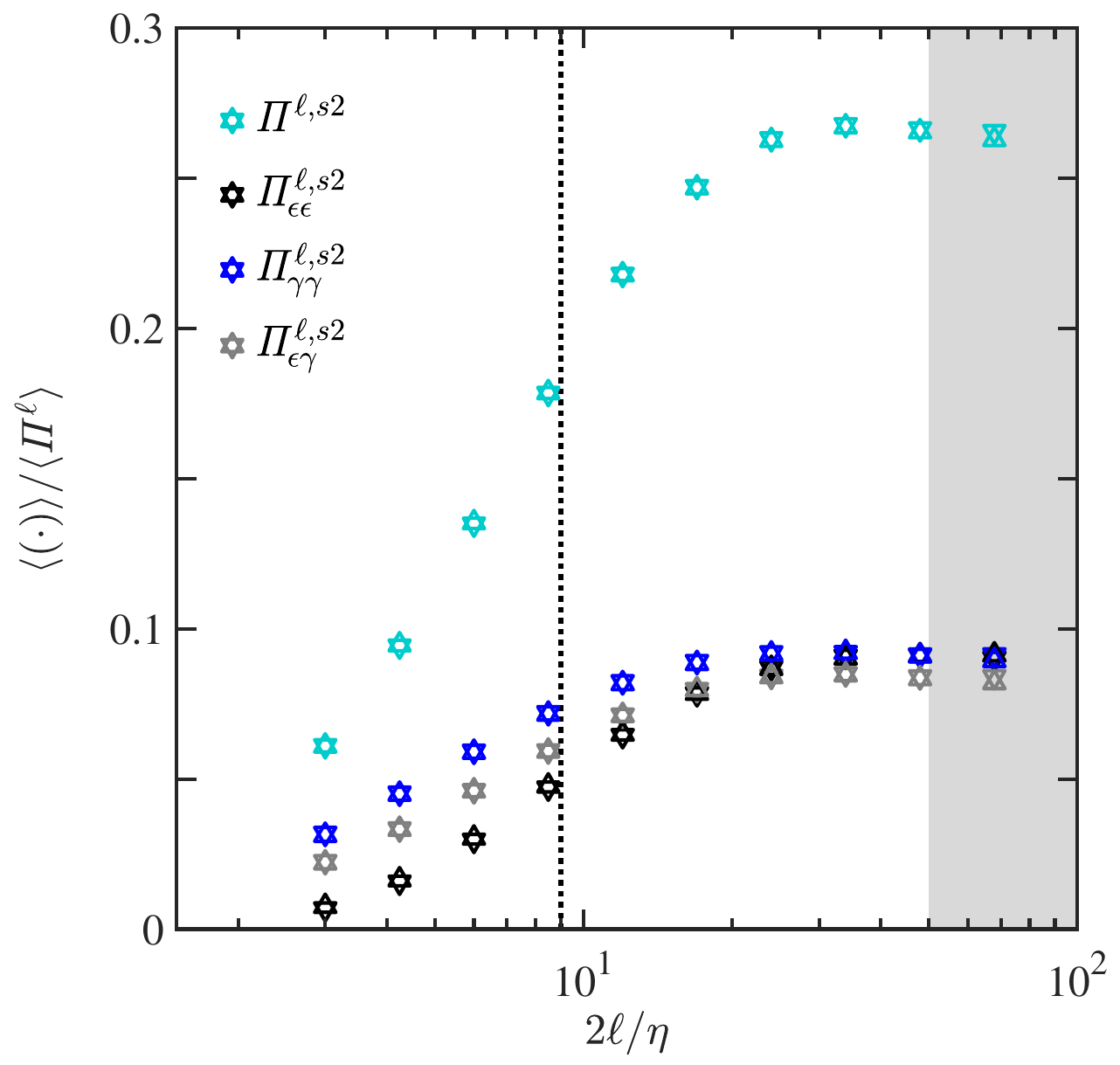}{}
    \subfigimg[width=0.49\linewidth,pos=ul,vsep=7pt,hsep=1.5pt]{(\textit{d})}{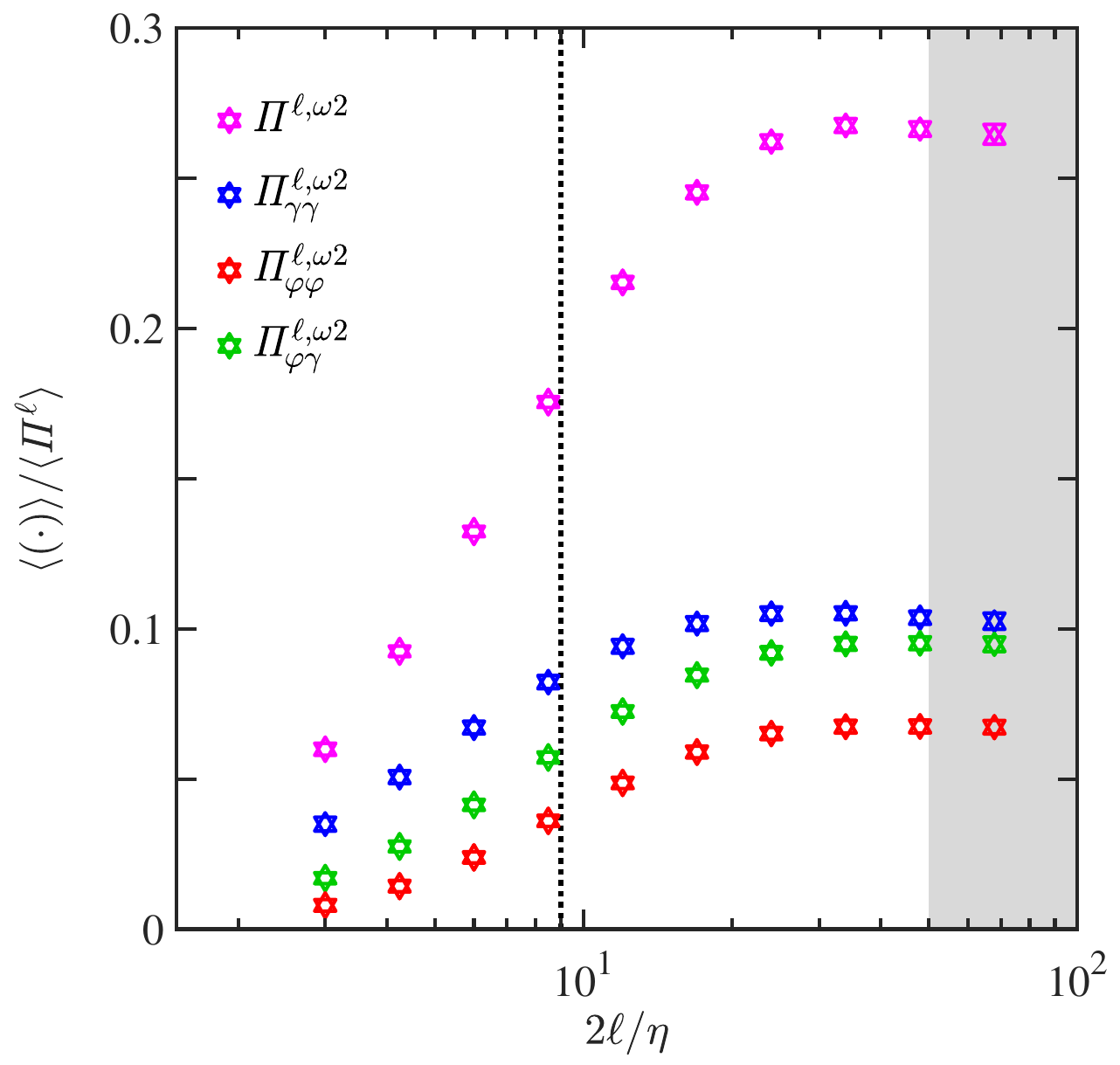}{}
    \caption{Normality-based contributions (as defined in \textsection \ref{sec:theory}) to the cascade rates associated with scale-local \copyedit{SS} (\textit{a}) and \copyedit{VS} (\textit{b}) and scale-\copyedit{non-local} \copyedit{SS} (\textit{c}) and \copyedit{VS} (\textit{d}). The vertical dotted lines represent the typical thickness of small-scale shear layers, $\delta_\gamma = 9\eta$, and the shaded regions capture the bottom of the inertial range for DNS400.}
    \label{fig:partCascDNS_loc}
\end{figure}

In \cref{fig:partCascDNS_loc}, we refine the results presented in \cref{fig:partCascDNS} by identifying scale-local and scale-\copyedit{non-local} contributions to energy transfer associated with \copyedit{SS} and \copyedit{VS}, as formulated in \textsection \ref{sec:theory}. Our analysis focuses primarily on the scale-local terms with three subscripts, as defined in (\ref{eq:iPi_s1_3term}) and (\ref{eq:iPi_w1_3term}), to facilitate detailed interpretations of the underlying energy transfer mechanisms. The scale-local and scale-\copyedit{non-local} terms with two subscripts, as defined in (\ref{eq:iPi_s1_2term}) and (\ref{eq:iPi_w1_2term}), are also shown for completeness and reflect a direct decomposition of the terms in \cref{fig:partCascDNS}. They are related to the terms with three subscripts through the expressions in (\ref{eq:2to3_s1}) and (\ref{eq:2to3_w1}).

As depicted in \cref{fig:partCascDNS_loc}(\textit{a}) the $\iPi^{\ell,s1}_{\epsilon\gamma\gamma}$ term provides the strongest contribution to scale-local \copyedit{SS}. It can be interpreted as the normal straining of the strain rate associated with shear layers at scale $\ell$. Its contribution is closely followed by that of $\iPi^{\ell,s1}_{\epsilon\epsilon\epsilon}$, which reflects how the self-interaction of normal straining amplifies strain rates. Together, these two terms are responsible for more than 90\% of the scale-local \copyedit{SS} throughout the subinertial range and the bottom of the inertial range. Their contribution dominates that of $\iPi^{\ell,s1}_{\gamma\gamma\gamma}$, which reflects how the self-interaction of shear straining, which is associated with shear layers, amplifies strain rates. 

As depicted in \cref{fig:partCascDNS_loc}(\textit{b}), the $\iPi^{\ell,\w1}_{\epsilon\gamma\gamma}$ term provides the strongest contribution to scale-local \copyedit{VS}. It represents the normal straining of the shear vorticity at scale $\ell$ and can be interpreted as a `shear layer stretching' term. Its contribution is considerably stronger than that of $\iPi^{\ell,\w1}_{\epsilon\varphi\varphi}$, which represents the normal straining of rigid rotation at scale $\ell$ and can be interpreted as a `vortex core stretching' term. As discussed for multiscale \copyedit{VS}, this suggests that prototypical models of stretched vortex tubes (e.g. Burgers vortex tubes) may provide an incomplete picture of \copyedit{VS}. Supplementing these models with contributions from stretched shear layers, which are often \copyedit{modelled} using Burgers vortex layers, is likely to more accurately capture the flow features that contribute to scale-local \copyedit{VS}. The inclusion of stretched shear layers is particularly important given that they also contribute significantly to scale-local \copyedit{SS} through the $\iPi^{\ell,s1}_{\epsilon\gamma\gamma}$ term.

While $\iPi^\ell$ vanishes in the limit of $\ell \to 0$, \copyedit{SS} and \copyedit{VS} are still relevant since they modulate the strength of unfiltered velocity gradients. Therefore, the normality-based analysis of these mechanisms at $\ell = 0$ provides insight into their contributions to the dynamics of velocity gradients at small scales. \autoref{tab:VS_SS} summarizes the average contributions of the normality-based constituents of these mechanisms, which can be interpreted as the relative contributions of the terms in (\ref{eq:iPi_s1_3term}) and (\ref{eq:iPi_w1_3term}) in the limit of $\ell \to 0$. Consistent with our scale-local energy transfer analysis, the $\epsilon\gamma\gamma$ terms, associated with the normal straining of shear layers, are the most significant contributors to \copyedit{SS} and \copyedit{VS} at $\ell = 0$. Importantly, normal straining contributes significantly more to the stretching of shear layers than shear straining, which is represented by the $\gamma\gamma\gamma$ terms. This highlights that the Burgers vortex layer forms a reasonable (albeit crude) model for small-scale shear layers, as its \copyedit{SS} and \copyedit{VS} are entirely attributed to the $\epsilon\gamma\gamma$ terms. \aref{sec:app:Burgers} provides further analysis of the \copyedit{SS} and \copyedit{VS} associated with Burgers vortex layers and tubes.

\begin{table}
\begin{center}
\def~{\hphantom{0}}
\begin{tabular}{lrrrrrrr}
Case   & SS: $\epsilon\epsilon\epsilon$ & SS: $\epsilon\gamma\gamma$ & SS: $\gamma\gamma\gamma$ & VS: $\epsilon\varphi\varphi$ & VS: $\epsilon\gamma\gamma$ & VS: $\gamma\gamma\gamma$ & VS: $\epsilon\varphi\gamma$ \\[0pt]
       &       &       &       &       &       &       &       \\[-4pt]
DNS315 & 0.349 & 0.536 & 0.115 & 0.123 & 0.536 & 0.115 & 0.226 \\
DNS400 & 0.354 & 0.531 & 0.115 & 0.125 & 0.531 & 0.115 & 0.229 \\
\end{tabular}
\caption{Normality-based contributions to \copyedit{SS} and \copyedit{VS} for the unfiltered DNS cases. The contributions to SS and VS are normalized by $\big\langle - \A{S}{}{ij}\A{S}{}{ik}\A{S}{}{jk} \big\rangle$ and $\big\langle - \A{S}{}{ij}\A{\iW}{}{ik}\A{\iW}{}{jk} \big\rangle$, respectively, such that they sum to unity for each mechanism.}
\label{tab:VS_SS}
\end{center}
\end{table}

Our results are also qualitatively consistent with the statistics reported by \citet{Wat2020} using the original triple decomposition \citep{Kol2004,Kol2007}. However, their approach yields non-zero contributions from terms analogous to the $\epsilon\epsilon\gamma$ \copyedit{SS} term and the $\gamma\varphi\varphi$ and $\gamma\varphi\gamma$ \copyedit{VS} terms, which are identically zero in our normality-based formulation. This reflects that their approach relies on a (potentially sensitive) optimization problem to identify a frame that maximizes an interaction scalar. By contrast, our normality-based approach has unambiguous foundations derived from the form of the VGT in the principal reference frame, as expressed in (\ref{eq:VGT_nor}).

 \subsubsection{\copyedit{The} LES cases: closure models and the artificial bottleneck effect}\label{sec:results:energy:LES}

Whereas our analysis of energy transfer for the DNS cases involves scales ($0 \leq 2\ell/\eta \leq 67.9$) primarily associated with the subinertial range, our analysis of the LES cases is constrained to the resolved scales ($2\ell \geq 2\ell_{LES} = 48\eta$), which are primarily associated with the inertial range. This constraint poses a challenge for computing the multiscale expansion of $\sigma^\ell_{ij}$, as formulated in (\ref{eq:res_VGT}), since we do not have access to the unresolved motions at scales $\ell < \ell_{LES}$. For the LES cases, we compute the residual stress tensor at scales $\ell \geq \ell_{LES}$ as
\begin{equation}
    \sigma^\ell_{ij} = \ol{\sigma^{\ell_{LES}}_{ij}}^{\ell_*} + \underbrace{\ol{\ol{u}_i^{\ell_{LES}} \ol{u}_j^{\ell_{LES}}}^{\ell_*} -  \ol{\ol{u}_i^{\ell_{LES}}}^{\ell_*} \ol{\ol{u}_j^{\ell_{LES}}}^{\ell_*}}_{\textstyle \tilde{\sigma}^\ell_{ij}},
\end{equation}
where $\ell_* = \sqrt{\ell^2 - \ell_{LES}^2}$. Here, the first term represents the contribution from the unresolved motions, as \copyedit{modelled} by (\ref{eq:model_vis}) and (\ref{eq:model_mix}) for Vis400 and Mix400, respectively, and $\tilde{\sigma}^\ell_{ij}$ represents the contribution from the resolved flow field. The term $\tilde{\sigma}^\ell_{ij}$ contributes to the interscale energy transfer associated with the resolved motions, $\tilde{\iPi}^\ell = -\A{\ol{S}}{\ell}{ij} \tilde{\sigma}^\ell_{ij}$, which can be expanded analogously to (\ref{eq:iPi_VGT}) as
\begin{equation}\label{eq:iPi_resolved}
    \tilde{\iPi}^\ell = -\A{\ol{S}}{\ell}{ij} \bigintsss_{\;\ell_{LES}^2}^{\ell^2} {\rm d}\theta^2 \left( \ol{\A{\ol{A}}{\theta}{ik}\A{\ol{A}}{\theta}{jk}}^{\phi} \right) = -\ol{\A{\ol{S}}{\ell_{LES}}{ij}}^{\ell_*} \bigintsss_{\;0}^{\ell_*^2} {\rm d}\theta_*^2 \left( \ol{\ol{\A{\ol{A}}{\ell_{LES}}{ik}}^{\theta_*} \ol{\A{\ol{A}}{\ell_{LES}}{jk}}^{\theta_*}}^{\phi} \right),
\end{equation}
where $\theta_* = \sqrt{\theta^2 - \ell_{LES}^2}$, $\phi = \sqrt{\ell^2 - \theta^2} = \sqrt{\ell_*^2 - \theta_*^2}$, and the remaining energy transfer, $\iPi^\ell - \tilde{\iPi}^\ell$, is supplied by the closure model. We use analogous expansions, denoted with tildes, to capture the resolved part of other multiscale cascade rates (e.g. $\tilde{\iPi}^{\ell,s}$, $\tilde{\iPi}^{\ell,\w}$ and $\tilde{\iPi}^{\ell,c}$). The resolved scale-\copyedit{non-local} contributions to the cascade rates associated with \copyedit{SS} and \copyedit{VS} are defined as $\tilde{\iPi}^{\ell,s2} = \tilde{\iPi}^{\ell,s} - \iPi^{\ell,s1}$ and $\tilde{\iPi}^{\ell,\w2} = \tilde{\iPi}^{\ell,\w} - \iPi^{\ell,\w1}$, respectively. In this section, we use these resolved cascade rates to provide insight into the structural origins of the artificial bottleneck effect in LES, as depicted in \cref{fig:spectra}(\textit{b}).

\begin{figure}
    \centering
    \subfigimg[width=0.49\linewidth,pos=ul,vsep=5.5pt,hsep=1.5pt]{(\textit{a})}{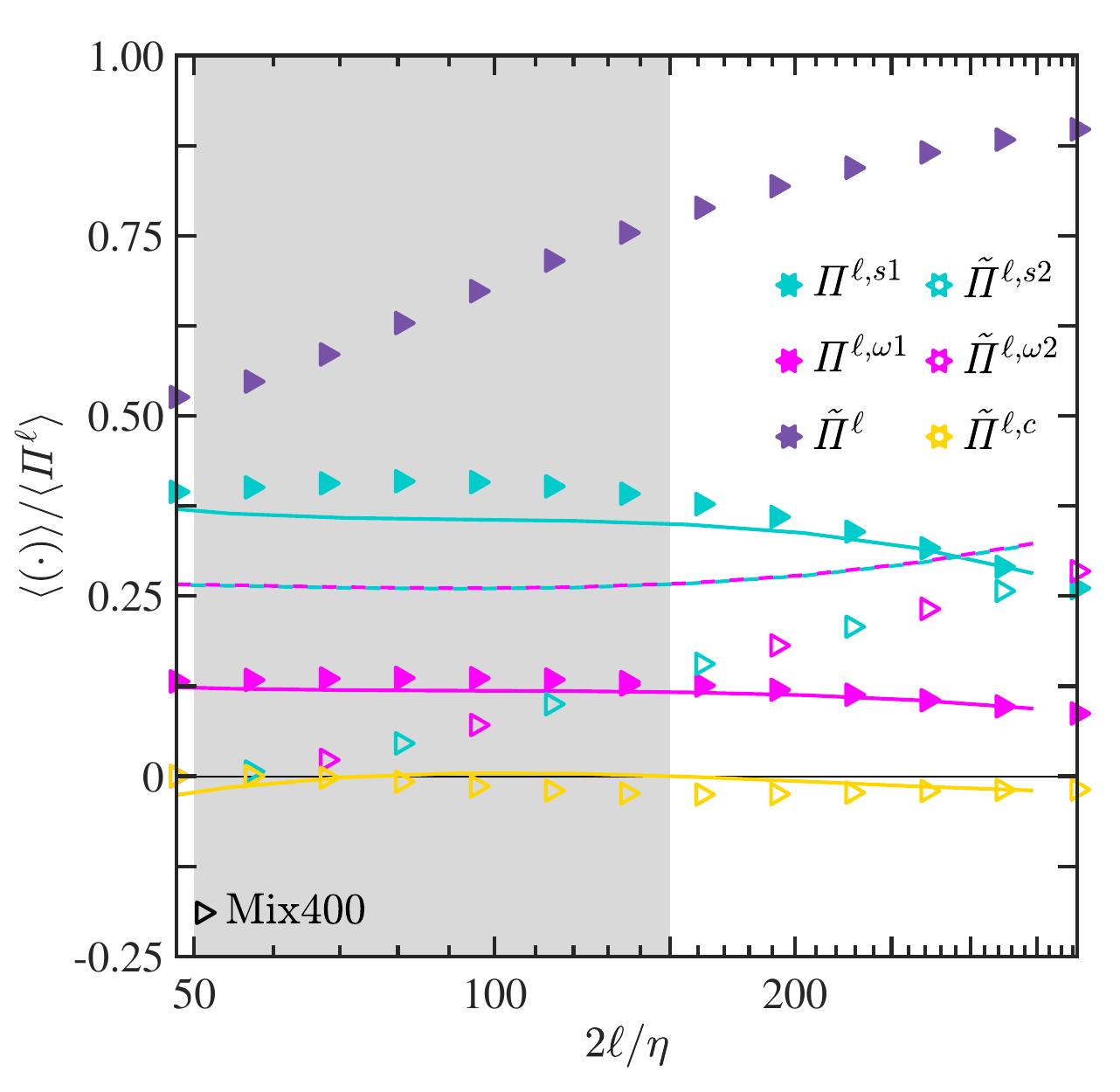}{trim=0pt 10pt 0pt 20pt,clip}
    \subfigimg[width=0.49\linewidth,pos=ul,vsep=5.5pt,hsep=1.5pt]{(\textit{b})}{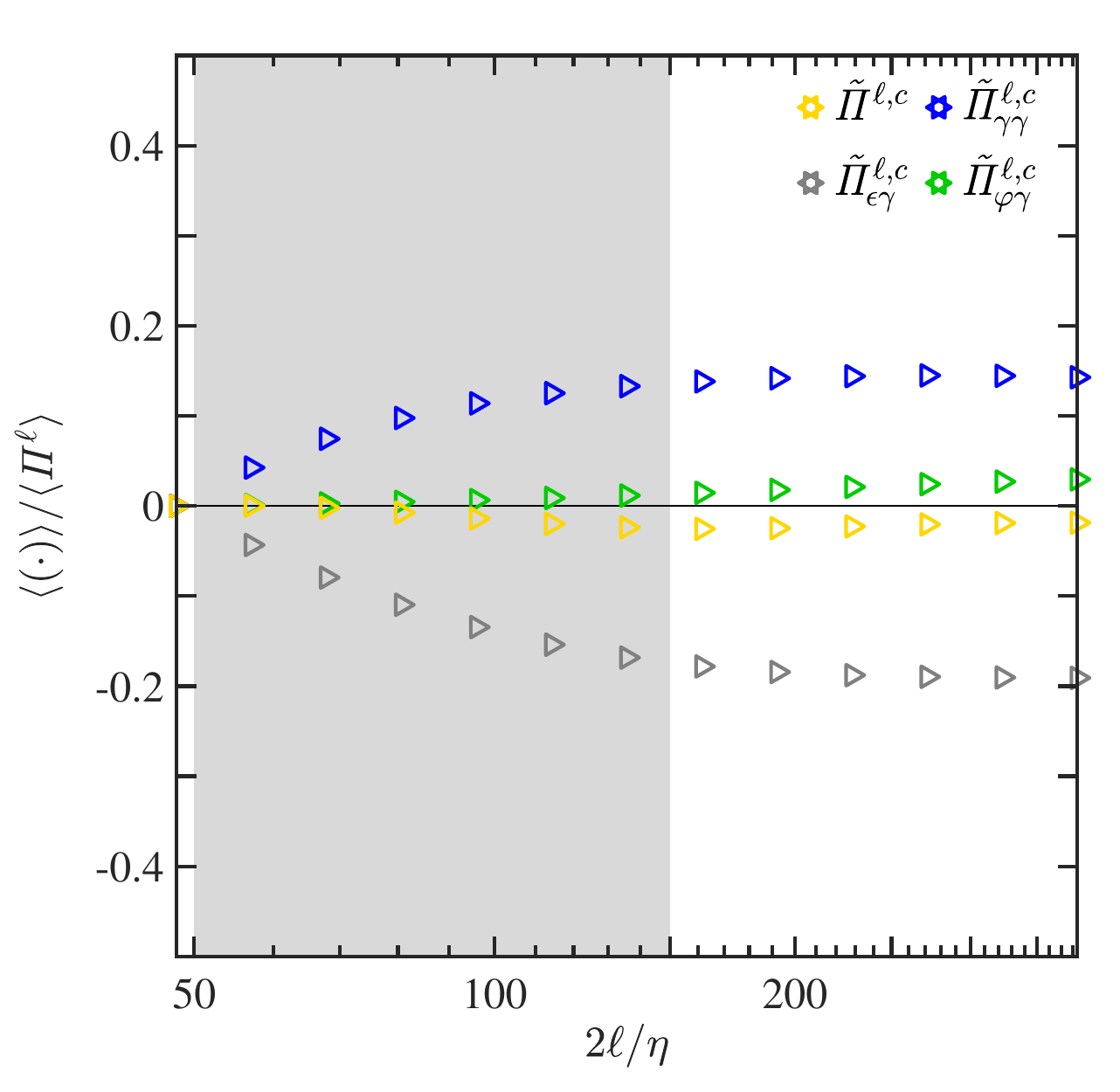}{trim=0pt 10pt 0pt 20pt,clip}
    \subfigimg[width=0.49\linewidth,pos=ul,vsep=5.5pt,hsep=1.5pt]{(\textit{c})}{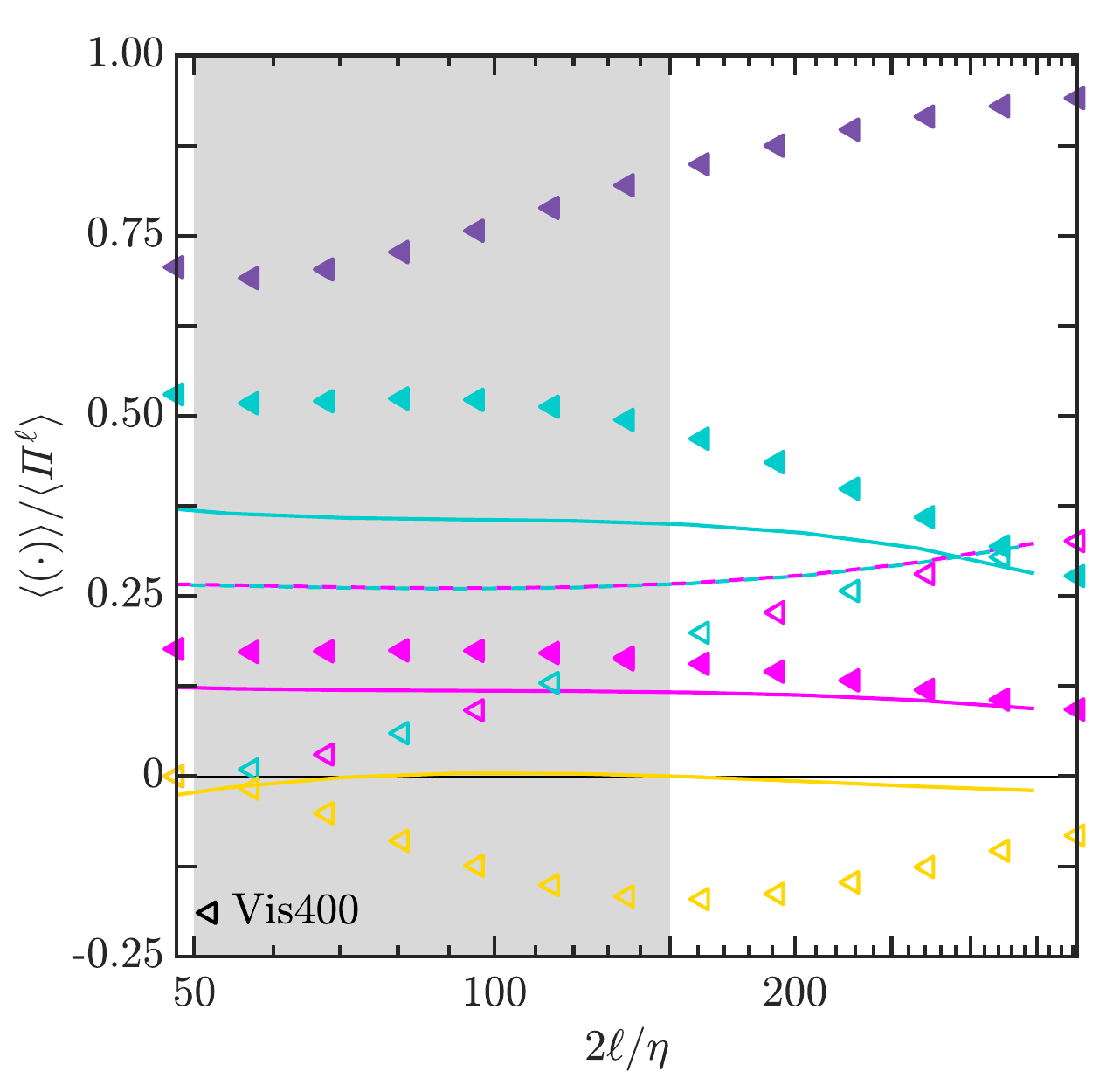}{trim=0pt 10pt 0pt 20pt,clip}
    \subfigimg[width=0.49\linewidth,pos=ul,vsep=5.5pt,hsep=1.5pt]{(\textit{d})}{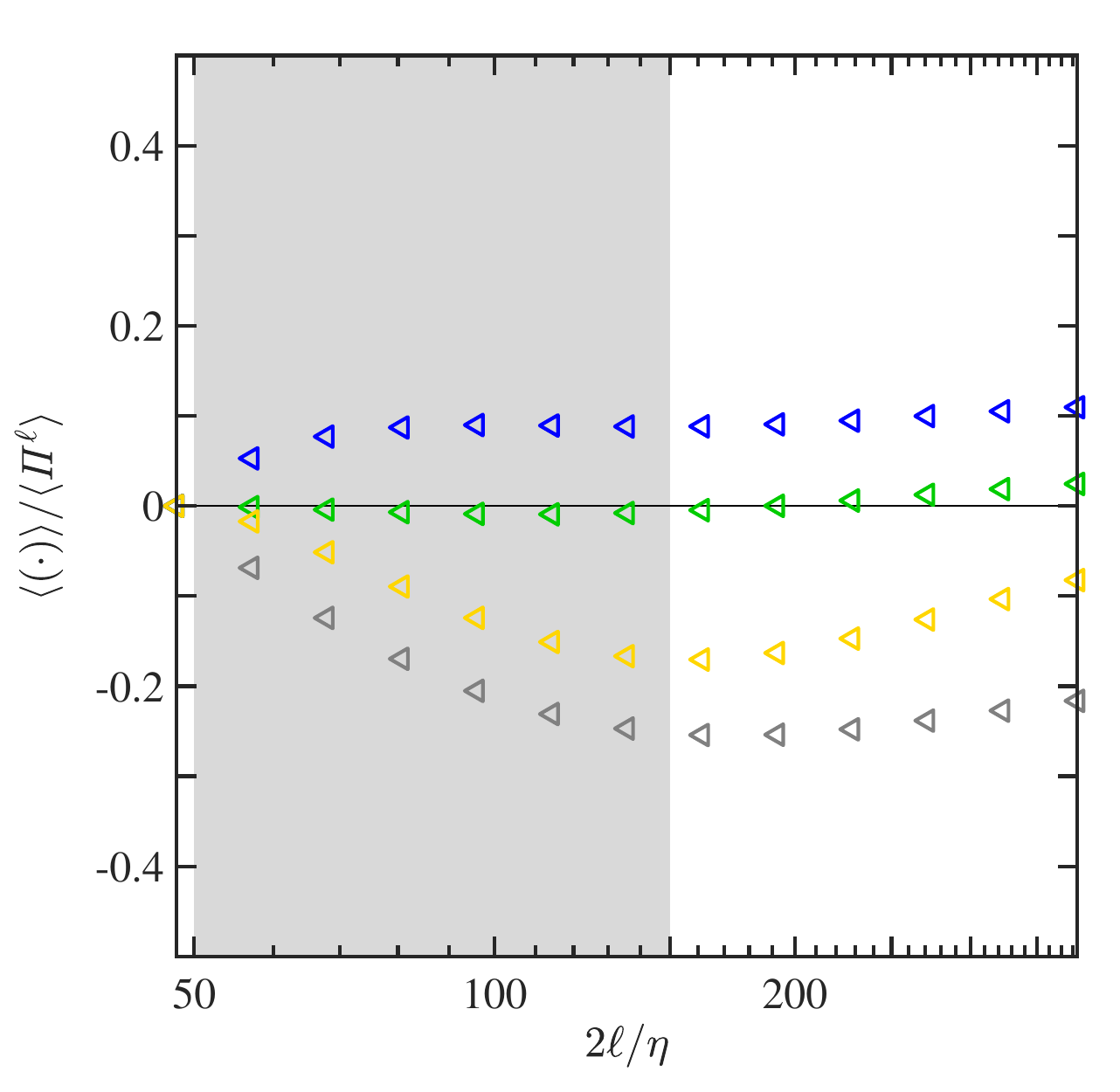}{trim=0pt 10pt 0pt 20pt,clip}
    \subfigimg[width=0.49\linewidth,pos=ul,vsep=5.5pt,hsep=1.5pt]{(\textit{e})}{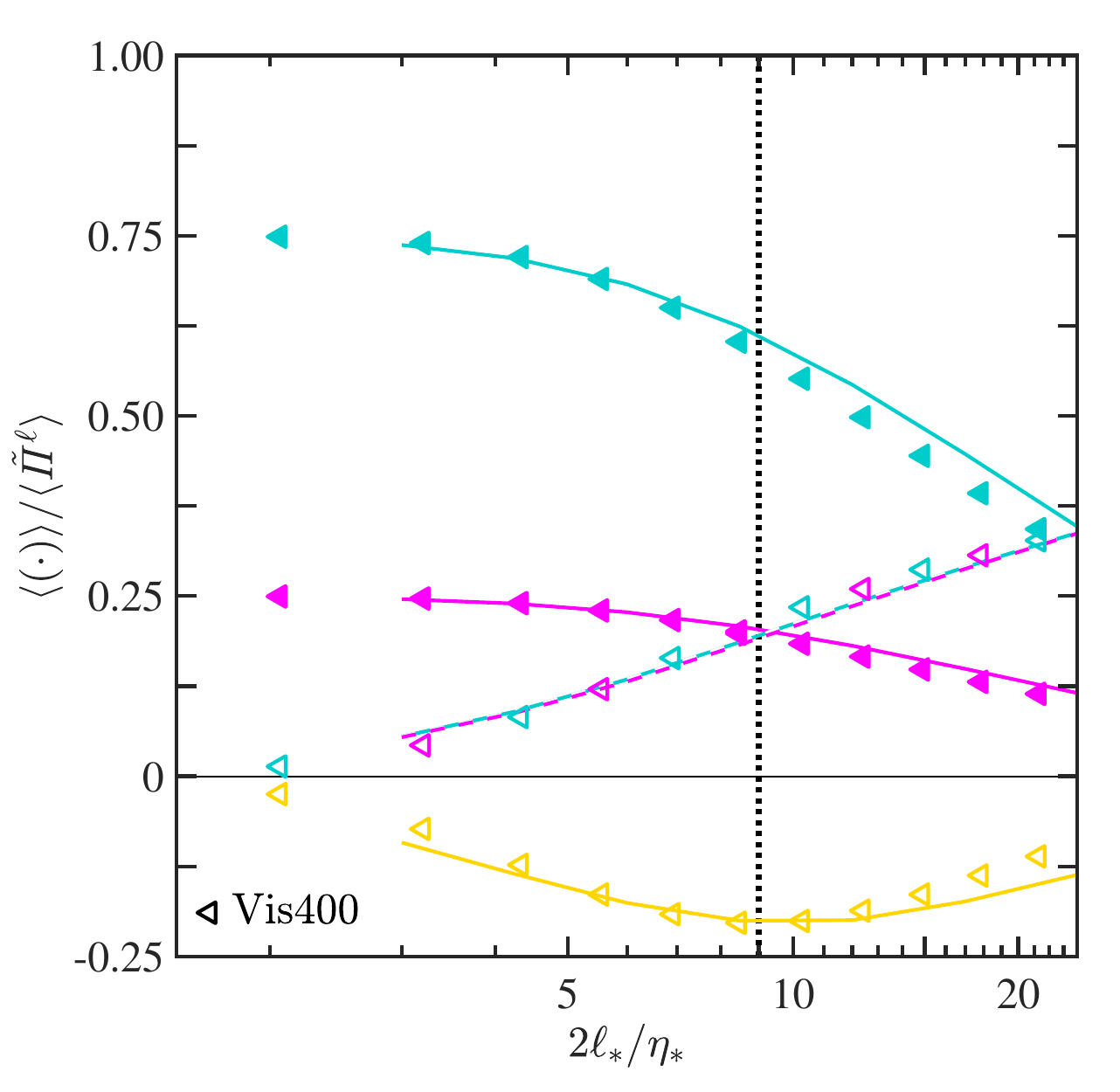}{trim=0pt 10pt 0pt 20pt,clip}
    \subfigimg[width=0.49\linewidth,pos=ul,vsep=5.5pt,hsep=1.5pt]{(\textit{f})}{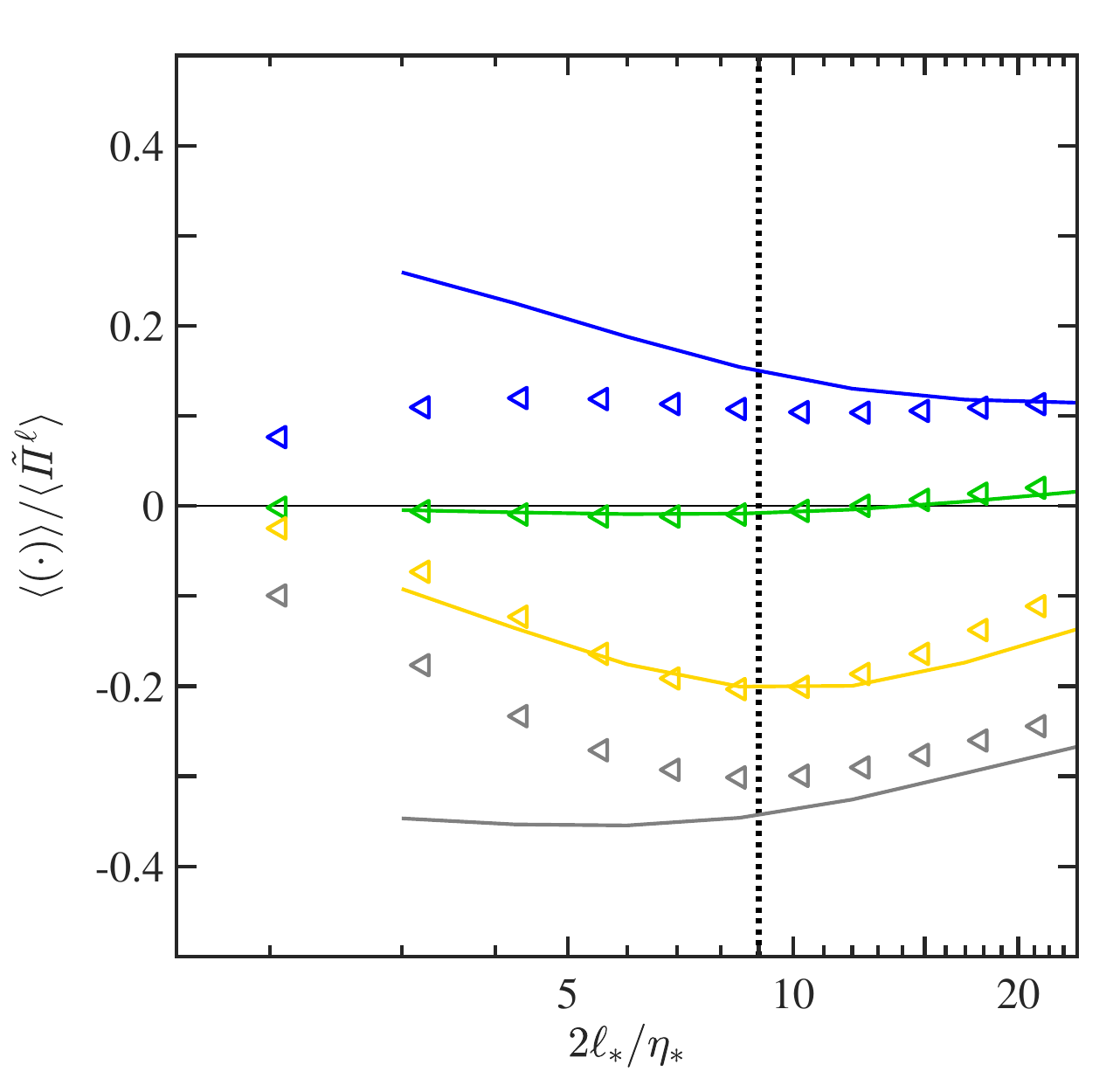}{trim=0pt 10pt 0pt 20pt,clip}
    \caption{(\textit{a},\textit{c}) Resolved symmetry-based scale-local and scale-\copyedit{non-local} cascade rates for Mix400 (\textit{a}) and Vis400 (\textit{c}) overlaid on curves that represent the DNS results of \citet{Joh2020,Joh2021} in the same style as \cref{fig:spectra}(\textit{a}). (\textit{b},\textit{d}) Normality-based contributions to $\tilde{\iPi}^{\ell,c}$ for Mix400 (\textit{b}) and Vis400 (\textit{d}). Panels (\textit{e}) and (\textit{f}) replot the Vis400 results from panels (\textit{c}) and (\textit{d}), respectively, normalized by $\left\langle \tilde{\iPi}^\ell \right\rangle$ and as a function of $2\ell_*/\eta_*$, where $\eta_* \approx 15\eta$. The curves (\textit{e},\textit{f}) represent a DNS at $Re_\lambda \approx 61$, which has a Kolmogorov scale of approximately $\eta_*$. The shaded regions approximate the inertial range and the vertical dotted lines represent $\delta_\gamma = 9\eta$.}
    \label{fig:partCascLES}
\end{figure}

\autoref{fig:partCascLES} shows the resolved symmetry-based cascade rates alongside the normality-based decomposition of the resolved strain--vorticity covariance term for the LES cases. For Mix400, the scale-local cascade rates associated with \copyedit{SS} and \copyedit{VS} are reasonably consistent with those of an analogously filtered DNS. This reflects that the scale-local terms at scales $\ell \geq \ell_{LES}$ are fully resolved and it is consistent with the partitioning results in \cref{fig:partFiltLES}(\textit{a}). By contrast, the corresponding scale-\copyedit{non-local} cascade rates do not match the filtered DNS and account for nearly all of the unresolved net energy transfer, $\left\langle \iPi^\ell \right\rangle - \left\langle\tilde{\iPi}^\ell \right\rangle$. There is virtually no backscatter associated with the resolved contribution from the strain--vorticity covariance term. This is consistent with the inertial range DNS results and reflects that the negative contribution from $\tilde{\iPi}^{\ell,c}_{\epsilon\gamma}$ is roughly balanced by the positive contribution from $\tilde{\iPi}^{\ell,c}_{\gamma\gamma}$. Therefore, the lack of a pronounced bottleneck effect for the mixed model is associated with the absence of backscatter from the $\tilde{\iPi}^{\ell,c}$ term \copyedit{that}, in particular, reflects the \copyedit{behaviour} of the shear layer terms.

For Vis400, the scale-local cascade rates are significantly stronger than those of the filtered DNS. Furthermore, the resolved cascade rate associated with the strain--vorticity covariance term produces significantly more backscatter than the negligible contribution observed for the filtered DNS. This backscatter reflects that the negative contribution from the $\tilde{\iPi}^{\ell,c}_{\epsilon\gamma}$ term significantly outweighs the positive contribution from the $\tilde{\iPi}^{\ell,c}_{\gamma\gamma}$ term, consistent with the DNS results in the subinertial range, as shown in \cref{fig:partCascDNS}(\textit{c}). Therefore, the pronounced artificial bottleneck effect produced by the eddy viscosity model is associated with backscatter from the $\tilde{\iPi}^{\ell,c}$ term \copyedit{that}, in this case, reflects that the resolved shear layer terms behave similarly to the shear layer terms in the subinertial range for DNS.

As was done for the partitioning in \cref{fig:partFiltLES}(\textit{b}), the resolved cascade rates for Vis400 can be interpreted as if they were produced by an unfiltered DNS at a lower Reynolds number. \autoref{fig:partCascLES}(\textit{e},\textit{f}) shows these cascade rates normalized by the total resolved cascade rate, $\left\langle \tilde{\iPi}^\ell \right\rangle$, instead of the total cascade rate, $\left\langle \iPi^\ell \right\rangle$, and plotted as a function of $2\ell_*/\eta_*$ instead of $2\ell/\eta$. Consistent with \cref{fig:partFiltLES}(\textit{b}), we select $\eta_* \approx 15\eta$ and compare these results \copyedit{with} those from a DNS at $Re_\lambda \approx 61$, which has a Kolmogorov scale of approximately $\eta_*$. For the symmetry-based results in \cref{fig:partCascLES}(\textit{e}), there is a remarkable collapse between the replotted Vis400 cascade rates and the $Re_\lambda \approx 61$ cascade rates, except at very large scales, where the effects of the forcing and the number of snapshots may be relevant. The fact that both the partitioning statistics and the energy transfer statistics for Vis400 collapse well for the same $\eta_*$ provides solid quantitative evidence that the eddy viscosity model behaves like an unfiltered DNS at a lower Reynolds number.

\begin{figure}
    \centering
    \includegraphics[width=0.49\textwidth]{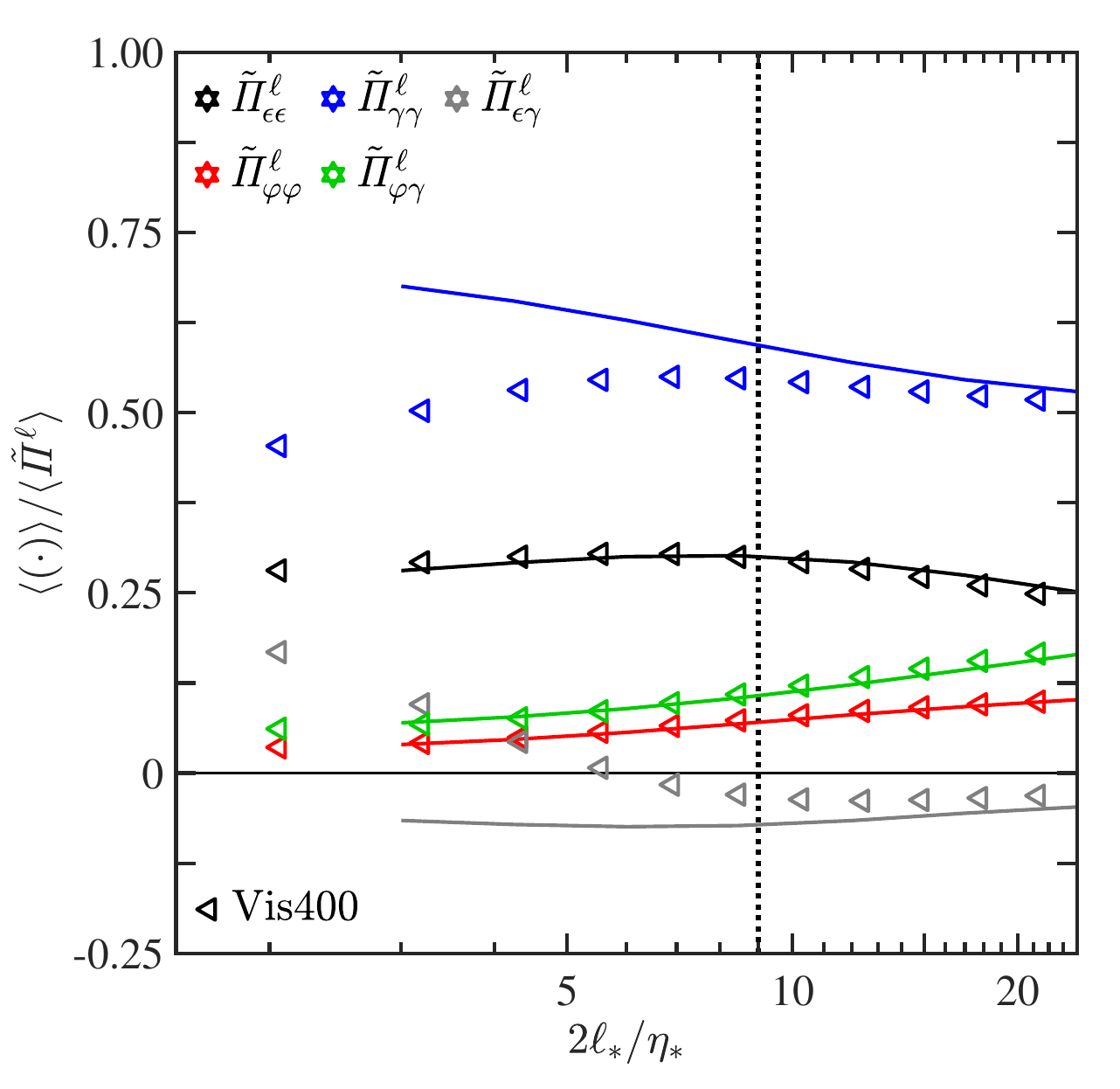}
    \caption{Resolved normality-based cascade rates for Vis400, plotted in the same style as \cref{fig:partCascLES}(\textit{e},\textit{f}).}
    \label{fig:partCascLES_nor}
\end{figure}

The collapse for the strain--vorticity covariance term highlights that the backscatter responsible for the artificial bottleneck effect in LES can be interpreted similarly to the backscatter responsible for the true bottleneck effect in DNS. Specifically, as shown in \cref{fig:partCascLES}(\textit{f}), both bottleneck effects are almost entirely the result of contributions from shear vorticity (i.e. shear layers). However, close to the LES filter width (i.e. at small $\ell_*$), the contributions of $\tilde{\iPi}^{\ell,c}_{\epsilon\gamma}$ and $\tilde{\iPi}^{\ell,c}_{\gamma\gamma}$ for Vis400 depart from those of $\iPi^{\ell,c}_{\epsilon\gamma}$ and $\iPi^{\ell,c}_{\gamma\gamma}$ for the $Re_\lambda \approx 61$ DNS. More generally, as shown in \cref{fig:partCascLES_nor}, the contributions from the cascade rates associated with shearing and the interaction between normal straining and shearing are the only ones that exhibit this departure. All other normality-based cascade rates collapse extremely well onto the $Re_\lambda \approx 61$ DNS results. We hypothesize that these differences reflect the imprint of the dynamic variations in the eddy viscosity field on the resolved motions since, if the eddy viscosity were truly constant, we would expect the eddy viscosity simulation results to collapse exactly onto those of an unfiltered DNS at a lower Reynolds number. Hence, while we observe remarkable similarities between these cases, the dynamic model does produce subtle differences that are appreciable at scales sufficiently close to the LES filter width. Despite these subtle differences, the striking similarity in the shear-layer origins of the true bottleneck effect in DNS and the artificial bottleneck effect produced by the eddy viscosity LES case provides insight that may aid the design of improved closure models.

\section{Concluding remarks}\label{sec:conc}

We have employed a normality-based analysis of filtered velocity gradients to identify the multiscale imprints of normal straining, pure shearing and rigid rotation in turbulent flows. Our analysis of Gaussian-filtered velocity gradients obtained from DNS data representing forced isotropic turbulence refines symmetry-based approaches for characterizing the structures and mechanisms underlying the energy cascade. Our concurrent analysis of resolved velocity gradients obtained from LES data characterizes how well closure models, including eddy viscosity and mixed models, capture these flow features.

Our normality-based approach provides a principled framework for distinguishing tube-like vortex cores, which are associated with rigid rotation, from sheet-like shear layers, which are associated with shearing. Partitioning multiscale velocity gradients based on this approach reveals that filtering induces key structural changes in the subinertial range of the cascade. Filtering mitigates the relative contribution of shearing in this range by smoothing transverse velocity gradients across shear layers. This effect persists until the filter width exceeds the empirical thickness of small-scale shear layers, whereafter the partitioning is relatively constant as a function of scale, including in the inertial range. The variations we observe would be obscured by an analogous symmetry-based analysis since the average strengths of the strain-rate and vorticity tensors are equipartitioned at all scales in homogeneous turbulence.

Our interscale energy transfer analysis shows that, in the subinertial and inertial ranges of the cascade, the forward energy transfer across a given scale is driven primarily by the straining of multiscale shear layers. The \copyedit{SS} associated with multiscale normal straining and the \copyedit{VS} associated with multiscale rigid rotation and shear--rotation interactions also contribute significantly to forward energy transfer. Our scale-local energy transfer analysis highlights that the normal straining of shear layers (i.e. shear layer stretching) is responsible for the majority of scale-local \copyedit{SS} and \copyedit{VS} at all scales considered. This suggests that supplementing prototypical models involving stretched vortex tubes (e.g. Burgers vortex tubes) with contributions from stretched shear layers (e.g. Burgers vortex layers) would more realistically capture the structure of energy transfer and velocity gradient amplification in turbulent flows.

Previous symmetry-based analyses \citep{Joh2020,Joh2021} pinpoint multiscale strain--vorticity covariance as a source of net backscatter that contributes to the bottleneck effect in the subinertial range of the energy cascade. Our analysis refines this notion by revealing that the net backscatter is almost entirely attributed to contributions from shear vorticity (i.e. shear layers). In particular, it is primarily associated with the interaction between normal straining and shear vorticity, which is tempered by the interaction between shear straining and shear vorticity. By contrast, the interaction between shear straining and rigid rotation, which can be associated with vortex tubes, provides a negligible net contribution. When combined with the fact that the backscatter is most significant at the scale associated with the thickness of small-scale shear layers, these results imply that shear layers are critical to the bottleneck effect. 

Broadly speaking, our partitioning and energy transfer analyses for the LES cases highlight that the mixed model effectively reproduces the statistical and structural flow features of an appropriately filtered DNS. By contrast, the eddy viscosity model instead reproduces flow features associated with an unfiltered DNS at a lower Reynolds number. Using the present (dynamic) eddy viscosity model, an LES at $Re_\lambda \approx 400$ reproduces the statistics associated with a DNS at $Re_\lambda \approx 61$, which has a Kolmogorov scale roughly \copyedit{15} times larger. In conjunction with the enhanced shearing at small scales identified by our partitioning analysis, this \copyedit{behaviour} helps explain why the eddy viscosity model produces prominent shear layer structures in the inertial range. It also reveals that the pronounced artificial bottleneck effect produced by this model in the inertial range has the same (shear layer) origins as the true bottleneck effect in the subinertial range for the DNS cases. Altogether, our LES analyses highlight the superior ability of the mixed model to mimic a filtered DNS. More generally, they provide a promising framework to assess (and potentially formulate) LES closure models. 

The partitioning and energy transfer statistics that we compute for forced isotropic turbulence exhibit excellent collapse at relatively high $Re_\lambda$. Given related analyses of unfiltered velocity gradients \citep{Aru2024,Aru2024b}, it is reasonable to expect a similar collapse for the statistics at small filter scales in appropriate regions of inhomogeneous flows. Characterizing how (and at what scale) inhomogeneities induce differences from the flow features we observe would provide insight into how turbulence is sustained in a broader class of flows. \rev{Beyond computing velocity gradient statistics, identifying precisely how flow structures contribute to those statistics would refine our understanding of the `recipe' for turbulence. In the spirit of \citet{She2024}, Burgers vortex layers and tubes of various sizes and strengths might be versatile building blocks for this task.} Our normality-based approach may also help identify unique imprints of flow structures that contribute to other vector gradients. For example, the ability to identify sheet-like structures may provide insight into the structures responsible for the current sheet thinning mechanism that dominates energy transfer in the inertial range of magnetohydrodynamic turbulence \citep{Cap2025}. Beyond identifying flow structures, characterizing the dynamical evolution of these structures (even empirically) would clarify their roles in fundamental processes that sustain turbulent flows. For example, a time-resolved analysis could clarify how shear layers contribute to the bottleneck effect and whether that is consistent with the vortex thinning hypothesis \citep{Joh2021}. Finally, our approach may help guide the development of more effective turbulence models. For example, it would be interesting to develop a model in the spirit of the stretched vortex subgrid-scale model \citep{Mis1997} that also incorporates contributions from subgrid-scale shear layers.


\begin{bmhead}[Funding.]
R.A. was supported by the Department of Defense (DoD) through the National Defense Science \copyedit{and} Engineering Graduate (NDSEG) Fellowship Program. This material is based upon work supported by the National Science Foundation under \copyedit{grant no.} CBET-2152373.
\end{bmhead}

\begin{bmhead}[Declaration of interests.]
The authors report no conflict of interest.
\end{bmhead}

\appendix

\section{Normality-based analysis of Burgers vortex layers and tubes}\label{sec:app:Burgers}

We apply the normality-based analysis of the VGT to the Burgers vortex layer and the Burgers vortex tube to provide insight into the structural organization of normal straining, pure shearing and rigid rotation in canonical vortical flows.

The Burgers vortex layer represents a canonical stretched shear layer. Its velocity field, $(u_x,u_y,u_z)$, in Cartesian coordinates, $(x,y,z)$, is given by
\begin{equation}\label{eq:Burges_layer_vel}
    u_x = -\alpha x, \quad u_y = U {\rm erf}\left( \frac{\sqrt{\alpha} x}{\sqrt{2\nu}} \right), \quad u_z = \alpha z,
\end{equation}
where $\alpha > 0$ is a strain rate parameter and $2U = u_y(x \to +\infty) - u_y(x \to -\infty)$ is the vortex layer strength (representing the circulation per unit length). The non-dimensional VGT for this flow is given by
\begin{equation}\label{eq:Burges_layer_VGT}
    \frac{1}{\alpha}\mathsfbi{A} = \begin{bmatrix} -1 & 0 & 0 \\ \frac{Re_U}{\sqrt{\pi}}{\rm exp}\left( -x_*^2 \right) & 0 & 0 \\ 0 & 0 & 1 \end{bmatrix},
\end{equation}
where $x_* = x\sqrt{\alpha/2\nu}$ is the non-dimensional transverse coordinate and $Re_U = U\sqrt{2/\alpha\nu}$ is the shear layer Reynolds number.

The Burgers vortex tube represents a canonical stretched vortex tube. Its velocity field, $(u_r,u_\vartheta,u_z)$, in cylindrical coordinates, ($r,\vartheta,z)$, is given  by
\begin{equation}\label{eq:Burges_vortex_vel}
    u_r = -\alpha r, \quad u_\vartheta = \frac{\mathit{\Gamma}}{2 \pi r}\left[ 1 - {\rm exp}\left( -\frac{\alpha r^2}{2\nu} \right) \right], \quad u_z = 2 \alpha z,
\end{equation}
where $\alpha > 0$ is again a strain rate parameter and $\mathit{\Gamma} > 0$ represents the circulation. The non-dimensional VGT for this flow is given by
\begin{equation}\label{eq:Burges_vortex_VGT}
    \frac{1}{\alpha}\mathsfbi{A} = \begin{bmatrix} -1 & -\frac{Re_\mathit{\Gamma}}{4 \pi r_*^2}\left[ 1 - {\rm exp}\left( -r_*^2 \right) \right] & 0 \\ -\frac{Re_\mathit{\Gamma}}{4 \pi r_*^2} \left[ 1 - {\rm exp}\left( -r_*^2 \right) \right] + \frac{Re_\mathit{\Gamma}}{2\pi}{\rm exp}\left( -r_*^2 \right) & -1 & 0 \\ 0 & 0 & 2 \end{bmatrix},
\end{equation}
in cylindrical coordinates, where $r_* = r\sqrt{\alpha/2\nu}$ is the non-dimensional radial coordinate and $Re_\mathit{\Gamma} = \mathit{\Gamma}/\nu$ is the circulation Reynolds number.

\begin{figure}
    \centering
    \subfigimg[width=0.49\linewidth,pos=ur,vsep=16pt,hsep=9pt]{(\textit{a})}{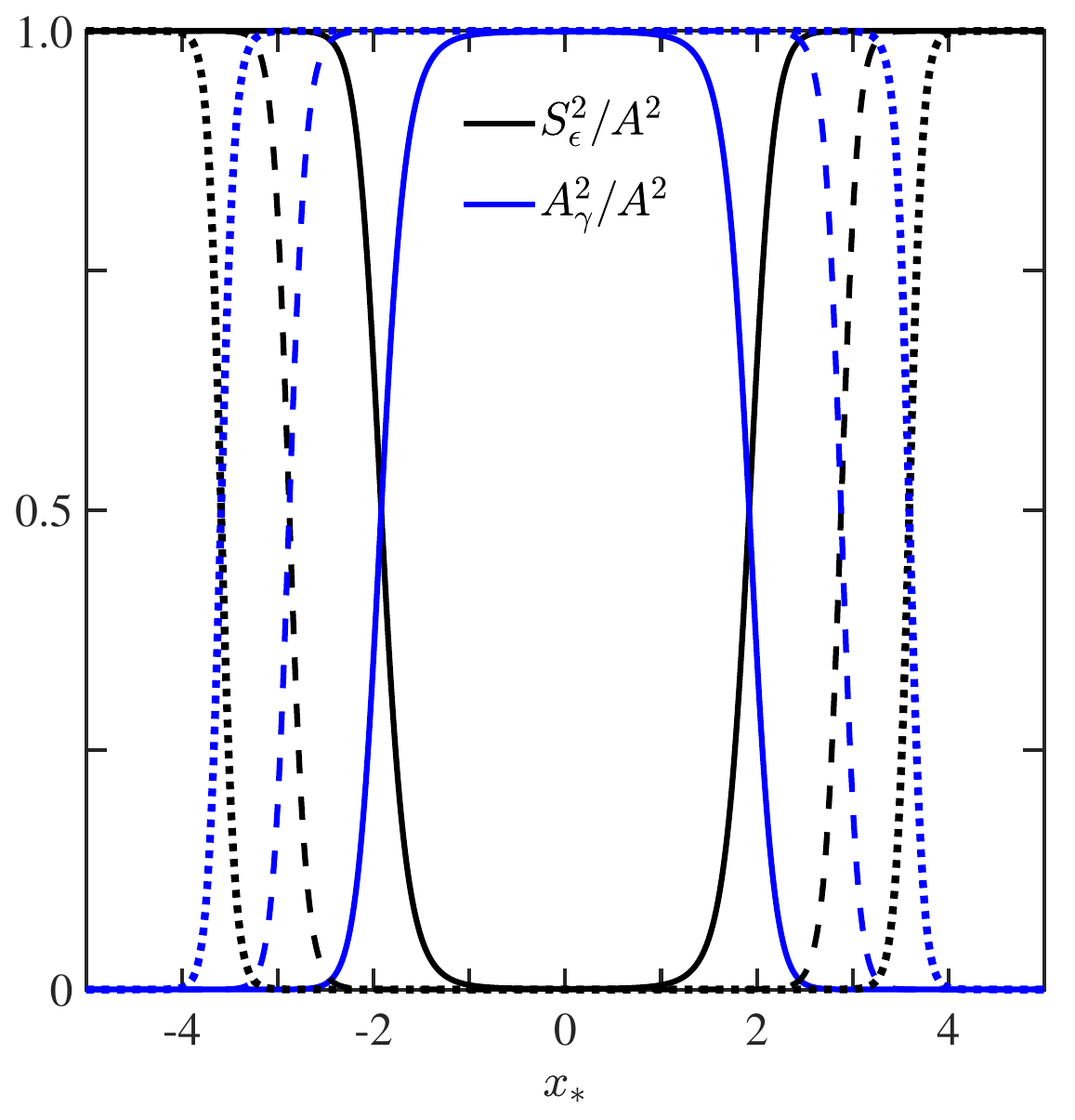}{}
    \\
    \subfigimg[width=0.49\linewidth,pos=ur,vsep=16pt,hsep=9pt]{(\textit{b})}{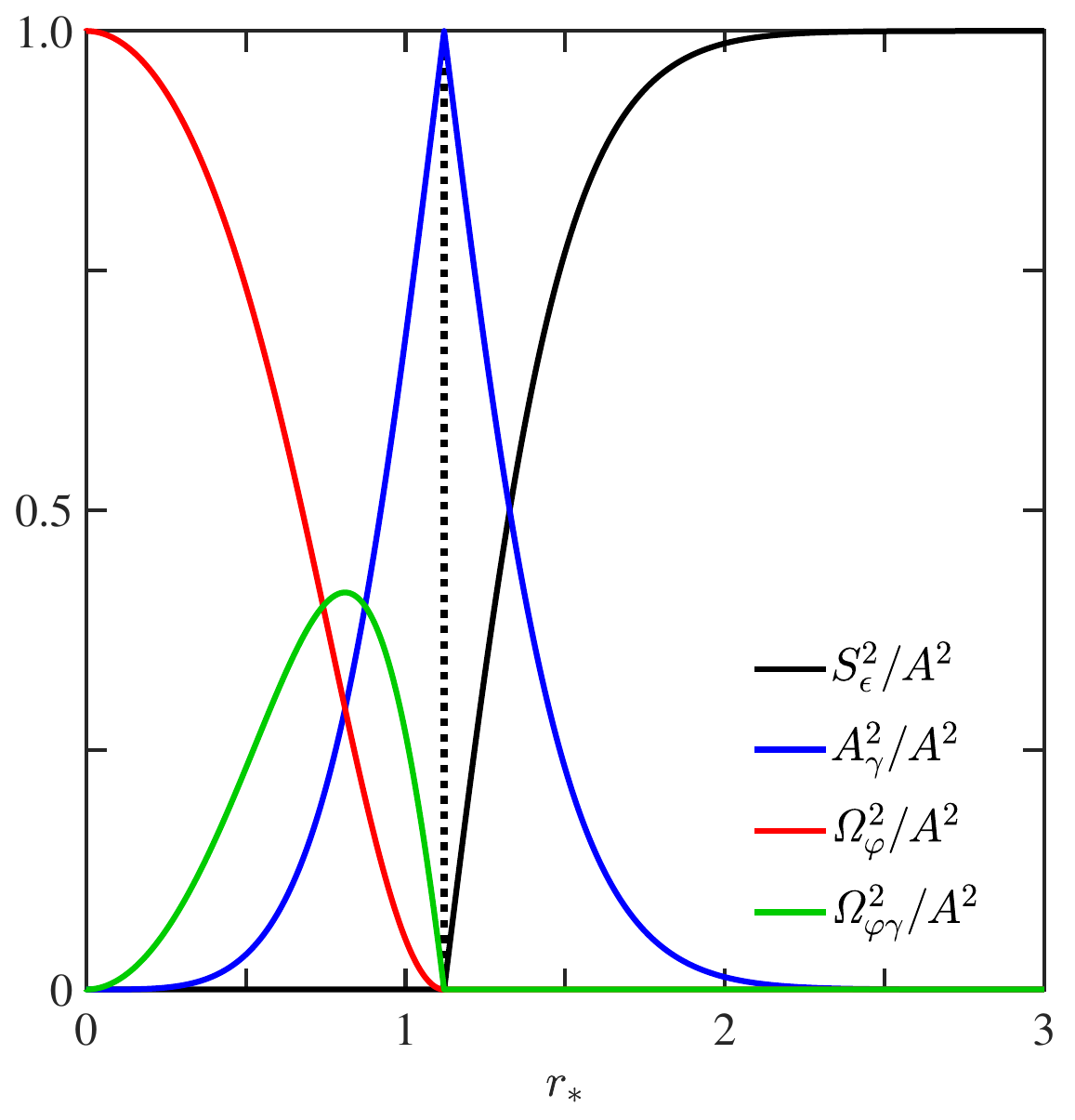}{}
    \subfigimg[width=0.49\linewidth,pos=ur,vsep=16pt,hsep=9pt]{(\textit{c})}{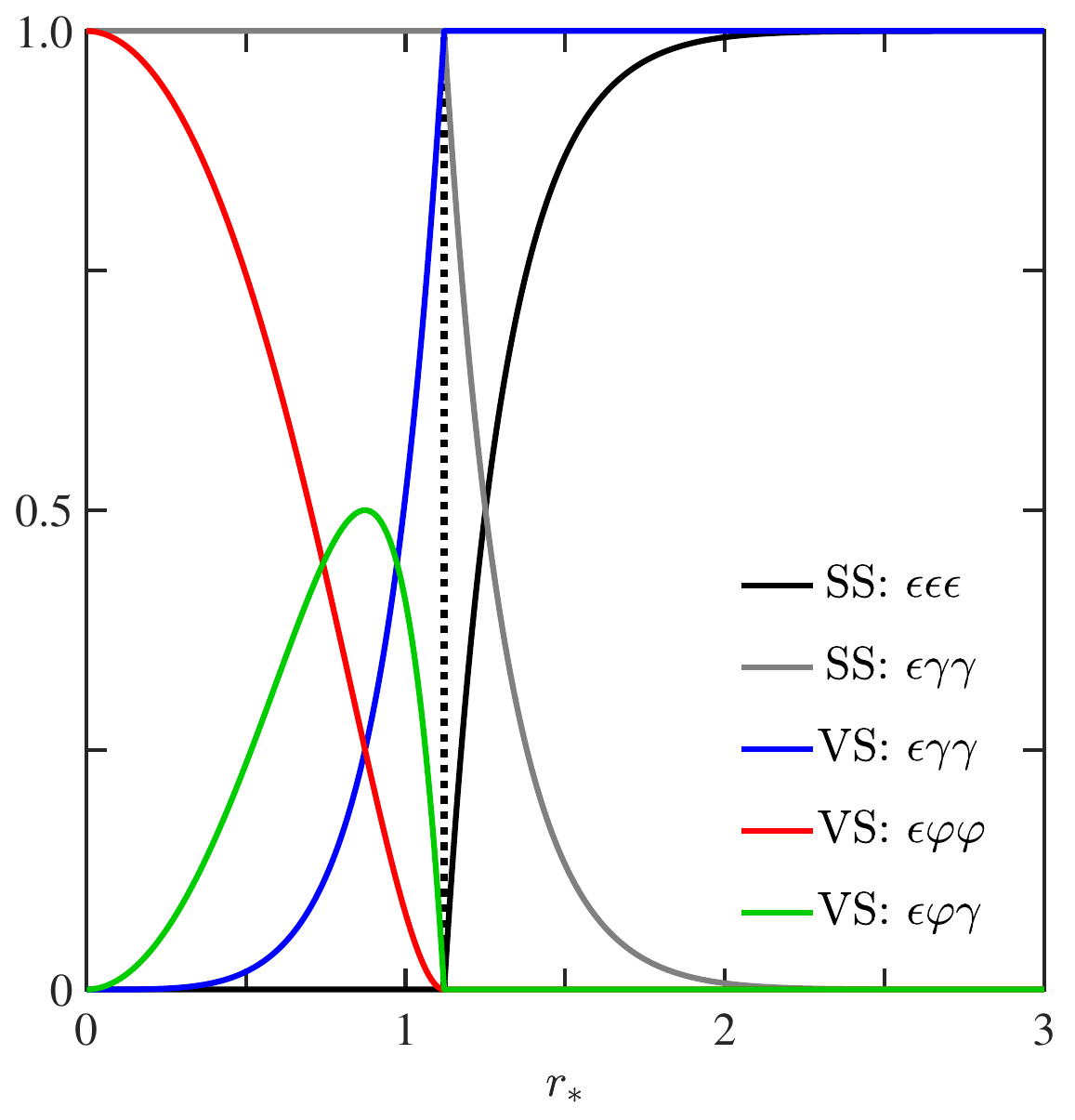}{}
    \caption{(\textit{a}) Velocity gradient partitioning for Burgers vortex layers at $Re_U = 10^2$ (solid), $Re_U = 10^4$ (dashed) and $Re_U = 10^6$ (dotted). (\textit{b}) Velocity gradient partitioning for the Burgers vortex tube in the limit of $Re_\mathit{\Gamma} \to \infty$. (\textit{c}) Normality-based contributions to \copyedit{SS} and \copyedit{VS} for the Burgers vortex tube from (\textit{b}). The vertical dotted lines (\textit{b},\textit{c}) represent the vortex boundary as identified by $Q = 0$ and $\mathit{\Delta} = 0$, where $Q$ is the second invariant of the VGT and $\mathit{\Delta}$ is the discriminant of the VGT.}
    \label{fig:partBurgers}
\end{figure}

\autoref{fig:partBurgers}(\textit{a}) shows the velocity gradient partitioning for Burgers vortex layers at various values of $Re_U$. Rigid rotation and shear--rotation correlations do not contribute to this flow since the eigenvalues of the VGT are real at all spatial locations. As expected, the partitioning is dominated by shearing inside the shear layer and normal straining far from the shear layer. The far-field \copyedit{behaviour} reflects that normal straining, which is purely symmetric, must dominate velocity gradients in regions where the vorticity is relatively weak (i.e. in nearly potential flow regimes). The value of $x_*$ where the partitioning transitions from the shearing regime to the normal straining regime increases with increasing $Re_U$, and it has been shown that $Re_U \lesssim 100$ for typical small-scale shear layers \citep{Wat2020}. Regardless of $Re_U$, \copyedit{SS} and \copyedit{VS} (not shown) are entirely associated with the $\epsilon\gamma\gamma$ terms (i.e. $-3\A{S}{\epsilon\vphantom{\gamma}}{ij}\A{S}{\gamma}{ik}\A{S}{\gamma}{jk}$ and $-\A{S}{\epsilon\vphantom{\gamma}}{ij}\A{\iW}{\gamma}{ik}\A{\iW}{\gamma}{jk}$) at all spatial locations. As shown in \textsection \ref{sec:results:energy:DNS}, these mechanisms provide the strongest relative contributions to scale-local \copyedit{VS} and \copyedit{SS} throughout the cascade.

\autoref{fig:partBurgers}(\textit{b}) shows the velocity gradient partitioning for a Burgers vortex tube in the limit of $Re_\mathit{\Gamma} \to \infty$. The inner core of this Burgers vortex tube is dominated by rigid rotation, whereas the region near the vortex boundary is dominated by pure shearing. This highlights that the vortex can be crudely described as a rigidly rotating inner core with a shear annulus wrapped around it. The shear--rotation correlations are strongest between the core and boundary regions, which is consistent with previous observations of vortex structures \citep{Aru2024}. As observed for the Burgers vortex layer, the velocity gradients are dominated by normal straining in the nearly potential flow regime far from the vortex core. At the vortex boundary, the partitioning profiles are continuous but not differentiable. This reflects that the eigenvalues of the VGT are not differentiable at this location, where they transition from a regime with one real eigenvalue and a pair of complex conjugate eigenvalues to a regime with three real eigenvalues. 

\autoref{fig:partBurgers}(\textit{c}) shows the relative contributions of normality-based \copyedit{SS} and \copyedit{VS} mechanisms for the Burgers vortex tube. The contributions of the $\epsilon\varphi\varphi$ and $\epsilon\varphi\gamma$ \copyedit{VS} terms are constrained to within the vortex boundary. They highlight that the `vortex core stretching' term is primarily associated with the inner vortex core and the shear--rotation interaction term is associated with the region between the core and the boundary. By contrast, the $\epsilon\gamma\gamma$ \copyedit{VS} term, which represents `shear layer stretching,' is primarily associated with the shear annulus near the vortex boundary and the region outside the vortex boundary. In the nearly potential flow regime far from the vortex core, \copyedit{SS} is negative and dominated by the $\epsilon\epsilon\epsilon$ (normal straining) term. However, within the vortex boundary, it is positive and dominated by the $\epsilon\gamma\gamma$ term in the limit of $Re_\mathit{\Gamma} \to \infty$.

\section{Collapse of the multiscale velocity gradient partitioning}\label{sec:app:collapse}

We \copyedit{analyse} the multiscale velocity gradient partitioning for various simulations of forced isotropic turbulence to identify how key features of the partitioning collapse in terms of $Re_\lambda$ and $\ell$. In addition to the simulations discussed in \cref{tab:sims}, we consider DNS cases at $Re_\lambda \approx 61$, $100$ and $160$. These simulations are conducted on grids of size $N_x^3 = 64^3$, $128^3$ and $256^3$ and have spatial resolutions of $k_{max}\eta \approx 1.3$, $1.4$ and $1.3$, respectively. Each of these supplementary DNS cases consists of 68 temporal snapshots that are spaced one large-eddy turnover time apart.

\begin{figure}
    \centering
    \subfigimg[width=0.49\linewidth,pos=ul,vsep=7pt,hsep=0pt]{(\textit{a})}{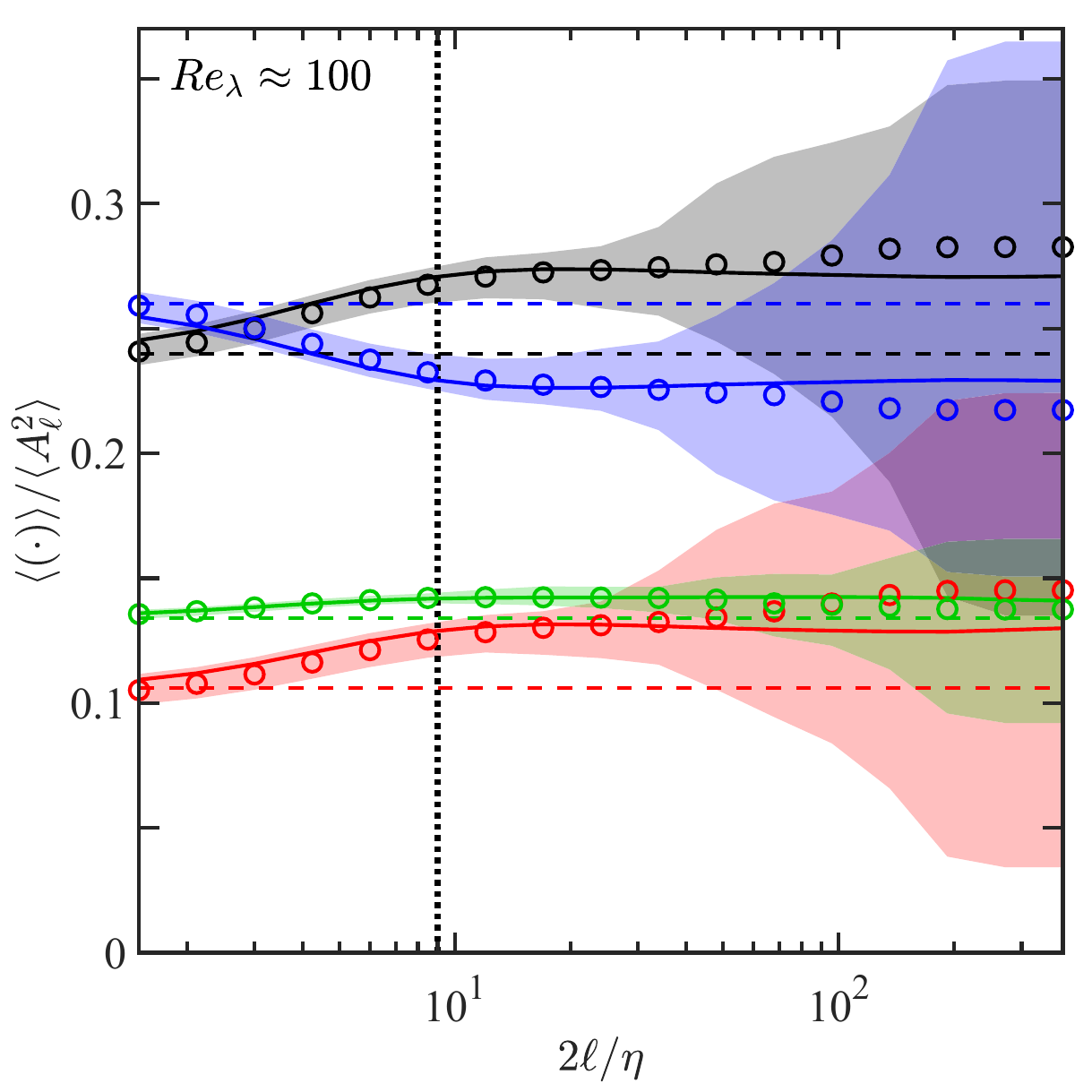}{}
    \subfigimg[width=0.49\linewidth,pos=ul,vsep=7pt,hsep=0pt]{(\textit{b})}{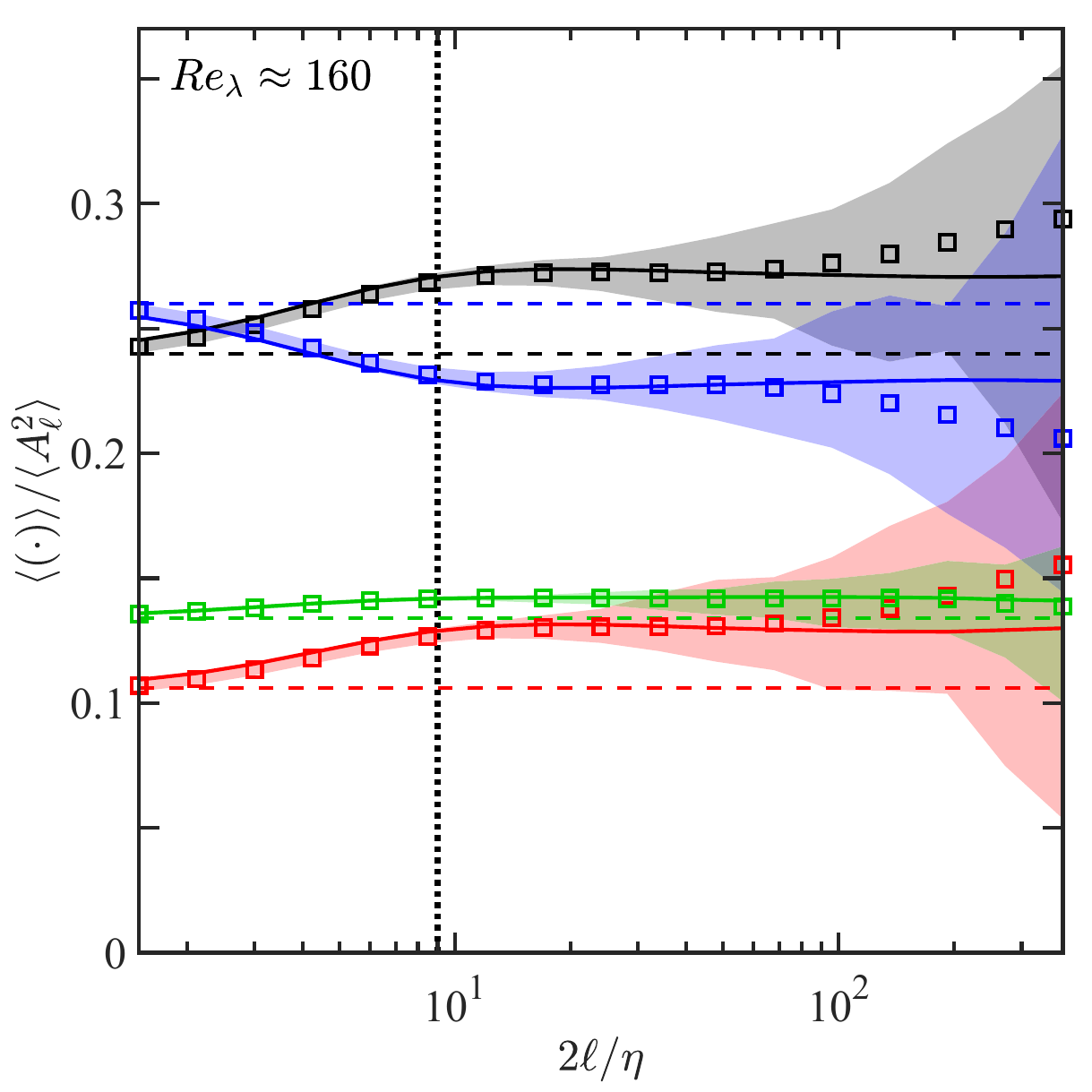}{}
    \subfigimg[width=0.49\linewidth,pos=ul,vsep=7pt,hsep=0pt]{(\textit{c})}{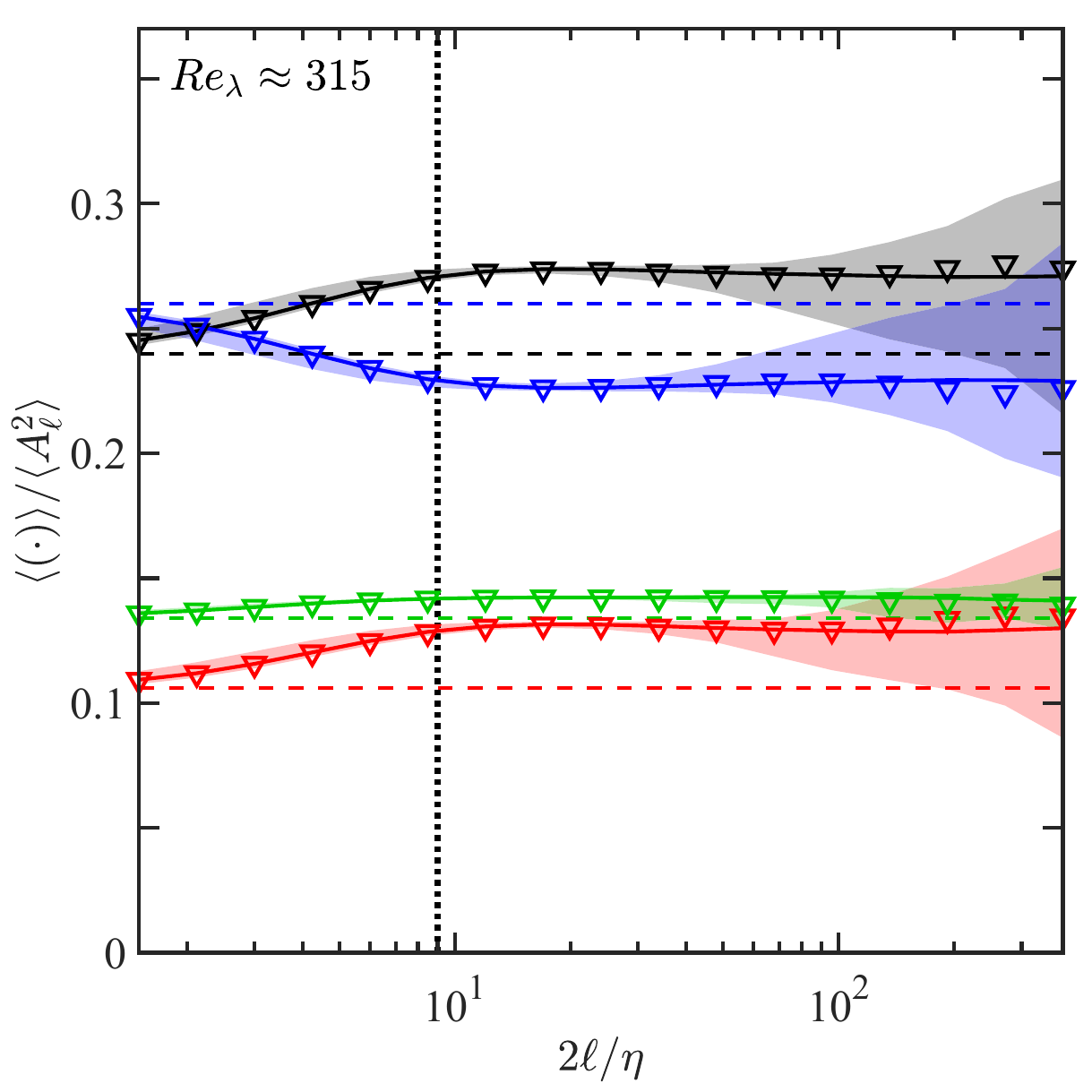}{}
    \subfigimg[width=0.49\linewidth,pos=ul,vsep=7pt,hsep=0pt]{(\textit{d})}{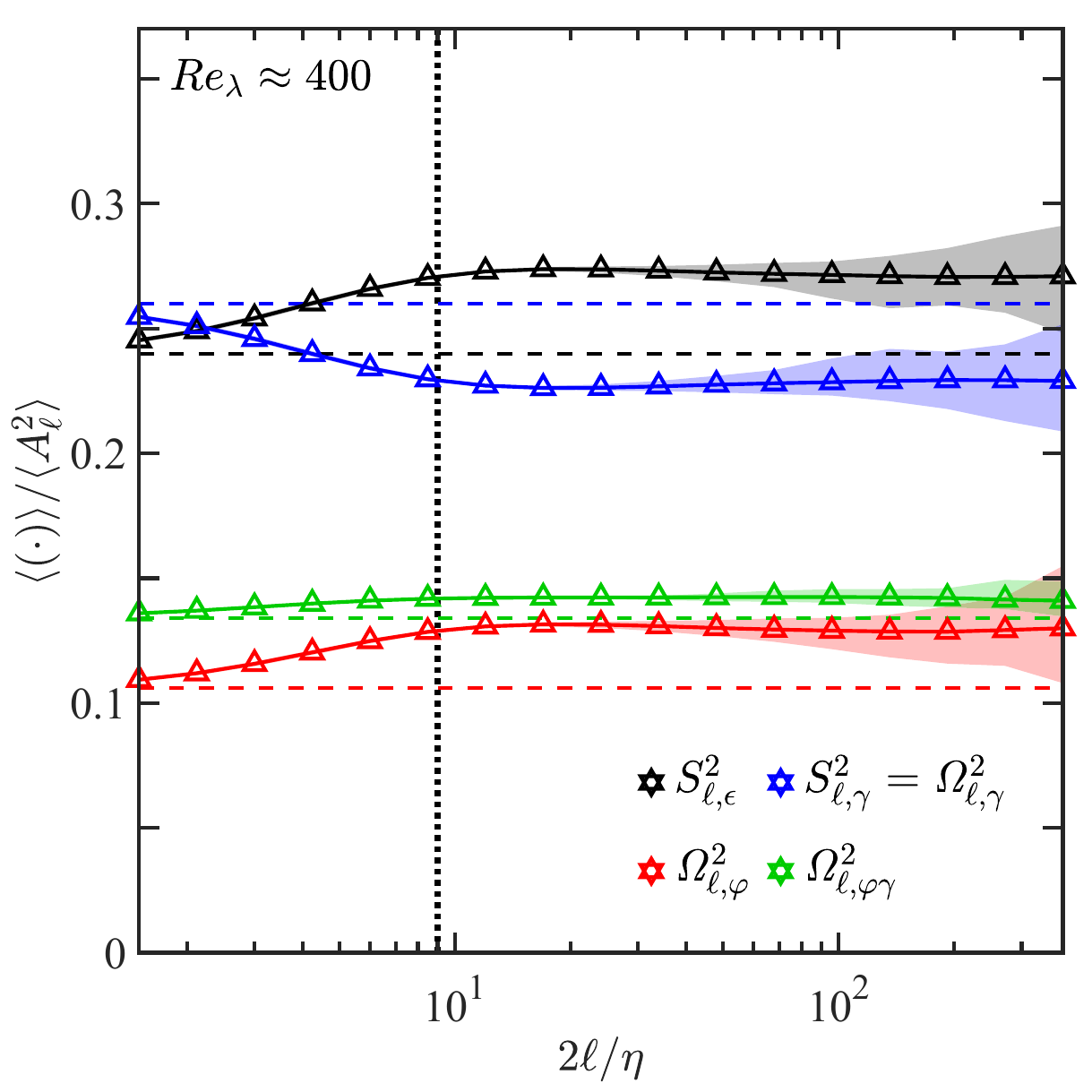}{}
    \caption{(\textit{a}) Partitioning of filtered velocity gradients for DNS cases at various Reynolds numbers. The symbols represent averages over all snapshots and the shading represents ranges for single-snapshot averages. In all panels, the solid curves represent the averaged partitioning at $Re_\lambda \approx 400$, the horizontal dashed lines represent the unfiltered partitioning in the high-$Re_\lambda$ limit, and the vertical dotted lines represent the typical thickness of small-scale shear layers, $\delta_\gamma = 9\eta$.}
    \label{fig:partFiltDNS_collapse}
\end{figure}

\autoref{fig:partFiltDNS_collapse} shows how the partitioning varies as a function of filter width for $100 \lesssim Re_\lambda \lesssim 400$. It shows that the multiscale partitioning statistics collapse at sufficiently high values of $Re_\lambda$. The collapse at small scales is consistent with previous findings that the unfiltered partitioning collapses well for $Re_\lambda \gtrsim 200$ \citep{Das2020}. \autoref{tab:collapse} further shows that the unfiltered partitioning statistics collapse to within $\pm 0.001$ for $Re_\lambda \gtrsim 315$. The collapse extends to larger scales as $Re_\lambda$ increases and, for $Re_\lambda \gtrsim 315$, it extends well into the inertial range (to within $\pm 0.002$), where the partitioning is also relatively insensitive to scale. The shaded regions in \cref{fig:partFiltDNS_collapse} capture the extent to which the spatially averaged partitioning statistics vary in time for each simulation. For all cases, the magnitude of these variations decreases as $\ell$ decreases since the degree of scale separation from the forcing increases. This scale separation, combined with an increasing number of grid points, also explains why the variations at a given scale become smaller as $Re_\lambda$ increases. For the $Re_\lambda \approx 315$ and $400$ simulations, which correspond to DNS315 and DNS400 in \cref{tab:sims}, our results suggest that a single snapshot is sufficient to obtain converged velocity gradient statistics in the range $0 \leq 2\ell/\eta \leq 67.9$. This helps justify our use of two snapshots to compute the interscale energy transfer statistics in that range of scales for DNS315 and DNS400 (see \textsection \ref{sec:results:energy}).

\begin{table}
\begin{center}
\def~{\hphantom{0}}
\begin{tabular}{lccccccc}
Type & Model & $Re_\lambda$ & $2\ell/\eta$ & $\frac{\big\langle S^2_{\ell,\epsilon} \big\rangle}{\big\langle A_\ell^2 \big\rangle}$ & $\frac{\big\langle A^2_{\ell,\gamma} \big\rangle}{\big\langle A_\ell^2 \big\rangle}$ & $\frac{\big\langle \iW^2_{\ell,\varphi} \big\rangle}{\big\langle A_\ell^2 \big\rangle}$ & $\frac{\big\langle \iW^2_{\ell,\varphi\gamma} \big\rangle}{\big\langle A_\ell^2 \big\rangle}$ \\[0pt]
       &                 &          &     &       &       &       &       \\[-4pt]
Random & \citet{Pop2000} & $400$    & $-$ & 0.279 & 0.442 & 0.139 & 0.140 \\
DNS    & $-$             & $\;\,$61 &   0 & 0.235 & 0.531 & 0.100 & 0.134 \\
DNS    & $-$             &      100 &   0 & 0.236 & 0.528 & 0.102 & 0.134 \\
DNS    & $-$             &      160 &   0 & 0.238 & 0.524 & 0.104 & 0.134 \\
DNS    & $-$             &      315 &   0 & 0.239 & 0.521 & 0.106 & 0.134 \\
DNS    & $-$             &      400 &   0 & 0.240 & 0.520 & 0.106 & 0.134 \\
DNS    & $-$             &      315 &  24 & 0.274 & 0.453 & 0.131 & 0.142 \\
DNS    & $-$             &      315 &  48 & 0.272 & 0.456 & 0.130 & 0.142 \\
DNS    & $-$             &      315 &  96 & 0.271 & 0.458 & 0.129 & 0.142 \\
DNS    & $-$             &      400 &  24 & 0.274 & 0.453 & 0.131 & 0.142 \\
DNS    & $-$             &      400 &  48 & 0.273 & 0.455 & 0.130 & 0.142 \\
DNS    & $-$             &      400 &  96 & 0.272 & 0.457 & 0.129 & 0.142 \\
LES    & Eddy viscosity  &      400 &  24 & 0.234 & 0.532 & 0.100 & 0.134 \\
LES    & Eddy viscosity  &      400 &  48 & 0.233 & 0.534 & 0.099 & 0.134 \\
LES    & Eddy viscosity  &      400 &  96 & 0.234 & 0.533 & 0.099 & 0.134 \\
LES    & Mixed           &      400 &  24 & 0.267 & 0.466 & 0.125 & 0.142 \\
LES    & Mixed           &      400 &  48 & 0.267 & 0.465 & 0.125 & 0.143 \\
LES    & Mixed           &      400 &  96 & 0.268 & 0.464 & 0.125 & 0.143 \\
\end{tabular}
\captionsetup{format=hang}
\caption{Partitioning statistics produced by random velocity gradients and various DNS and LES cases. Each LES case represents a separate simulation where $\ell = \ell_{LES}$ represents the LES filter width.}
\label{tab:collapse}
\end{center}
\end{table}

\autoref{tab:collapse} also provides a complementary view to the multiscale partitioning analysis for the LES cases in \textsection \ref{sec:results:VGT:LES}. Here, instead of plotting the multiscale partitioning statistics for a single LES, we consider the partitioning at the LES filter scale, $\ell_{LES}$, and vary $\ell_{LES}$ across three different simulations. These simulations employ LES filter widths of $2\ell_{LES}/\eta = 24$, $48$ and $96$ and have grid sizes of $N_x^3 = 256^3$, $128^3$ and $64^3$, respectively. They each have a spatial resolution of $k_{max}\ell_{LES} \approx 3.0$, which is comparable to a DNS resolution of $k_{max}\eta \approx 1.5$ \citep{Kam2024}. Each of these supplementary LES cases consists of 68 temporal snapshots that are spaced half of a large-eddy turnover time apart. For all values of $\ell_{LES}$, the partitioning statistics for the eddy viscosity simulations resemble those of the unfiltered DNS at $Re_\lambda \approx 61$, which has a Kolmogorov scale roughly \copyedit{15} times larger. Further, the partitioning statistics for the mixed model simulations resemble those of the $Re_\lambda \approx 400$ DNS filtered at scale $\ell_{LES}$ and, as such, are relatively insensitive to $\ell_{LES}$. These results support our main conclusions regarding the \copyedit{behaviour} of the eddy viscosity and mixed models over a broad range of LES filter widths. Although not shown, our conclusions are robust to different model formulations, including those considered by \citet{Kam2024}.

Finally, we compare the partitioning statistics \copyedit{with} those \copyedit{produced by} random velocity gradients. The random case consists of 20 snapshots of size $N_x^3 = 1024^3$, where each snapshot represents a synthetic, divergence-free velocity field constructed from Fourier modes with Gaussian-random complex weights. These weights are rescaled to obey the model energy spectrum described in section 6.5.3 of \citet{Pop2000} with a Kolmogorov constant of $C = 1.5$ and a viscous roll-off parameter of $\beta = 5.2$. Interestingly, the inertial range partitioning statistics are more similar to those associated with the random case than they are to the unfiltered partitioning statistics. Moreover, the partitioning statistics for the random case, which are constant to within $\pm 0.002$ for $0 \leq 2\ell/\eta \leq 384$, do not capture the enhanced relative contribution of shearing in the subinertial range. This suggests that the imprint of viscous-scale shear layers can be viewed as a defining feature of turbulence associated with the incompressible Navier--Stokes equations. The imprint may be encoded in the phases of the Fourier modes of the velocity field, which are uncorrelated for the random case.


\bibliographystyle{jfm}
\bibliography{references}

\end{document}